\documentclass{article}

\PassOptionsToPackage{numbers}{natbib}

\usepackage[main,final]{iaseai26}

\usepackage[utf8]{inputenc} % allow utf-8 input
\usepackage[T1]{fontenc}    % use 8-bit T1 fonts
\usepackage{booktabs}       % professional-quality tables
\usepackage{nicefrac}       % compact symbols for 1/2, etc.
\usepackage{microtype}      % microtypography

\usepackage{graphicx}   % for \includegraphics and \scalebox
\usepackage{adjustbox}  % optional but useful if you later use width=1.05\textwidth

\usepackage{amsmath,amssymb,amsfonts}
\usepackage{algorithmic}
\usepackage{graphicx}
\usepackage{url}
\usepackage{textcomp}
\usepackage{tabularx} 
\usepackage{colortbl}
\usepackage{xfp}
\usepackage{svg}
\usepackage[font=footnotesize]{subcaption}
\usepackage{pdflscape}

\usepackage{enumitem}
\usepackage{hyperref}

\newcommand{\ro}[1]{\textcolor{black}{#1}}

\usepackage{cellspace}
\definecolor{blue10}{rgb}{0.1,0.2,0.8}
\definecolor{blue20}{rgb}{0.2,0.4,0.8}
\definecolor{blue30}{rgb}{0.3,0.6,0.8}
\definecolor{blue40}{rgb}{0.4,0.8,0.8}
\definecolor{blue50}{rgb}{0.5,1.0,1.0}
\definecolor{red10}{rgb}{0.8,0.1,0.1}
\definecolor{red20}{rgb}{0.8,0.2,0.2}
\definecolor{red30}{rgb}{0.8,0.3,0.3}
\definecolor{red40}{rgb}{0.8,0.4,0.4}
\definecolor{red50}{rgb}{0.8,0.5,0.5}
\newcommand{\coloredcell}[2]{
    \pgfmathsetmacro{\lightness}{min(max(10 + 80*abs(#1), 10), 90)} % Calculate lightness from 10 to 90
    \ifdim #1 pt > 0 pt
        \cellcolor{blue\roundedlightness} #2 % Apply blue shades for positive values
    \else
        \cellcolor{red\roundedlightness} #2  % Apply red shades for negative values
    \fi
}

\newcommand{\cellvaluerho}[2]{\celltextp{#2}\colortone{\thetone{\multconst}{#1}}$#1$ ($#2$)}

\newcommand{\cellvaluetau}[3]{\celltextp{#3}\colortone{\thetone{\multconst}{#1}}$#2$ ($#3$)}

\title{\LARGE \bf
Why Automate This? Exploring Correlations Between Desire for Robotic Automation, Invested Time and Well-Being
}

\author{%
Ruchira Ray\thanks{Now at the University of Edinburgh.}\\
University of Texas at Austin \\
\texttt{ruchiraray@utexas.edu}
\And
Leona Pang \\
University of Texas at Austin \\
\texttt{leonapang@utexas.edu}
\And
Sanjana Srivastava \\
Stanford University \\
\texttt{sanjana2@stanford.edu}
\And
Li Fei-Fei \\
Stanford University \\
\texttt{feifeili@stanford.edu}
\And
Samantha Shorey \\
University of Pittsburgh \\
\texttt{samshorey@pitt.edu}
\And
Roberto Mart{\'i}n-Mart{\'i}n \\
University of Texas at Austin \\
\texttt{robertomm@utexas.edu}
}

\begin{document}

\maketitle
\thispagestyle{empty}
\pagestyle{empty}

%%%%%%%%%%%%%%%%%%%%%%%%%%%%%%%%%%%%%%%%%%%%%%%%%%%%%%%%%%%%%%%%%%%%%%%%%%%%%%%%
\begin{abstract}

Understanding the motivations underlying the human inclination to automate tasks is vital for developing robots that fit seamlessly into daily life. Accordingly, we ask: are individuals more inclined to automate activities based on the time they consume or the feelings experienced while performing them? This study explores these preferences and whether they vary across social groups, specifically gender category and income level. Leveraging data from the BEHAVIOR-1K dataset, the American Time-Use Survey, and the American Time-Use Survey Well-Being Module, we investigate the relationship between the desire for robot automation, time spent, and associated feelings: Happiness, Meaningfulness, Sadness, Painfulness, Stressfulness, or Tiredness. Our key findings show that, despite common assumptions, time spent on activities does not strongly predict automation preferences; instead, happiness and pain are the strongest indicators. We also identify differences by gender and economic level: Women prefer to automate stressful activities, whereas men prefer to automate those that make them unhappy; mid-income individuals prioritize automating less enjoyable and meaningful activities, while low and high-income show no significant correlations. We hope our research helps motivate the design of robots that align with user priorities, moving domestic robotics toward more socially relevant solutions. All data and an interactive tool are publicly available at \url{https://robin-lab.cs.utexas.edu/why-automate-this/}.

\end{abstract}

%%%%%%%%%%%%%%%%%%%%%%%%%%%%%%%%%%%%%%%%%%%%%%%%%%%%%%%%%%%%%%%%%%%%%%%%%%%%%%%%
\section{Introduction}
\label{s:intro}
% Robots offer a future in which people are alleviated from the labor of physical activities that they routinely perform but would ideally prefer not to do. However, it is unclear which domestic activities should be automated and why. In workplaces, the procurement of automated technologies is often a top-down decision made by management based primarily on economical factors. The home is instead a domain where technology adoption is driven by the choices of everyday consumers who will alleviate their own duties. 

\begin{figure}
    \centering
    \includegraphics[width=0.52\linewidth]{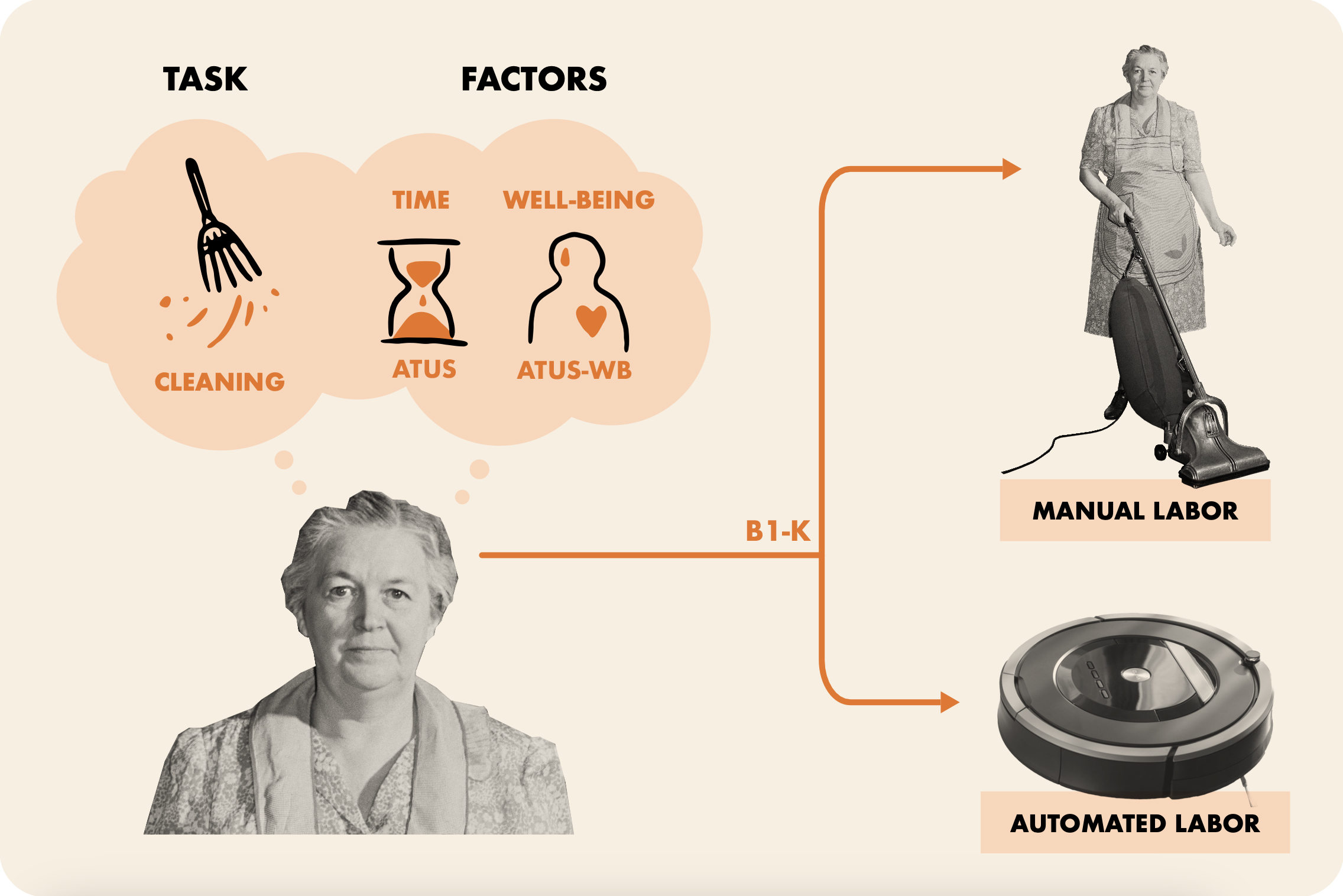}
 \caption{\textbf{What drives the desire for robotic automation?} Do people want to perform everyday household tasks themselves, or do they want a robot to do it? What motivates the decision to delegate tasks: a desire to save time or to avoid unpleasant experiences? To answer this question, we perform a statistical analysis with three data sources, B1K~\cite{li2023behavior}, ATUS~\cite{atus} and ATUS-WB~\cite{atus-wb}.(Figure by Helen Tseng. Original image sources:  
 \cite{miltsova2018robotic, rothstein1942wiegel})}
 \label{fig:pull}
    \vspace{-1em}
\end{figure}

Robots offer a future in which people are alleviated from the labor of physical activities that they routinely perform, but would ideally prefer not to do. Unlike workplace automation, which is often driven by economic factors, the adoption of technology at home depends on consumer choices. 
To align household robotics research with user needs, Li et al.~\cite{li2023behavior} established BEHAVIOR-1K, a benchmark that includes a ranked list of the activities people want robots to perform for them. The benchmark reveals \textit{which} tasks people most want automated, not \textit{why}: Are these tasks time-consuming or do they evoke unpleasant feelings (Fig.~\ref{fig:pull})?

% Seeking to align household robotics research efforts with the needs and desires of users, pioneering work by Li et al.~\cite{li2023behavior} established BEHAVIOR-1K, a benchmark that includes a ranked list of the activities people want robots to perform for them. The benchmark reveals \textit{which} activities people most want automated, but not \textit{why} they are preferred options for automation: do people want robots to take over the activities they spend the most time doing or the activities they find the least enjoyable or meaningful (Fig.~\ref{fig:pull})?  

As an initial step towards understanding what motivates automation in the home, we analyze information from three publicly available datasets: 1) the BEHAVIOR-1K (B1K) survey, reporting on the desire for robotic automation based on thousands of individual surveys, 2) the American Time-Use Survey (ATUS~\cite{atus}), a nationally representative survey conducted by the US Department of Labor, reporting on the time Americans spend on different everyday activities, and 3) the ATUS Wellbeing Module (ATUS-WB~\cite{atus-wb}), an additional dataset reporting on the feelings Americans experience when performing ATUS activities. Our contrastive analysis addresses the following research questions:
\begin{itemize}[leftmargin=*]
    \item \textbf{RQ1-} Does the average time spent (T) on an activity predict the desire for robotic automation (DA)? 
    \item \textbf{RQ2-} Which feelings experienced while performing an activity are the strongest predictors of the desire for robotic automation? Happiness (H), Meaningfulness (M), Painfulness (P), Sadness (B), Stressfulness (S), or Tiredness (Z).
\end{itemize}
% \todo{check with hypotheses}

% We also investigate potential differences between gender and income groups, as household robots will automate activities more likely to be performed by women due to social divisions of labor~\cite{federici2013reproduction}. In the United States, women are primarily responsible for the category of activities ATUS defines as ``housework'': tasks like laundry, sweeping the floor, and putting away groceries. On average, American women dedicate twice as much time to these chores and are twice as likely to perform them on any given day compared to men, who are more likely to perform other types of activities such as yard work~\cite{atus}. These delineations reflect long-standing cultural connotations of domestic tasks as ``masculine'' or ``feminine''~\cite{gelber1997yourself} and are starkly evident in the way housework is divided within heterosexual partnerships~\cite{brenan2020women}. Previous research also suggests that gender is not the only relevant trait. In light of this, we seek to understand if or how these differing experiences affect what activities people prioritize for automation:

We also investigate differences between gender and income groups, as household robots will automate tasks more likely to be performed by women due to social divisions of labor~\cite{federici2013reproduction}. In the U.S., women are primarily responsible for the category of activities ATUS defines as ``housework'' (laundry, sweeping the floor, putting away groceries) and spend twice as much time on these chores as men, who more often do other activities like yard work~\cite{atus}. These delineations reflect long-standing cultural connotations of domestic tasks as ``masculine'' or ``feminine''~\cite{gelber1997yourself} that are evident in how housework is divided in heterosexual partnerships\cite{brenan2020women}. American mothers spend roughly the same amount of time on housework regardless of their economic strata~\cite{glynn2018unequal}, however in areas with high-income inequality, differences in time spent among high-earning and low-earning women are more pronounced~\cite{schneider2017income}.

We therefore seek to understand if or how these differing experiences affect what activities people prioritize for automation:
\begin{itemize}[leftmargin=*]
    \item \textbf{RQ3-} Do gender-based differences -- in terms of the average time spent on an activity or the feelings associated with performing an activity -- yield differences in desire for robotic automation? 
    \item \textbf{RQ4-} Do income-based differences -- in terms of the average time spent on an activity or the feelings associated with performing an activity -- yield differences in desire for robotic automation? 
\end{itemize}

% Lastly, we inquire into activities that one group shows a greater desire to automate than others. We explore what we might learn about this activity (in terms of how time-consuming it is or how it impacts well-being) by looking more deeply at the data we have for that demographic group.

% Our study reveals that, even though emerging household technologies are typically marketed as ``time-saving devices''~\cite{bose1984household}, there is no correlation between the average time people invest in an activity and their desire to automate it. However, the happiness experienced during an activity presents a strong (negative) correlation with the desire to automate it. Painfulness of an activity also emerges as a strong positive predictor of the desire for automation for the general population. Interestingly, sadness and tiredness do not predict the general population's desire for automation.

Our study reveals that, while household technologies are often marketed as “time-saving devices”~\cite{bose1984household}, there is no correlation between time spent on an activity and the desire to automate it. Instead, lower happiness during an activity strongly predicts a higher desire for robotic automation, as does painfulness; sadness and tiredness do not. Psychological theories of affective regulation suggest that individuals are motivated to seek positive experiences and avoid negative ones \cite{kahneman1999well,carver2001self,gross2015emotion}. We therefore frame the desire for robotic automation as a form of affective regulation: people may wish to offload activities that are unpleasant, not merely time-consuming.
Examining the data by gender, we find that men prioritize automating activities they are less happy doing, while women prioritize automating activities that are linked to stress. %are more likely to desire automation for activities they are less happy doing, while women favor automating those linked to stress. 
Both men and women prioritize automating activities that they spend comparatively less time on than their counterparts. Further, the mid-income group prioritizes automating activities that are experienced as less meaningful.
%We observe a strong negative correlation between differences in time spent and the desire for robot automation, suggesting that both groups tend to want to automate activities they spend comparatively less time on. Further, for mid-income individuals, the desire for robot automation also correlates negatively with happiness and meaningfulness.

% Examining the data by gender, we find that men tend to exhibit a stronger desire for automation in activities they are less happy doing, whereas women favor automating activities that are associated with stress. We observe a strong negative correlation between differences in time spent and the desire for automation; suggesting that men and women tend to want to automate activities they spend comparatively less time on. 
% Further, income-level analysis reveals that for mid-income individuals, the desire to automate correlates negatively with happiness and meaningfulness.

% Additionally, socioeconomic analysis reveals that for high-income individuals, the ranking of activities based on the desire to automate correlates positively happiness whereas for low-income individuals, it correlates positively with meaningfulness.

Additionally, we open-source all our data, including the B1K survey results, along with an interactive tool for data processing, statistical analysis, and visualization, available on a public website\footnote{\url{https://robin-lab.cs.utexas.edu/why-automate-this/}}. ATUS is a rich source of critical demographic data, yet it is not easy to parse; we hope that our tool lowers the barrier for others to replicate our results and further use it to guide future robotics research.

\textbf{
Statement of Contributions:} This paper presents the first data-driven investigation into why people wish to automate household tasks, analyzing how time use and emotional experience relate to automation preferences across three large datasets: BEHAVIOR-1K, ATUS, and ATUS-WB. Building on prior work that identified what people want robots to do, we provide quantitative analysis exploring the emotional and demographic trends shaping automation preferences, offering insights for designing robots that align with user priorities. All data, code, and an interactive tool for dataset exploration and visualization are made publicly available.

%We also provide a comprehensive literature study about the role of AI in technology broadly and in home robotics, specifically.

% While American mothers of different socioeconomic groups spend roughly the same amount of time on housework regardless of their economic strata~\cite{glynn2018unequal}, in areas with high-income inequality, differences in time spent among high-earning and low-earning women are more pronounced~\cite{schneider2017income}.

\section{Literature Review}
\label{s:lr}
\textit{Automating Household Activities:}
In the absence of artificial general intelligence, the specific task a robot accomplishes is a fundamental design decision -- shaping functionality, behavior, and the parameters of plausible events~\cite{lee2007autonomous}. While routine household activities may seem simple, they are technically complex and require a detailed understanding of task steps. Cakmak \& Takayama's aggregation of digital ``chore-lists'' indicates that cleaning tasks make up nearly half of all domestic chores~\cite{cakmak2013towards}. Cleaning tasks are united by the need to apply a tool to a stationary surface, yet each require different tool-use skills. HRI researchers have taken various approaches to collecting the information to support automation, including crowd-sourcing human explanations of task procedures~\cite{puig2018virtualhome,porfirio2023crowdsourcing} and end-user demonstration~\cite{otero2008human}. 

The survey portion of the benchmark BEHAVIOR-1K~\cite{li2023behavior} finds that participants' highest priorities for automation are nearly all classified as \textit{housework} by the ATUS coding rules, with interior cleaning tasks (such as mopping the floor and scrubbing bathroom fixtures) at the top.  This aligns with prior studies showing cleaning is the most common task people envision for robots~\cite{bugmann2011can,dautenhahn2005robot}. People also tend to prefer robots for hands-on, physical tasks~\cite{takayama2008beyond}. Importantly, these priorities reflect not just time spent on tasks but also their unpleasantness, difficulty, level of dirt, or danger~\cite{sung2009sketching,schneiders2021domestic, kang2020first}. 

% high-priority tasks for automation were characterized not just through the amount of time participants spent doing them but through their unpleasant quality~\cite{sung2009sketching,schneiders2021domestic}. Factors such as the level of dirt, the difficulty, or the danger associated with a task also influence purchasing intentions of household robots, varying between cleaning, cooking, and laundry~\cite{kang2020first}. People also prefer robots over humans for practical, hands-on occupations that involve working with real-world materials~\cite{takayama2008beyond} and indicate criteria that may shape decisions about which tasks to delegate to robots.

\textit{Household Activities as Household Experiences:}
Bell and Kay argue that designing technologies for the home requires pivoting from the dominant, commercial view where household activities are ``simple, beginning-to-end processes''~\cite{kitchenman} and instead accounting for experience. Principles for feminist HRI also calls for moving beyond metrics of efficiency (which drive the development of robots in industrial contexts) towards users' bodily sensations and emotions ~\cite{winkle2023feminist}. We take up these calls by examining how experiences with household activities shape the desire for robotic automation.

\textit{Time Used for Household Activities:}
Decades of studies utilizing time use data, such as ATUS, show that women spend significantly more time on household activities than men~\cite{bianchi2012housework,sayer2005gender}. Further the type of household activities performed by women are often more time-consuming and repetitive, reinforcing the constant demand on women’s time~\cite{budig2004activity}. Women are also more likely to engage in multiple household tasks simultaneously, adding to their cognitive and physical load~\cite{bittman2003does}. Despite women's persistent responsibility for housework across income levels~\cite{hook2017women}, lower-income households face greater burdens due to limited resources to outsource domestic labor and increased time demands, leading to heightened stress for women and men~\cite{aguiar2007measuring,sullivan2013we}.

% Women are expected to maintain their responsibility for housework regardless of their economic status~\cite{hook2017women}. However, higher-income households often have the financial resources to outsource domestic labor through services such as cleaning, cooking, and childcare~\cite{aguiar2007measuring}. Men in lower-income households perform more domestic work than men in higher-income households~\cite{sullivan2013we}. The greater time commitments of lower-income households may exacerbate the stress and burden associated with managing both work and household responsibilities for men and women.

\textit{Impact of Household Robots on Time Use and Well-Being:}
The division of household labor raises important questions about whether the desire for robotic automation may be related to the social assignment of domestic activities. Automation is expected to significantly reduce the time spent on unpaid domestic work~\cite{hertog2023future}. For instance, robotic vacuum cleaners have been shown to decrease time spent on cleaning,~\cite{sung2010domestic} fulfilling the commonly espoused goal of easing domestic burdens. Accordingly, international time use data suggests automation of household chores is likely to have a greater impact on women~\cite{hertog2023future}. Tasks traditionally considered masculine (such as lawn maintenance) see less demand for automation and existing robotic solutions are less frequently adopted~\cite{de2015makes}. 

As Lee and Šabanović~\cite{koreanhome} observe, domestic technologies are shaped by social dynamics, not just functional needs. HRI studies demonstrate that the use of household robots is shaped by gendered, cultural norms around domestic responsibility \textit{and }technical expertise~\cite{koreanhome,sung2008housewives,fink2013living,forlizzi2006service}. Though women traditionally dedicate more time to household tasks such as vacuuming, researchers find that men and children began taking on cleaning roles when robotic vacuums were introduced~\cite{forlizzi2006service,sung2008housewives}. This aligns with broader patterns in which men more often configure digital household technologies~\cite{rode2010roles}. By shifting household dynamics and automating certain tasks, robots may help rebalance aspects of domestic responsibility, the burden of which has social and psychological impacts\footnote{However, these gains in equality (in terms of time) may come at the cost of upholding stereotypes around technical expertise.}~\cite{hochschild2012second}.

% Beyond considering the potential of automation to save people time, scholars who study ‘the future of work’ emphasize the importance of preserving humans’ capacity to perform meaningful tasks that provide significance, fulfillment and a sense of purpose~\cite{bailey2017mismanaged,rosso2010meaning}. Within the domestic context, this perspective opens questions about the aspects of daily life that provide individuals with personal satisfaction or pleasure. Taken together, these bodies of scholarship point to a need for HRI to develop a more in-depth understanding of the desire for automation. What opportunities for increasing (or persevering) well-being are provided by automating (or not automating) specific domestic activities?

Beyond saving time, scholars of the future of work stress the importance of preserving meaningful tasks that provide fulfillment and purpose~\cite{bailey2017mismanaged,rosso2010meaning}. Within the domestic context, this opens questions about which daily activities bring satisfaction. Together, these perspectives highlight the need for HRI to better understand the desire for robotic automation: what opportunities for well-being arise from automating (or not automating) specific domestic tasks?

\section{Methods}
\label{s:m}
% \todo{mention how we got household activities --> how is filtering done}
% Why 
To address our research questions (RQ1-RQ4), we conduct an exploratory, statistical analysis of three publicly available datasets: BEHAVIOR-1K (B1K)~\cite{li2023behavior}, the American Time Use Survey (ATUS)~\cite{atus}, and the ATUS Well-being Module (ATUS-WB)~\cite{atus-wb}. B1K contains information about the desire for robotic automation; ATUS contains information about the time spent on activities; and ATUS-WB contains information about the feelings experienced during different activities (Happiness (H), Meaningfulness (M), Painfulness (P), Sadness (B), Stressfulness (S), and Tiredness (Z)).

%To address our research questions (RQ1-RQ4), investigate the correlations between the desire for automation and time spent on activities, as well as various well-being metrics such as happiness, meaningfulness, sadness, tiredness, stressfulness, and pain. [SS - this says the same thing as the first sentence in the Hypothes paragraph so I cut it.]

% \todo{repeat of gender info and percentages}
% What 
% How
% \todo{recheck}
% For the analysis, we also focus specifically on a category of activities grouped by ATUS as "household activities" (Category Code 20000), which includes activities such as interior cleaning, laundry, food and drink preparation, yard care, pet care, and vehicle repair. Although ATUS refers to this category as "household activities," we included a wider range of activities that may occur within the home, such as leisure activities, and collectively referred to this expanded category as "household chores."

% Activities like sleeping, which did not appear in BEHAVIOR-1K as automatable tasks, were excluded from this calculation.

\subsection{Hypotheses}
Our analysis aims to test working hypotheses about the desire for robotic automation (DA) and its correlations with time spent (T) and various well-being metrics:
%These hypotheses are grounded in existing research that suggests relationships between the time spent on tasks, well-being outcomes (such as happiness, meaningfulness, stressfulness, and tiredness), and the desire for automation.

% \todo{check wording of H1}
\textbf{(H1)} \textit{There is a positive correlation between the desire for robotic automation and the time people spend on a task}. We hypothesize that this general relationship T-DA remains invariant among social subgroups based on gender (G) and income level (I). However, the specific activities that each subgroup spends more/less time on varies: 
% We test (a) The desire for automation is direct/positively related to the time people spend on a task, and, thus, knowing the time spent on a task is a good predictor for the desire for automation of an activity. 
% Prior research suggests that tasks requiring significant time investment are often perceived as more burdensome, leading to a higher desire for automation ~\cite{}.
\textbf{(H1b)} \textit{A relative increment in time spent by a social subgroup on a specific activity (compared to the general population and other subgroups) is directly related to a relative increment in the desire for robotic automation for that activity (also compared to the general population and other subgroups).} 
%The assumption that the time spent on activities motivates the desire for automation has been prevalent in the community~\cite{}.
% Social subgroups may experience different levels of time burden, which could increase their desire for automation for time-heavy tasks~\cite{}.

\textbf{(H2)} \textit{There is a correlation between the desire for robotic automation of an activity and the feelings experienced when performing this activity.}
This correlation will be \textbf{(H2.1) negative} with respect to \textbf{happiness}, \textbf{(H2.2) negative} with respect to \textbf{meaningfulness}, \textbf{(H2.3) positive} with respect to \textbf{painfulness}, \textbf{(H2.4) positive} with respect to \textbf{sadness}, \textbf{(H2.5) positive} with respect to \textbf{stressfulness}, and \textbf{(H2.6) positive} with respect to \textbf{tiredness}. 
% \todo{finish this}.
In addition to the general trends observed in the entire population, we hypothesize that \textit{if a social subgroup experiences difference in feelings from an activity (compared to the general population and other subgroups), their desire to automate that activity will correspondingly shift.} The change will be \textbf{(H2.1b) negative} with respect to \textbf{happiness}, \textbf{(H2.2b) negative} with respect to \textbf{meaningfulness}, \textbf{(H2.3b) positive} with respect to \textbf{painfulness}, \textbf{(H2.4b) positive} with respect to \textbf{sadness}, \textbf{(H2.5b) positive} with respect to \textbf{stressfulness}, and \textbf{(H2.6b) positive} with respect to \textbf{tiredness}.

\subsection{Research Design}
% Stack Version

% \begin{figure}[t]
% \centering
% \includegraphics[width=0.6\columnwidth,trim={0cm 6cm 0cm 0cm},clip]{figures/pc1b.pdf}%
% \\
% \includegraphics[width=0.6\columnwidth,trim={0cm 6.5cm 0cm 0cm},clip]{figures/pc2b.pdf}%
% % \begin{subfigure}[t]{0.88\columnwidth}
% %     \centering
% %     \includegraphics[width=0.99\textwidth,trim={0cm 5cm 0cm 0cm},clip]{figures/pc2b.pdf}%
% %     \caption{Gender Distribution}
% % \end{subfigure}%
% % \\%
% % \begin{subfigure}[t]{0.48\columnwidth}
% %     \centering
% %     \includegraphics[width=0.99\textwidth,trim={1.5cm 0cm 3cm 1.2cm},clip]{figures/pc2.pdf}%
% %     \caption{Income Distribution}
% % \end{subfigure}%
% \caption{\textbf{Dataset Demographic Distribution}: Distribution of gender (top) and income levels (bottom) across B1K, ATUS and ATUS-WB datasets; NB=Non-binary; ATUS and ATUS-WB have almost the same demographics; B1K is slightly different but with similar trends, e.g., most participants are of middle income class}
% \vspace{-1em}
% \label{fig:ds-demo}
% \end{figure}

\begin{figure}[t]
    \centering
    % ---------- Left plot ----------
    \begin{subfigure}[t]{0.45\columnwidth}
        \centering
        \includegraphics[width=\textwidth,trim={0cm 6cm 0cm 0cm},clip]{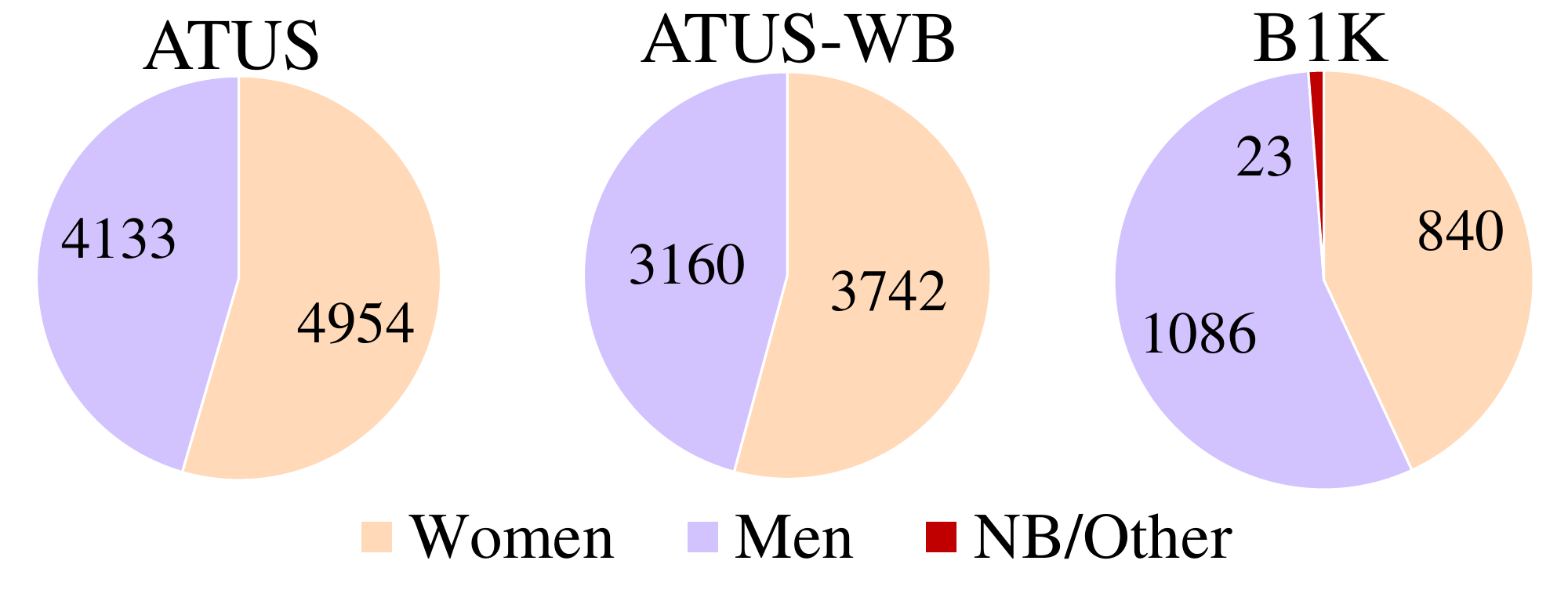}
        \caption{Gender distribution}
    \end{subfigure}
    \hfill
    % ---------- Right plot ----------
    \begin{subfigure}[t]{0.45\columnwidth}
        \centering
        \includegraphics[width=\textwidth,trim={0cm 6.5cm 0cm 0cm},clip]{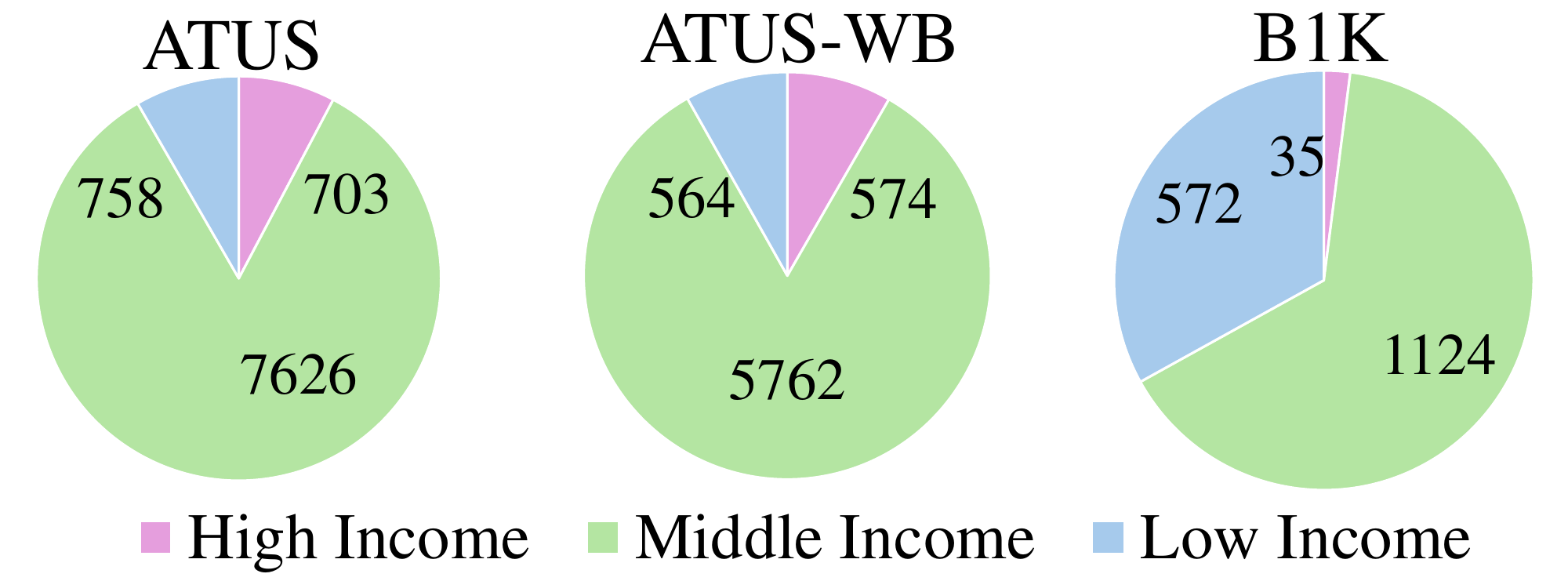}
        \caption{Income distribution}
    \end{subfigure}

    \caption{\textbf{Dataset Demographic Distribution:} Distribution of gender (left) and income levels (right) across B1K, ATUS, and ATUS-WB datasets; NB=Non-binary; ATUS and ATUS-WB have almost identical demographics; B1K is slightly different but shows similar trends (e.g., most participants are of middle-income class).}
    \vspace{-1em}
    \label{fig:ds-demo}
\end{figure}

% \todo{update demographic percentages}
\subsubsection{Datasets}
% In the following, we specify the content, the acquisition process, and other characteristics of the three datasets we analyzed: BEHAVIOR-1K dataset (B1K)~\cite{li2023behavior}, the American Time Use Survey (ATUS)~\cite{atus}, and the ATUS Well-being (ATUS-WB) Module~\cite{atus-wb}.

% what-------------------------------------------------------
% 1. Content
The BEHAVIOR-1K dataset (B1K)~\cite{li2023behavior} comprises responses from 1,949 participants collected via Amazon Mechanical Turk.
% 2. Acquiring process
\ro{Participants were asked, ``\textit{On a scale of 1 to 10, rate how much you want a robot to do this activity for you?}''} Responses were recorded on an independent Likert scale, with 1 representing ``less beneficial'' and 10 representing ``most beneficial.'' Each participant rated 50 different tasks from approximately 2,000 tasks sourced from time-use surveys and WikiHow articles. Participant demographics are summarized in Fig.~\ref{fig:ds-demo}.

% 1. Content
% The American Time Use Survey (ATUS)~\cite{atus} provides information on how Americans use their time daily and has been performed annually by the Bureau of Labor Statistics since 2003. 
% 2. Acquiring process
% Respondents are asked to record tasks performed that day in minute intervals in a diary. Respondents provide detail for each activity including what they were doing, start and end time, who else was present, and other demographic information like name, sex, birth date, number of household members, age of each household member, income, and so on. 

The American Time Use Survey (ATUS)~\cite{atus} captures how Americans spend their time daily, logging activities by minute intervals with details such as activity type, duration, and demographic information (e.g., sex, income, number of household members). This survey has been performed annually by the Bureau of Labor Statistics (BLS) since 2003. 
% "spent your time yesterday, [yesterday's day & date], from 4:00 in the morning until 4:00 AM this morning. I'll need to know where you were and who else was with you. If an activity is too personal, there's no need to mention it."

% \todo{how to get mean}

% .. [describe how data is processed] calculate the allocation of time among various activities.  --> described later

% 3. Demographics

% 1. Content
The American Time Use Well-Being Module contains information related to how people felt during selected activities.
% 2. Acquiring process
The ATUS Well-Being Module is administered to a subset of ATUS respondents. Selected respondents are asked to rate how they felt during three activities they completed the previous day. Ratings are on a scale of 0 to 6 across six feelings: Happy, Meaningful, Pain, Sad, Stress and Tired. Our study used all 6 of the available factors from the ATUS-WB module. These were selected by the BLS and are not further defined in the documentation. The BLS let respondents interpret questions by their “common understanding” of the terms.

%After the main ATUS interview, selected respondents are asked additional questions about the feelings experienced during 3 randomly chosen activities from their previous day's time diary (with certain exclusions like sleeping, grooming, and personal activities).

% Demographic:

% % how--------------------------------------------------------

% To enable comparative analysis across the B1K, ATUS, and ATUS-WB datasets, we standardized activity categorization. B1K offers detailed task data (e.g., \textit{changing sheets}), while ATUS and ATUS-WB present higher-level activity categories (e.g., \textit{interior cleaning}). We aligned B1K tasks with ATUS activities, manually coding tasks without direct equivalents in ATUS (e.g., from WikiHow) based on similarity. Since ATUS uses a similar coding method for time-use diaries, we leveraged its extensive category definitions. We excluded \textit{work, main job} across all datasets, as our analysis does not contend with automation of wage-labor. 
B1K offers detailed task data (e.g., \textit{changing sheets}), while ATUS and ATUS-WB present higher-level activity categories (e.g., \textit{interior cleaning}). To enable comparative analysis across the datasets, B1K tasks were manually coded using the extensive category definitions provided by ATUS for analysis of time-use diaries. We excluded \textit{work, main job} across all datasets, as our analysis does not contend with automation of wage labor.

ATUS includes the category code \textit{Household Activities} \textit{(20000}) for tasks like cleaning, laundry, food preparation, yard care, and vehicle repair. For visual clarity, we use this subset in our main plots, while the full analysis, dataset curation, and alignment details are provided in the Appendix \ref{appendix}.

\newcommand{\highoffig}{0.385} % controls height of each subplot

\begin{figure*}[t!]
\centering
\scalebox{1.0}{
\begin{minipage}{\textwidth}
\centering
% remove space between plots
\setlength{\tabcolsep}{1pt} % no space between tabular columns
\renewcommand{\arraystretch}{0} % remove vertical padding

\begin{tabular}{@{}cccccccc@{}} % 8 plots, no inter-column space
\includegraphics[height=\highoffig\columnwidth]{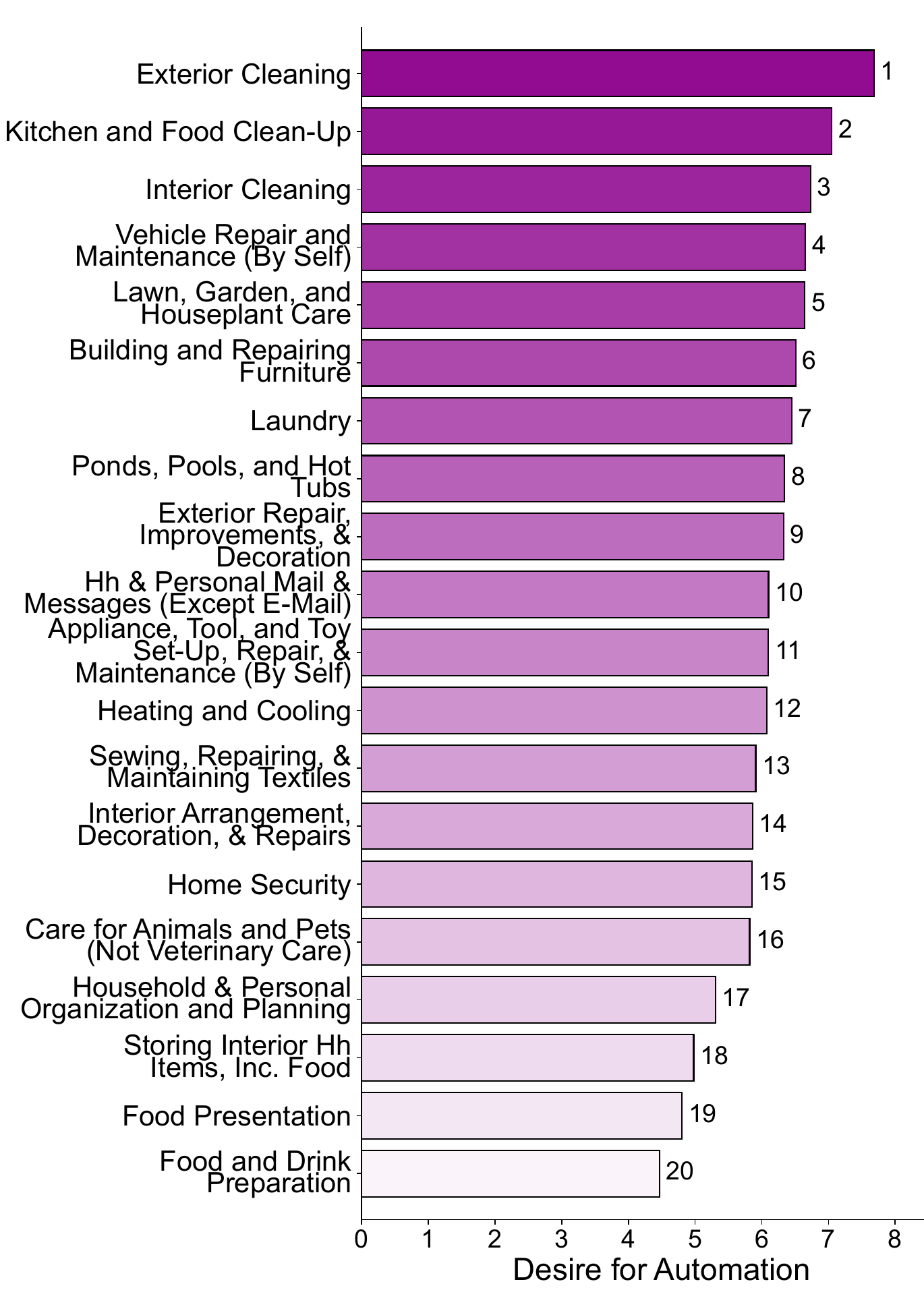} &
\includegraphics[height=\highoffig\columnwidth]{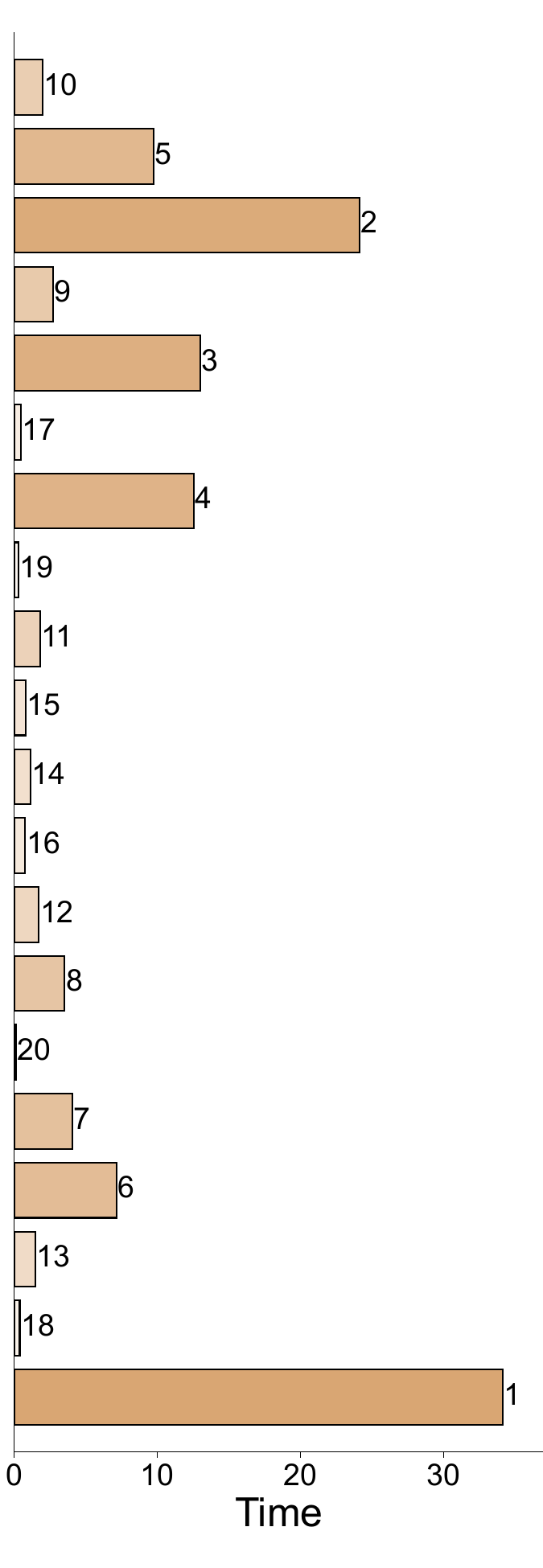} &
\includegraphics[height=\highoffig\columnwidth]{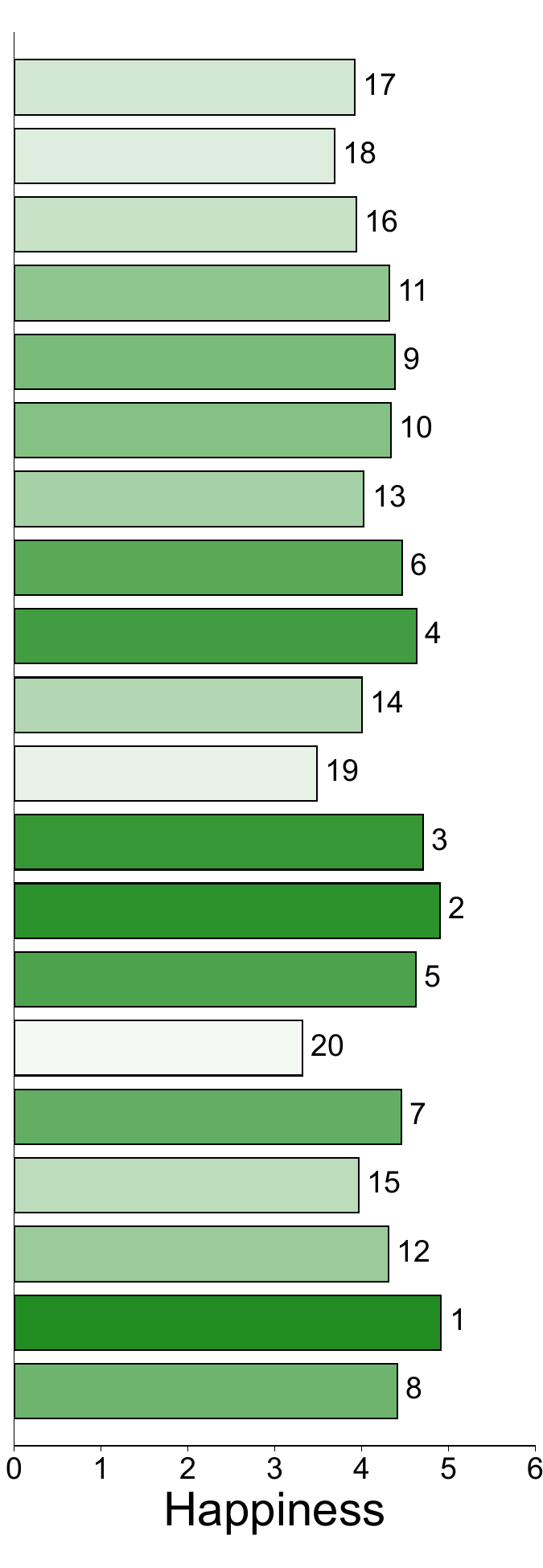} &
\includegraphics[height=\highoffig\columnwidth]{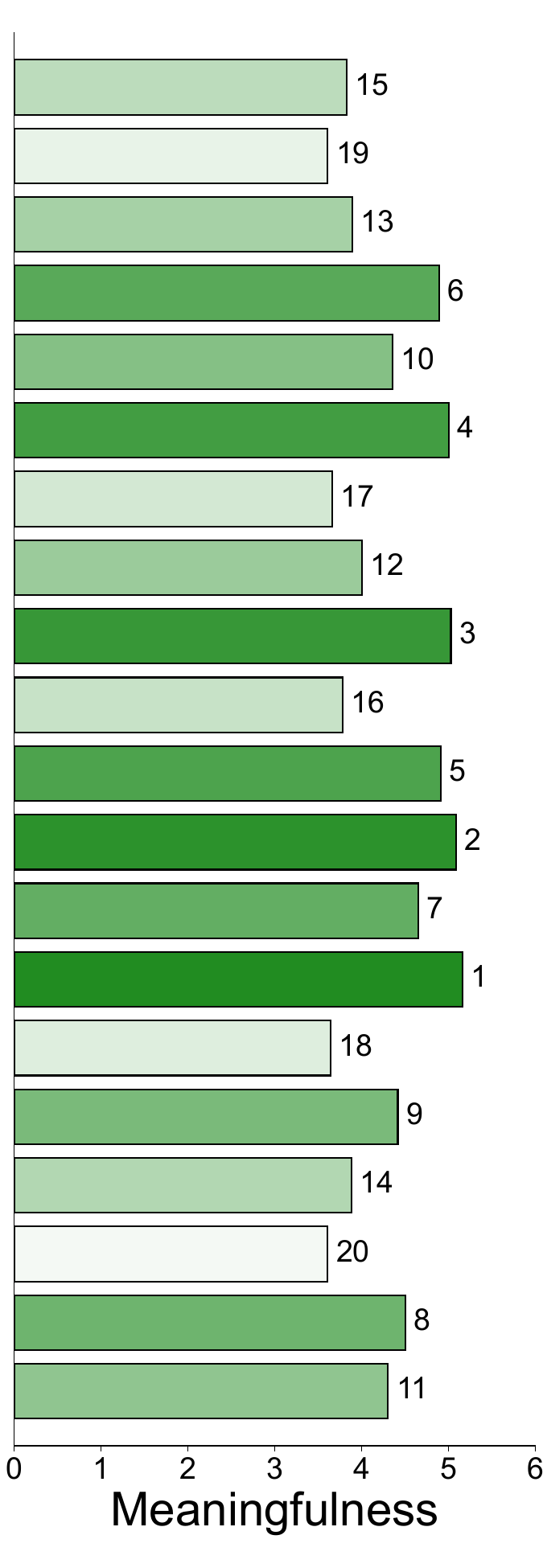} &
\includegraphics[height=\highoffig\columnwidth]{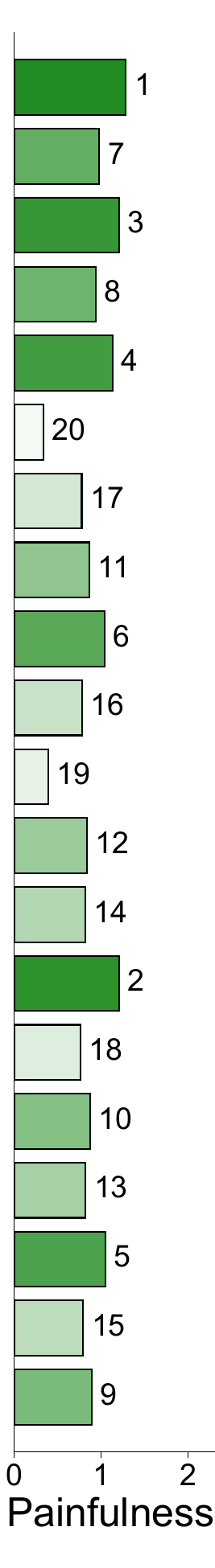} &
\includegraphics[height=\highoffig\columnwidth]{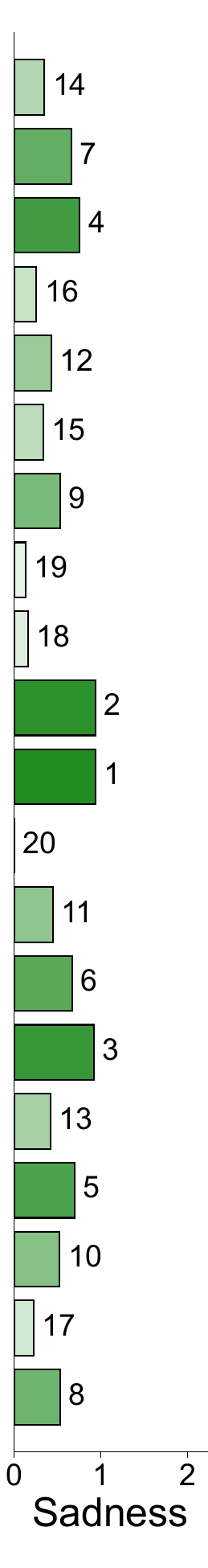} &
\includegraphics[height=\highoffig\columnwidth]{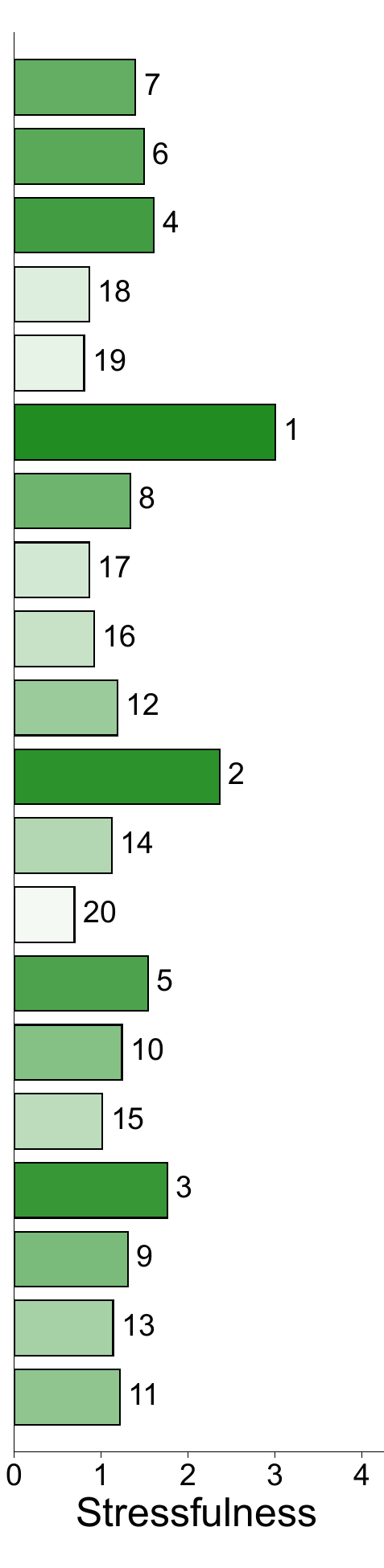} &
\includegraphics[height=\highoffig\columnwidth]{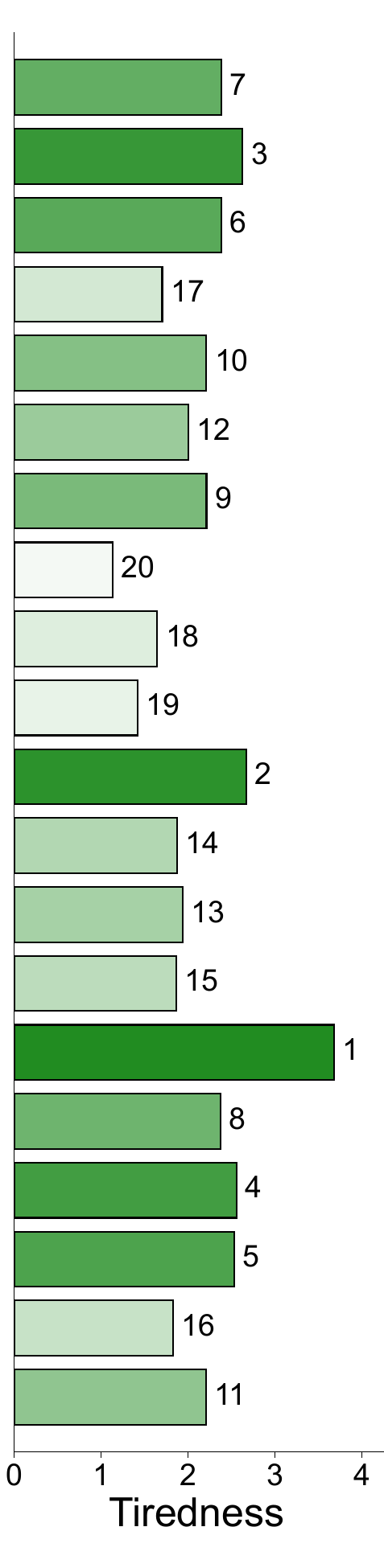}
\end{tabular}

\vspace{-0.4em}
\caption{\textbf{Desire for Robotic Automation (DA), Time Spent (T), Happiness (H), Meaningfulness (M), Painfulness (P), Sadness (B), Stressfulness (S), and Tiredness (Z) for the General Population (GP) in the subset of \textit{Household Activities}}.
All ranks are ordered by DA. Darker color indicates higher rank; numbers on each bar indicate rank position. Visually, no clear trends are observable, except for a slight negative correlation between DA and H and a positive one between DA and P.}
\label{fig:all-for-all}
\vspace{-1em}
\end{minipage}
}
\end{figure*}

\subsubsection{Dependent Variables} 
Our dependent variables are Desire for Robotic Automation (DA), Average Time Spent (T), and activity feelings: Happiness (H), Meaningfulness (M), Painfulness (P), Sadness (B), Stressfulness (S), and Tiredness (Z):

\textbf{Desire for Robotic Automation (DA):} The DA score per activity was then computed by averaging the total scores of all tasks within each activity.

% The mean DA score was derived from the B1K dataset. At the task-level, the mean DA score for a specific task was calculated by dividing the total sum of scores for that task by the number of occurrences. To align with our ATUS dataset we coded the discrete tasks of B1K into activities using the extensive category definitions and documentation provided by ATUS. The mean DA score per activity was then determined by dividing the total sum of scores for all tasks within an activity by the number of tasks in that activity.

\textbf{Average Time Spent on Activities (T):} 
% The average time data was derived from the American Time Use Survey (ATUS) dataset. We focused on activities common to both the ATUS and B1K datasets, excluding activities like sleeping which did not appear in B1K as an activity that could be automated. To compute the average time spent on each activity, we aggregated the total time recorded for activities and divided this by the number of entries in the dataset. 
% This calculation provided a mean time value per activity over the entire dataset.
The average time spent was derived from the ATUS dataset, focusing on activities shared with B1K while excluding non-automatable ones like sleeping. Average time per activity was calculated by dividing the total recorded time by the number of entries. 

There are two approaches for determining average time spent on activities. One averages over all participants, capturing how much time people spend on an activity daily--- this provides a true population-level average, which we utilize in our research. The other approach averages only among those who did the activity, reflecting task duration but not overall prevalence. For example, ATUS indicates that the average time spent by the entire population on \textit{Golfing} is approximately 0.02 hours, accounting for both participants and non-participants. Alternatively, for those who do golf, the average time spent is about 2.97 hours per day.  

% \roberto{we do what in our analysis?}
% \todo state which type of average we used and why.  I see you imply that we went with entire population in paragraph below, but I'd state it shortly and disntinctly here with a rationale. 

\textbf{Activity Well-Being Happiness (H), Meaningfulness (M), Painfulness (P), Sadness (B), Stressfulness (S), Tiredness (Z) Score:} 
% To calculate the mean well-being score associated with each activity, we used the ATUS Well-being Module. Consistency in activity-level comparison was maintained by selecting a subset of the Well-being Module data that corresponded to the activities included in our time calculation. The mean well-being scores for an activity were computed by dividing the total sum of each metric by the number of occurrences of that activity within the module dataset.
The mean well-being score associated with each activity is calculated using the ATUS-WB Module. Consistency in activity-level comparison was maintained by selecting a subset of ATUS-WB data that corresponded to the activities included in our time calculation. The scores for an activity are computed by averaging the total sum of each metric by the number of its occurrences within the module.

% It is important to note that the well-being scores are only available for individuals who actually who took part of the ATUS Well-Being study. Therefore, when calculating these metrics, we are effectively working with two distinct populations: one group includes all respondents, for whom we calculate the Desire for Automation (DA) and average time spent on activities, while the second group includes only those for whom well-being metrics are measured. 

% Explain ranking
% \begin{figure}
% \centering
% \includegraphics[width=\linewidth]{figures/fake_comparison2.png}
% \caption{\textbf{Household activities (Males vs. Females)}: Comparison of Desire for Automation; (left) Household activities ranked by desired for automation from male participants, darker color indicates higher rank; (right) Household activities ranked by desired for automation from female participants, darker color indicates higher rank;}
% \label{fig:da-mvsf-hh}
% \end{figure}

\begin{figure}[t!]
\centering
\includegraphics[height=0.47\columnwidth]{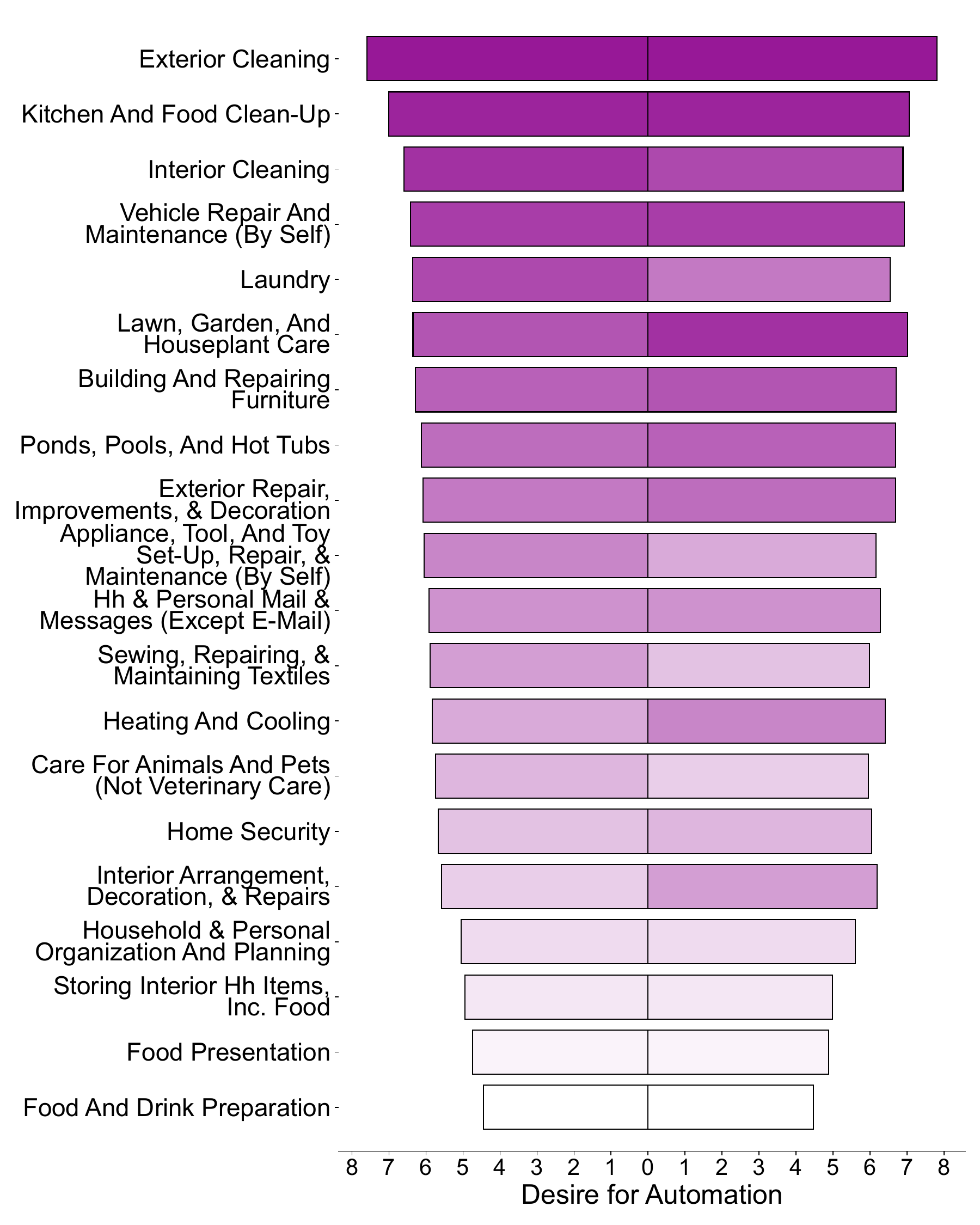}\includegraphics[height=0.47\columnwidth]{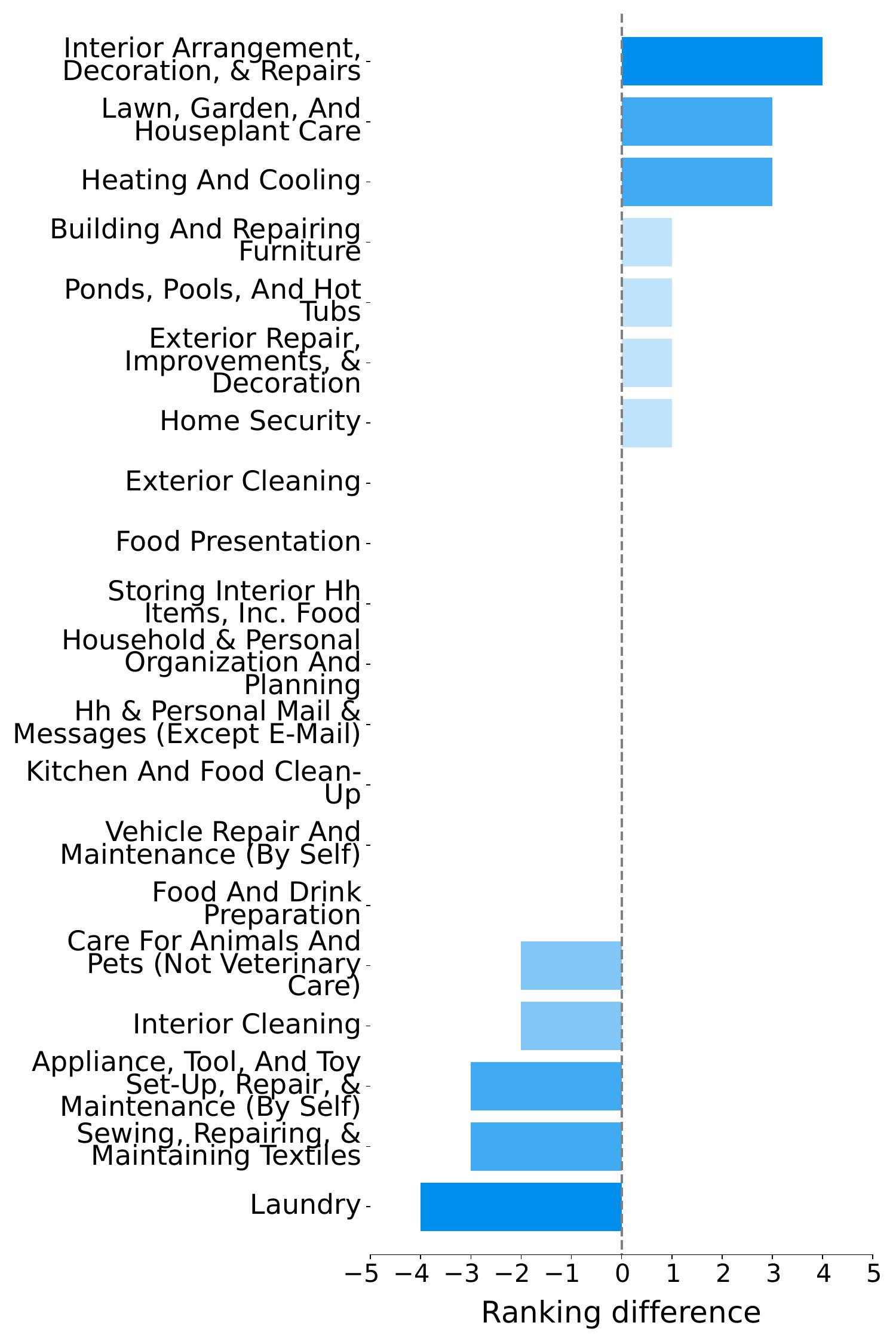}
\caption{\textbf{Comparison of The Desire for Robotic Automation of Household Activities (Men vs. Women)}: (Left) Household activities ranked by desire for robotic automation from men on the left and corresponding DA for women on the right, a darker color indicates higher rank; (Right) $\mathit{RankPosMen}-\mathit{RankPosWomen}$: Ordered differences in ranking positions between men and women, a darker color indicates larger absolute difference. A positive value for activity indicates that the activity ranks further away from the top for men than for women, corresponding to $\mathit{RankPosMen}>\mathit{RankPosWomen}$.}
\vspace{-1.5em}
\label{fig:male-female}
\end{figure}

\subsubsection{Independent Variables} 
Our independent variables consist of subgroups that represent two demographic categories: gender subgroups (men and women) and income levels (low, middle, and high-income levels). These help us to isolate and analyze any distinct preferences and correlations within each demographic group.

\textbf{Gender:}
The B1K~\cite{li2023behavior} collected gender data, with participants identifying as women (43.41\%), men (55.50\%), non-binary (0.83\%), or ``other'' (0.26\%) ATUS adheres to a binary classification of sex, recognizing only male (53.8\%) and female (46.2\%). This binary classification does not represent the full spectrum of gender identities, potentially resulting in mismeasurement and misrepresentation of participants.  
In this paper, we analyze both the ``male/female'' response as reported in ATUS and the ``man/woman'' response as reported in B1K surveys as the independent variable ``gender category:'' men/women. Though these responses may reflect two different aspects of one's identity (biological sex vs. gender), our decision was driven by the need for comparability across the three datasets. We discuss the variable \textit{Gender} further in our limitations section (Sec.~\ref{s:d}) and in Appendix~\ref{sec:gendervar}.

\textbf{Income:}
% Explain how the data was originally organized: ATUS has 14 levels, Behavior has 9. Then to make the data sets comparable we create our own categorization.
We divided ATUS and B1K participants into high, low, and mid-income levels based on household size, the poverty line, and income data from 2021. The upper boundary for the low-income category was taken from historical poverty thresholds from the U.S. Census Bureau~\cite{census2021}. The lower boundary for the high-income category was informed by the Pew Research Center’s analysis~\cite{pewResearch2021}. The income distribution of BEHAVIOR 1K shows 64.9\% mid-income, 33.0\% low-income, and 2.0\% high-income groups. Further explanation can be found in Appendix~\ref{sec:subgrouping} and on our website.

\subsection{Measures}

To investigate the motivating factors behind the desire for robotic automation, we compare the ranked lists of activities based on the Desire for Robotic Automation to ranked lists based on time spent, and the ATUS-WB variables: Happiness, Meaningfulness, Painfulness, Sadness, Stressfulness and Tiredness (details on how activity-level rankings are constructed are provided in Appendix~\ref{sec:creating-rankings}). Fig.~\ref{fig:all-for-all} depicts the ranked activities for the entire population using the order of the Desire for Robotic Automation for the other variables. 
We use ranked lists since the absolute values for each variable are non-comparable (desire for robotic automation score vs. time [minutes] vs. well-being scores).
To compare the ranked lists, we used non-parametric correlation tests: Spearman's rho and Kendall's tau. %~\cite{puth2015effective}
We performed these analyses on the general population and the subgroups. 
For all analyses, we established the null hypothesis (\(H_0\)) as the absence of a correlation between the two variables being compared, and the alternative hypothesis (\(H_1\)) as a correlation exists. 
We considered a \(p\)-value of less than 0.05 as the threshold for statistical significance. If the \(p\)-value fell below this threshold, it indicated sufficient evidence to reject the null hypothesis in favor of the alternative.
We follow the general convention by Cohen~\cite{cohen2013statistical} used in social sciences and consider a large correlation when the absolute value of the correlation coefficients are $>0.5$, medium if they are around $0.3$ and small if it is close to $0.1$. \ro{However, correlations only characterize statistical pairwise co-occurrence and do not imply causation. By identifying correlations, we provide a critical understanding of related events that could guide further causal experimental research.}

We complement the comparative analysis between absolute ranks with relative rank-difference analysis for social subgroups. Here, we determine the difference in ranking position for each activity between multiple subgroups or between a subgroup and the general population, providing a positive integer label $l$ if the activity ranks $l$ positions higher in the first analyzed subgroup than in the second and $-l$ if the activity ranks $l$ positions lower. With these labels, we can create a new ranked list where the top activity is the one with the largest positive ranking changes, and the last is the one with the largest negative. An example can be seen in Fig.~\ref{fig:male-female} with a comparison between DA for men and women (full ranking differences can be found on our website). Comparing these ranking changes between lists based on different variables (e.g., desired for automation vs. time), we can infer if the differences observed in one list correlate with the changes in another. For example, we can infer if activities with the most disparate ranking between men and women correspond to activities in which there is also a high degree of difference in the time-spent. 

\section{Results}
\label{s:r}
We first compare the Desire for Robotic Automation (DA) to the time spent on activities, followed by comparing DA to well-being metrics. Table~\ref{tab:davsx} summarizes correlations between the DA and the other dependent variables ---Time spent (T), activity Happiness (H), Meaningfulness (M), Painfulness (P), Sadness (B), Stressfulness (S), and Tiredness (Z). Table~\ref{tab:darel} reports our results of analyzing the relative rank changes between DA and the other dependent variables for social subgroups ---women (WN), men (MN), high (HI), middle (MI) and low income (LI)--- when compared to the general population (GP).

\subsection{Time as predictor for desire for robotic automation} 

Our analysis does not find a positive or negative correlation between the time spent on activities and the desire to automate them, neither in the general population nor in social subgroups (first row in Table~\ref{tab:davsx}).  
In all cases, the null hypothesis cannot be rejected ($p > 0.05$), indicating that T and DA may be completely independent.
Fig.~\ref{fig:all-for-all} depicts this lack of correlation graphically for the general population and the subset of \textit{household activities}. The left-most plot shows the activities ordered by DA. The second-left plot depicts the time spent on the activities, using the same DA-based ordering: no clear pattern is visible. These results contradict our first hypothesis (\textbf{H1}).

Analyzing relative changes in rankings between DA and T for social subgroups (Table~\ref{tab:darel}, first row), we observe a strong negative correlation between men and the general population, and between men and women. This suggests that when activities rank \textbf{higher} in DA for men (compared to the general population), they rank lower in T for men (compared to the general population). This trend is similar for ranked activities between men and women. This challenges hypothesis \textbf{H1b} for men: we assumed that a higher relative ranking in DA corresponds to a higher relative ranking in T. 

%Original version of first sentence in the paragraph above: When observing the differences in the ranking based on DA for different social groups and comparing those differences to the differences in the rankings based on time 

\newcommand\multconst{100.0}
\newcommand{\celltextp}[1]{\ifdim#1pt>0.05pt \color{black}\normalfont\else\color{black}\bfseries\boldmath\fi}

\newcommand{\colortone}[1]{\ifdim#1pt>0pt\cellcolor{blue!#1}\else\cellcolor{red!\fpeval{-#1}}\fi}

\DeclareExpandableDocumentCommand{\thetone}{ m m }{% 
  \fpeval{ ( #1 * #2 ) }%
}

\begin{table}[t!]
\centering
\caption{Rank Correlation of Desire for Automation}
\vspace{1.0\baselineskip} 
\label{tab:davsx}
\resizebox{0.75\columnwidth}{!}{
\begin{tabular}{|l|l|c|c|c|c|c|c|}
\hline
 & & \textbf{GP} & \textbf{WN} & \textbf{MN} & \textbf{HI} & \textbf{MI} & \textbf{LI} \\ \hline
\textbf{DA-T} 
    & $\rho$
    & \cellvaluerho{0.03}{.83}  
    & \cellvaluerho{-0.09}{.54}  
    &  \cellvaluerho{0.03}{.83}  
    &  \cellvaluerho{-0.05}{.72}  
    &  \cellvaluerho{0.02}{.90}  
    & \cellvaluerho{-0.12}{.35}  \\ 
    & $\tau$
    & \cellvaluetau{0.03}{0.03}{.73}   
    & \cellvaluetau{-0.09}{-0.06}{.50}  
    &  \cellvaluetau{0.03}{0.02}{.86}   
    & \cellvaluetau{-0.05}{-0.04}{.72} 
    & \cellvaluetau{0.02}{0.02}{.87}   
    & \cellvaluetau{-0.12}{-0.09}{.38}  \\ \hline
\textbf{DA-H}     
    & $\rho$         
    & \cellvaluerho{-0.36}{.01}  
    &  \cellvaluerho{-0.27}{.07}  
    &  \cellvaluerho{-0.36}{.01}  
    & \cellvaluerho{-0.10}{.56}  
    & \cellvaluerho{-0.29}{.04}  
    & \cellvaluerho{-0.26}{.12}  \\ 
    & $\tau$
    & \cellvaluetau{-0.36}{-0.26}{.01}  
    & \cellvaluetau{-0.27} {-0.20}{.05} % recheck as p =0.048 
    &\cellvaluetau{-0.36}{-0.24}{.02}   
    & \cellvaluetau{-0.10}{-0.09}{.45}
    & \cellvaluetau{-0.29}{-0.19}{.04}  
    & \cellvaluetau{-0.26}{-0.18}{.11}  \\ \hline
\textbf{DA-M}     
    & $\rho$         
    & \cellvaluerho{-0.21}{.15}
    & \cellvaluerho{-0.21}{.17}  
    & \cellvaluerho{-0.25}{.08} 
    & \cellvaluerho {-0.23}{.20} 
    & \cellvaluerho {-0.31}{.03} 
    & \cellvaluerho {0.02}{.88} \\ 
    & $\tau$
    & \cellvaluetau{-0.21}{-0.15}{.13}  
    & \cellvaluetau{-0.21}{-0.15}{.14}
    & \cellvaluetau{-0.25}{-0.18}{.06} 
    & \cellvaluetau{-0.23}{0.13}{.27} 
    & \cellvaluetau{-0.31}{-0.22}{.02}  
    & \cellvaluetau{0.02}{0.01}{.92} \\ \hline
\textbf{DA-P}     
    & $\rho$         
    & \cellvaluerho{0.30}{.04}
    & \cellvaluerho{-0.07}{.63}
    & \cellvaluerho{0.27}{.07} 
    & \cellvaluerho{0.01}{.97}
    & \cellvaluerho{0.22}{.13}
    & \cellvaluerho{0.27}{.10}  \\ 
    & $\tau$
    & \cellvaluetau{0.30}{0.23}{.02}  
    & \cellvaluetau{-0.07}{-0.05}{.60} 
    & \cellvaluetau{0.27}{0.16}{.11}  
    & \cellvaluetau{0.01}{0.03}{.82}
    & \cellvaluetau{0.22}{0.15}{.13}
    & \cellvaluetau{0.27}{0.20}{.09} \\ \hline
\textbf{DA-B}     
    & $\rho$         
    & \cellvaluerho{0.08}{.57}
    & \cellvaluerho{0.07}{.64}
    & \cellvaluerho{-0.12}{.43}
    & \cellvaluerho{0.02}{.93}
    & \cellvaluerho{0.01}{.96}
    & \cellvaluerho{0.09}{.58} \\ 
    & $\tau$
    & \cellvaluetau{0.08}{0.06}{.56}
    & \cellvaluetau{0.07}{0.04}{.71}
    & \cellvaluetau{-0.12}{-0.06}{.52}
    & \cellvaluetau{0.02}{0.02}{.89}
    & \cellvaluetau{0.01}{-0.01}{.94}
    & \cellvaluetau{0.09}{0.08}{.51}\\ \hline
\textbf{DA-S}      
    & $\rho$        
    & \cellvaluerho{0.22}{.13} 
    & \cellvaluerho{0.32}{.03}
    & \cellvaluerho{0.02}{.88}
    & \cellvaluerho{0.13}{.46}
    & \cellvaluerho{0.11}{.44} 
    & \cellvaluerho{0.21}{.21} \\ 
    & $\tau$
    & \cellvaluetau{0.22}{0.16}{.11}  
    & \cellvaluetau{0.32}{0.22}{.04}
    & \cellvaluetau{0.02}{0.01}{.92}
    & \cellvaluetau{0.13}{0.12}{.32}
    & \cellvaluetau{0.11}{0.07}{.47}
    & \cellvaluetau{0.21}{0.13}{.25} \\ \hline

    \textbf{DA-Z}    
    & $\rho$          
    & \cellvaluerho{0.07}{.62} 
    & \cellvaluerho{0.12}{.44}
    & \cellvaluerho{-0.03}{.84}
    & \cellvaluerho{0.11}{.52}
    & \cellvaluerho{-0.01}{.96}
    & \cellvaluerho{0.26}{.12} \\ 
    & $\tau$
    & \cellvaluetau{0.07}{0.04}{.67}   
    & \cellvaluetau{0.12}{0.07}{.49}  
    & \cellvaluetau{-0.03}{-0.01}{.92} 
    & \cellvaluetau{0.11}{0.10}{.41} 
    & \cellvaluetau{-0.01}{-0.01}{.90}  
    & \cellvaluetau{0.26}{0.17}{.14} \\ \hline
\end{tabular}
}
% \vspace{-1em}
\end{table}

\subsection{Well-Being factors as predictors for desire for robotic automation} 
\textbf{Happiness:}
Looking at the correlation between DA and H (second row, Table~\ref{tab:davsx}), we observe a medium negative correlation, indicating a low level of H experienced during an activity correlates to a high desire to automate it and vice versa. 
This is clear for the general population, men, mid-income subgroups and partially for women, where the probability of the null hypothesis is very low. Other subgroups show a similar trend, but the correlation is weak to medium and we cannot reject the null hypothesis. Based on these results, we consider a lack of Happiness a good indicator of the desire for robotic automation, especially for the general population, mid-income individuals and men (\textbf{H2.1}).

Analyzing relative changes in rankings between DA and H for social subgroups (Table~\ref{tab:darel}, second row), the clearest pattern is between the high-income group and the general population, where we measure a strong positive correlation. This indicates that if an activity ranks higher in DA for the high-income subgroup than for the general population, it also ranks higher in H. For this group (and less marked, between middle-income and the general population and between high and middle-income), our results contradict our hypothesis (\textbf{H2.1b}) that activities ranking higher in DA will rank lower in Happiness.

%Original intro sentence: Observing the differences in ranking based on DA for different social groups and comparing those differences to the differences in the rankings based on happiness

\textbf{Meaningfulness:} Regarding the correlation between DA and M (third row, Table~\ref{tab:davsx}), the general population and most social subgroups tend to show a weak negative correlation, which supports our hypothesis (\textbf{H2.2}). We see an exception for the low-income subgroup, where the correlation is nearly zero. For all, except mid-income, the null hypothesis (no correlation) cannot be rejected.

When analyzing relative changes in rankings between DA and M for pairs of social groups (third row, Table~\ref{tab:darel}), we do not observe clear patterns. An exception can be observed between the low-income group and the general population, where there is a medium positive correlation, indicating that if activities rank higher in DA for the low-income group (relative to the general population), they tend to also be ranked as more meaningful. Our hypothesis to explain changes in relative rankings as negatively correlated (\textbf{H2.2b}) to changes in M is not empirically supported.

\textbf{Painfulness:} For the general population and several social subgroups, we find a moderate positive correlation between DA and P (fourth row, Table~\ref{tab:davsx}), though the null hypothesis can only be rejected for the general population. 
Exceptions are women and high-income subgroups, where we measure a minimal correlation and the null hypothesis cannot be rejected. Thus, our hypothesis about a positive correlation explaining the relationship between DA and P (\textbf{H2.3}) holds for the general population.

When observing relative changes in DA rankings between pairs of social groups and their relationship to relative changes in the corresponding P rankings, we do not observe a clear pattern (fourth row, Table~\ref{tab:darel}). Additionally, for most pairwise correlation analyses, the null hypothesis cannot be rejected. We do not find empirical support for our hypothesis about a positive correlation explaining changes in relative rankings (\textbf{H2.3b}).

% Regarding income groups, the primary observation is that correlations between Pain and DA are also not statistically significant. Overall, pain is a meaningful predictor of automation desire for the general population but shows varied significance across subgroups.

\textbf{Sadness:}
We find no clear correlation between DA and B either for the general population or for social subgroups (fifth row, Table~\ref{tab:davsx}), suggesting that Sadness is not a good predictor of automation desire (\textbf{H2.4}). The null hypothesis (no correlation) cannot be rejected.

Similarly, when observing the relative rank changes in DA for pairs of social groups and their correlation to rank changes in Sadness (row fifth, Table~\ref{tab:darel}), we do not observe strong affinity, empirically invalidating our hypothesis of a positive relative correlation (\textbf{H2.4b}).

\begin{table}[t!]
\centering
\caption{Relative Rank Correlation Analysis}
\vspace{1.0\baselineskip} 
\label{tab:darel}
\resizebox{0.75\columnwidth}{!}{
\begin{tabular}{|l|l|c|c|c|c|c|c|}
\hline
 & & \textbf{WN-GP} & \textbf{MN-GP} & \textbf{MN-WN} & \textbf{HI-GP} & \textbf{MI-GP} & \textbf{LI-GP} \\ \hline
\textbf{DA-T} & $\rho$ 
    & \cellvaluerho{-0.18}{.20}  
    & \cellvaluerho{-0.43}{.00}  
    & \cellvaluerho{-0.42}{.00}  
    & \cellvaluerho{-0.16}{.30}  
    & \cellvaluerho{0.11}{.44}  
    & \cellvaluerho{0.05}{.72} \\ 
    & $\tau$ 
    & \cellvaluetau{-0.18}{-0.14}{.18}  
    & \cellvaluetau{-0.43}{-0.31}{.00}  
    & \cellvaluetau{-0.42}{-0.29}{.00}  
    & \cellvaluetau{-0.16}{-0.12}{.27}  
    & \cellvaluetau{0.11}{0.08}{.47}  
    & \cellvaluetau{0.05}{0.04}{.74} \\ \hline
\textbf{DA-H} & $\rho$ 
    & \cellvaluerho{-0.02}{.91} 
    & \cellvaluerho{-0.10}{.51} 
    & \cellvaluerho{-0.13}{.42}  
    & \cellvaluerho{0.32}{.07} 
    & \cellvaluerho{0.23}{.13} 
    & \cellvaluerho{-0.02}{.92}\\ 
    & $\tau$ 
    & \cellvaluetau{-0.02}{-0.01}{.93} 
    & \cellvaluetau{-0.10}{-0.08}{.48} 
    & \cellvaluetau{-0.13}{-0.08}{.49}  
    & \cellvaluetau{0.32}{0.25}{.05}  
    & \cellvaluetau{0.23}{-0.18}{.12}  
    & \cellvaluetau{-0.02}{-0.01}{.94} \\ \hline
\textbf{DA-M} & $\rho$ 
    & \cellvaluerho{0.06}{.67}  
    & \cellvaluerho{0.01}{.93}  
    & \cellvaluerho{0.09}{.56} % f m 
    & \cellvaluerho{0.04}{.84}  
    & \cellvaluerho{0.09}{.54}  
    & \cellvaluerho{0.32}{.05} \\ 
    & $\tau$ 
    & \cellvaluetau{0.06}{0.05}{0.66}  
    & \cellvaluetau{0.01}{0.02}{.88}  
    & \cellvaluetau{0.09}{0.07}{.52}  %f m
    & \cellvaluetau{0.04}{0.04}{.77}  
    & \cellvaluetau{0.09}{0.07}{.53}  
    & \cellvaluetau{0.32}{0.23}{.06} \\ \hline
\textbf{DA-P} & $\rho$ 
    &  \cellvaluerho{0.31}{.04}
    &  \cellvaluerho{0.00}{.99}
    &  \cellvaluerho{0.21}{.17}
    &  \cellvaluerho{-0.25}{.16}
    &  \cellvaluerho{0.14}{.36}
    &  \cellvaluerho{0.27}{.11}\\ 
    & $\tau$ 
    &  \cellvaluetau{0.31}{0.21}{.06}
    &  \cellvaluetau{0.00}{0.00}{.97}
    &  \cellvaluetau{0.21}{0.13}{.23}
    &  \cellvaluetau{-0.25}{-0.15}{.22}
    &  \cellvaluetau{0.14}{0.09}{.41}
    &  \cellvaluetau{0.27}{0.20}{.10}\\ \hline
\textbf{DA-B} & $\rho$ 
    & \cellvaluerho{0.15}{.33}
    & \cellvaluerho{-0.27}{.07}
    & \cellvaluerho{-0.22}{.14}
    & \cellvaluerho{0.16}{.38}
    & \cellvaluerho{-0.03}{.83}
    & \cellvaluerho{0.08}{.63}\\ 
    & $\tau$ 
    &  \cellvaluetau{0.15}{-0.09}{.41}
    &  \cellvaluetau{-0.27}{-0.19}{.09}
    &  \cellvaluetau{-0.22}{-0.16}{.15}
    &  \cellvaluetau{0.16}{0.09}{.46}
    &  \cellvaluetau{-0.03}{-0.03}{.76}
    &  \cellvaluetau{0.08}{0.06}{.60}\\ \hline
\textbf{DA-S} & $\rho$ 
    &  \cellvaluerho{-0.07}{.64}
    &  \cellvaluerho{0.11}{.47}
    &  \cellvaluerho{0.00}{.99}
    &  \cellvaluerho{-0.01}{.92}
    &  \cellvaluerho{0.00}{.98}
    &  \cellvaluerho{-0.11}{.51}\\ 
    & $\tau$ 
    &  \cellvaluetau{-0.07}{-0.04}{.75}
    &  \cellvaluetau{0.11}{0.07}{.50}
    &  \cellvaluetau{0.00}{0.00}{.99}
    &  \cellvaluetau{-0.01}{0.00}{.97} 
    &  \cellvaluetau{0.00}{-0.01}{.92}
    &  \cellvaluetau{-0.11}{-0.08}{.51}\\ \hline
\textbf{DA-Z} & $\rho$ 
    &  \cellvaluerho{0.19}{.22}
    &  \cellvaluerho{0.10}{.52}
    &  \cellvaluerho{0.22}{.16}
    &  \cellvaluerho{0.00}{.97}
    &  \cellvaluerho{-0.14}{.35}
    &  \cellvaluerho{0.07}{.70}\\ 
    & $\tau$ 
    &  \cellvaluetau{0.19}{0.14}{.22}
    &  \cellvaluetau{0.10}{0.08}{.48}
    &  \cellvaluetau{0.22}{0.16}{.15} 
    &  \cellvaluetau{0.00}{0.00}{.97}
    &  \cellvaluetau{-0.14}{-0.10}{.38}
    &  \cellvaluetau{0.07}{0.04}{.74}\\ \hline
\end{tabular}
}
% \vspace{-1em}
\end{table}

\textbf{Stressfulness:} We find a weak positive correlation between DA and S for the general population and most social subgroups (sixth row, Table~\ref{tab:davsx}). However, the null hypothesis can only be rejected for women, which also shows the strongest correlation. This supports our hypothesis that more stressful activities correlate with a higher desire for robotic automation, but only for women (\textbf{H2.5}).

When observing the relative rank changes in DA for pairs of social groups and their correlation with rank changes in stress (sixth row, Table~\ref{tab:darel}), there are no strong correlations, contradicting our hypothesis of a positive correlation (\textbf{H2.5b}).

\textbf{Tiredness:}
We find that Z has a weak, non-significant positive relationship with DA in the general population and in the social subgroups (seventh row, Table~\ref{tab:davsx}). For most subgroups, the null hypothesis probability is high, indicating that DA and Z being uncorrelated cannot be ruled out. Therefore, Tiredness is not a good predictor of the desire to automate, contrary to our initial hypothesis (\textbf{H2.6}).

When we observe the relative rank changes in DA for pairs of social groups and their correlation to rank changes in Z (seventh row, Table~\ref{tab:darel}), we do not observe strong affinity with some correlations being weak and positive and others weak and negative, but with the null hypothesis being significant for all of them. Our hypothesis of a positive relative correlation was not supported (\textbf{H2.6b}).

\section{Discussion}
\label{s:d}
In the following, we analyze our research questions (RQ1-4) based on the results in the previous section. Our primary analysis is a statistical evaluation of activity ranks, complemented with illustrative or unexpected trends for particular activities.

\textit{RQ1: Does the average time-spent predict the desire for robotic automation?}
Our results contradict our first hypothesis, showing that time spent on an activity is \textbf{not} an indicator of the desire for robotic automation (\textbf{H1}). %Only when observing the change in rank in the desire for automation between men and the general population and comparing to rank changes based on time spent we observe a negative correlation (\textbf{H1b}). 
This contrasts prior work~\cite{ray2008people} that assumes people want to automate activities where they invest more time, motivating the search for other factors such as well-being. Remarkably, the activity with the lowest desire for robotic automation among the household subset is where people invest the most time: \textit{food and drink preparation}. This raises questions about whether the numerous robotics efforts to automate cooking in industrial contexts would be successfully adopted in the home. %adapted for the domestic is illuminating given the numerous robotics efforts to automate cooking. 
%ExplainThe relatively high ranking in happiness among the household activities (8th) could explain these results.

\textit{RQ2: Which feelings are the strongest predictors of the desire for robotic automation?}
Only the happiness and pain that activities bring to the general population are solid indicators for the desire to automate them (\textbf{H2.1} and \textbf{H2.3}). For many activities, the well-being scores for painfulness, sadness, stressfulness, and tiredness are close to zero, suggesting that these feelings were not evoked, rather than being low, which explains in part the lack of correlations. The meaningfulness and stress that the activities elicit are also weak to medium correlated ---negatively (\textbf{H2.2}) and positively (\textbf{H2.5}), respectively---however we cannot reject the hypothesis that there is no correlation between them ($p>0.05$).
% Both the sadness and tiredness felt during activities are not good predictors of how much the general population desires to automate them (\textbf{H2.4} and \textbf{H2.6}); this is counterintuitive as one would expect that tasks that make people feel tired would be strongly correlated with tasks they want to see automated. 
Interestingly, while \textit{food and drink preparation} is associated with relatively high happiness and low desire for robotic automation among the household subset (last), \textit{kitchen and food cleaning} ranks very low in happiness (18th) with a high desire for robotic automation: people seem to enjoy cooking but not cleaning afterwards, a task that robotics research could focus on.

\textit{RQ3: Do gender-based differences yield differences in the desire for robotic automation?}
While we observe significant differences in the time spent on activities and the desire to automate them based on gender, we only observe a clear pattern between the differences in time men spend on activities and their automation desires. Specifically, there is a strong negative correlation with respect to the general population and women (\textbf{H1b}), suggesting that \textit{men exhibit a higher desire to automate activities they spend less time on when compared to women and the general population} and vice versa. We also observe that the correlation between the stress generated by the activities is more strongly (positively) correlated with the desire women have to automate them, which indicates that, for this social subgroup, stress is a good predictor: \textit{women want to automate activities that stress them} (\textbf{H2.5}). Whereas \textit{for men, low happiness serves as a more relevant predictor} (\textbf{H2.1}). 

% While we observe clear differences in the time spent on activities and the desire to automate them based on gender, we only observe a clear pattern between the differences in time spent on activities by men and the desire they have to automate them: there is a strong negative correlation with respect to the general population and women (\textbf{H1b}), suggesting that men tend to show a higher desire to automate the activities they spend less time on than women or general population and vice versa. We also observe that the correlation between the stress generated by the activities is more strongly (positively) correlated with the desire women have to automate them, which indicates that, for this social subgroup, stress is a good predictor: women want to automate tasks that stress them. 

Remarkably, the largest differences between men and women in DA occur in activities associated with their stereotypical roles: men rank activities such as \textit{laundry} and \textit{sewing} much higher than women (largest negative rank difference in Fig.~\ref{fig:male-female}, right), corresponding to activities on which they spend much less time. In contrast, women rank activities such as \textit{repairs}, \textit{lawn/garden care} and \textit{heating and cooling} higher than men in DA (largest positive rank difference in Fig.~\ref{fig:male-female}, right), corresponding to activities that rank lower than men in their spent time. A strong negative correlation between these differences indicates that the way men and women differ in how they prioritise automating activities corresponds inversely to the differences in the time they spend: \textit{they want to automate the activities they spend less time on}, perhaps because they want \textit{someone or something} to do these for them.

\textit{RQ4: Do socioeconomic differences yield differences in the desire for robotic automation?}
Overall, the trends observed in the general population tend to hold for the mid-income subgroup (fifth column of Table~\ref{tab:davsx}). There are some exceptions: while happiness and pain are good indicators of DA for the general population, happiness and meaningfulness serve as better indicators for middle-income individuals.
When analyzing the way rank changes for different income subgroups, we do not observe clear patterns based on time spent (\textbf{H1b}) or well-being factors (\textbf{H2.Xb}) except in two surprising cases: 1) Differences in DA ranks between the high-income group and the general population are positively correlated with between-group differences in happiness rankings -- indicating they want to automate activities that make them happier (Table~\ref{tab:darel}, second row, fourth column). And 2) the differences in DA ranks between the low-income group and the general population correlate positively with the differences in meaningfulness ranks: activities that are more meaningful for individuals of the low-income group show a higher DA (Table~\ref{tab:darel}, third row, sixth column).

\textit{Study Limitations:} Firstly, the theoretical constructs used regarding the gendered division of labour assume heteronormative household structures. Also, our use of a binary gender classification reflects a broader issue in AI and HCI, where queer experiences are marginalised through classification~\cite{queerinai2023queer}. Further, the demographic distribution of participants, particularly regarding race, limits our ability to conduct intersectional analyses that account for overlapping forms of oppression~\cite{costanza2020design,mccullough2024}, which is crucial given that women of colour disproportionately make up the housekeeping workforce and often shoulder a double burden when these activities become wage labor~\cite{banerjeedomestic}. In particular, the number of high-income participants is significantly low. Since ATUS data is drawn from the U.S. population, our findings may not generalise to other cultural contexts. Finally, as an exploratory study engaging in open-science using three existing datasets, future work would benefit from collecting data from the same participants to directly relate time use, emotional experiences, and automation preferences.%Finally, as an exploratory study using three existing datasets, ideally, the same participants would report all values.

% The ideal scenario is having the same participants answer questions about the time they spend and their desire for automation, which would offer a better understanding of individuals' time utilization, emotional experiences, and their relation to automation preferences.

% Ideally, the same participants would provide data on automation preferences, time spent and emotional experiences, offering deeper insights into time utilization, feelings, and automation desires.

% The ideal scenario is having the same participants answer questions about the time they spend and their desire for automation, which would offer a better understanding of individuals' time utilization, emotional experiences, and their relation to automation preferences.

% \todo{look into}
% The demographic profile of BEHAVIOR-1K shows a predominantly white (75\%) participant pool, with most respondents falling into the 30-40 age range (\todo{how many?}) and identifying as men (55.7\%).

% This contrasts with data from the Behavior dataset, where individuals answer regardless of their actual engagement, potentially skewing the results. 
% context:
% In analyzing time spent on activities like "Golfing," two methods for calculating the mean can be employed. ATUS indicates that the average time spent by the entire population on "Golfing" is approximately 0.02 hours, accounting for both participants and non-participants. Alternatively, for those who do golf, the average time spent is about 2.97 hours per day

\section{Conclusion}
\label{s:c}
% We presented an original study of the underlying reasons behind the desire for automation, assessing how these tendencies change across gender and income groups. Our analyses integrated and analyzed data from three datasets, BEHAVIOR-1K, ATUS, and ATUS Well-Being Module. 
% Our analysis did not find a correlation between the desire for automation and time spent on activities. We did find a correlation between the desire for automation and the well-being metrics happiness and pain. Significant trends indicated differences among genders and income levels. We open source all our data and analysis tools for the community to further study other relevant correlations. 

We present an original study of the underlying reasons behind the desire for robotic automation, assessing how these tendencies change across subgroups with respect to time spent and feelings evoked. Our study integrated and analyzed data from three datasets, BEHAVIOR-1K, ATUS, and ATUS Well-Being Module. No correlation was found between automation desire and time spent, but happiness and pain were significant predictors. Significant trends indicated differences among genders and income levels. We open-source our data and an interactive analysis tool to enable reproducibility and support further robotics research.
% show that there is a high positive correlation between the desire for automation across gender and income, reflecting an overarching agreement on what tasks should be automated.

\clearpage
\section*{Acknowledgements}
This project was partially funded by Good Systems, an initiative of the University of Texas at Austin.
% \end{thebibliography}

\bibliographystyle{unsrtnat} 
\bibliography{main}

% \bibliographystyle{ieeetr} 
% % Change this to your preferred BibTeX style

% \bibliography{main}

\clearpage
\section{Appendix}
\label{appendix}
% \newpage

\subsection{Open Source B1K Dataset and Online Tool}

\begin{figure}[t!]
    \centering
    \includegraphics[width=1.0\linewidth]{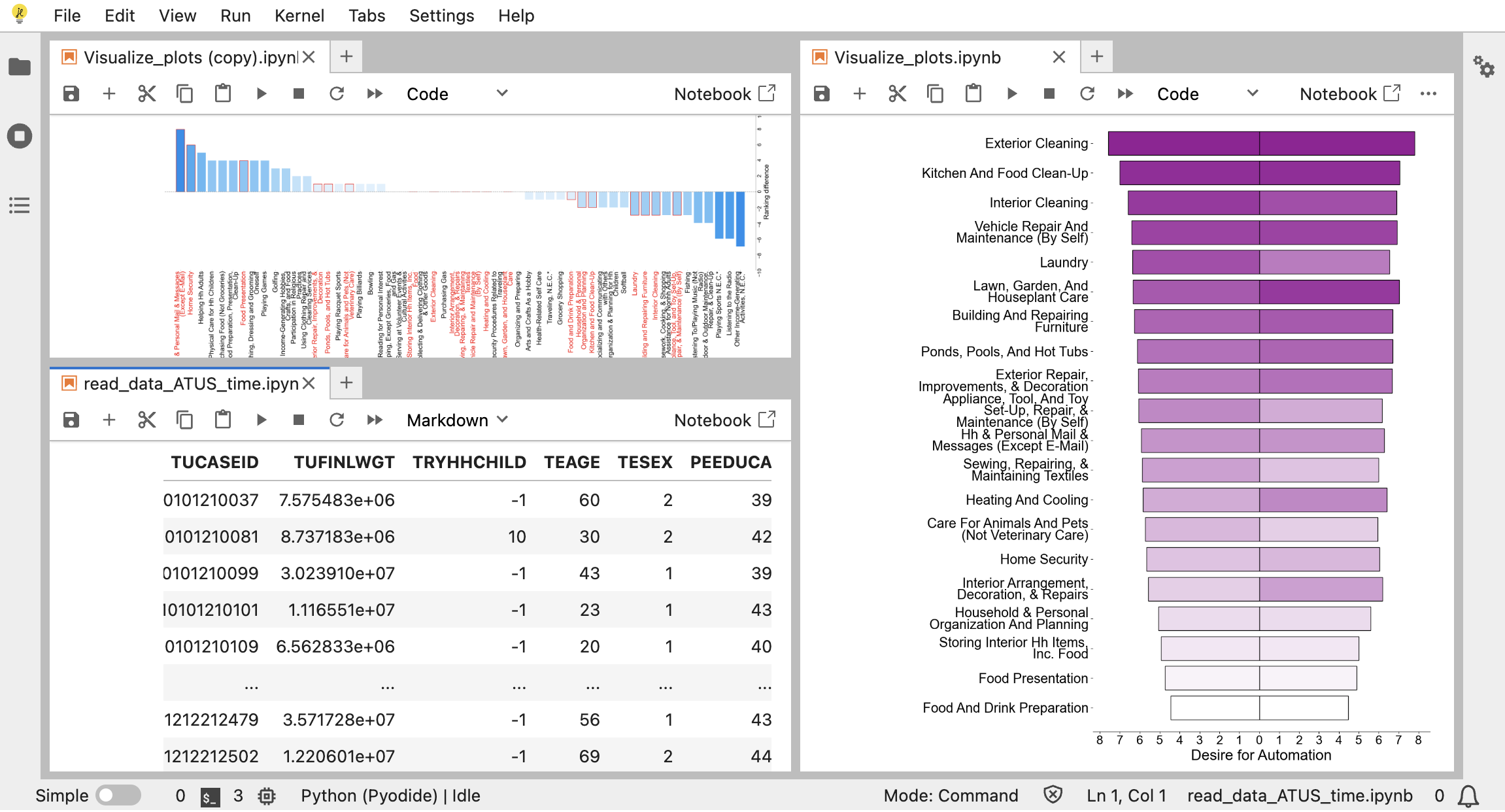}
    \caption{\textbf{Screenshot of our online open-source tool for parsing and visualizingfrom B1K, ATUS, and ATUS-WB datasets.} As part of this work, we release the full B1K survey and an open-source Python-based tool to reproduce and extend our analyses. We hope the tool facilitates future research at the intersection of social science and robotics.}
    \label{fig:tool}
\end{figure}

As part of this work, we open source the full set of responses to the BEHAVIOR-1K survey (1949 participants). 
Together with the data, we open source an online tool at \url{https://robin-lab.cs.utexas.edu/why-automate-this/} to reproduce our results and extend them to other statistical analyses using B1K, ATUS, and ATUS-WB. ATUS is a vast source of statistical information but its accessibility is limited: we provide parsing, visualizing and operating tools based on python to investigate and use its content.
Our tool (see Fig.~\ref{fig:tool}) is locally hosted on a server that can run Python code. It aims to simplify the complete analytical process stated in our methods. The tool has dedicated modules for reading ATUS and BEHAVIOR data, allowing users to import and manipulate the dataset. The application also includes functionality for visualizing data using a variety of plots, such as demographic visualizations and variable-based activity ranking comparisons. It also allows users to do correlation analysis on dependent variables across many subgroups. 

\subsection{Demographics of the Datasets}

% 3. Demographics
\textbf{B1K:} The demographic profile of BEHAVIOR-1K shows a predominantly white (75\%) participant pool identifying as men (55.7\%). Women represent 43.1\% of participants, with 1.2\% identifying as Non-Binary or Other genders. Income distribution within this dataset shows 64.9\% mid-income, 33.0\% low-income, and 2.0\% high-income participants (see below for how income groups are defined). Only 5.74\% of participants identified as having a disability.

\textbf{ATUS:} For our analysis, we focused on the 22,546 entries from ATUS 2021, which has a gender distribution of 45.5\% male and 54.5\% female and a racial composition that is predominantly white (68.2\%). ATUS data also provides information on income levels, with 83.9\% in the mid-income range, 7.7\% in the low-income range, and 8.4\% in the high-income range. 

\textbf{ATUS-WB:} The gender distribution of ATUS-WB respondents is 45.8\% male and 54.2\% female. In terms of income, 83.5\% fall in the mid-income range, 8.2\% in the low-income range, and 8.3\% in the high-income range.

\subsection{Dataset Alignment and Social Subgrouping} 
\label{sec:subgrouping}

To enable comparative analysis across the B1K, ATUS, and ATUS-WB datasets, we standardized activity categorization. B1K offers detailed task data (e.g., \textit{changing sheets}), while ATUS and ATUS-WB present higher-level activity categories (e.g., \textit{interior cleaning}). We aligned B1K tasks with ATUS activities, manually coding tasks without direct equivalents in ATUS (e.g., from WikiHow) based on similarity. Since ATUS uses a similar coding method for time-use diaries, we leveraged its extensive category definitions. We excluded \textit{work, main job} across all datasets, as our analysis does not contend with automation of wage labor.

For the ATUS data, we separated responses by gender using the ''TESEX`` variable and by income based on the number of household members and the ''HEFAMINC'' variable. When gender was specified, the dataset was filtered to retain only entries that matched the numeric code corresponding to male or female in the ''TESEX'' column. 

% We considered income levels categorized into ''low,'' ''mid,'' and ''high''. This was based by the thresholds set by the U.S. Census Bureau~\cite{census2021} and the Pew Research Center’s analysis~\cite{pewResearch2021}. This threshold is dependent on number of household members. The upper boundary for the low-income category was taken from historical poverty thresholds from the U.S. Census Bureau~\cite{census2021}. The lower boundary for the high-income category was informed by the Pew Research Center’s analysis~\cite{pewResearch2021}. We calculated the number of household members by grouping the ATUS dataset by household ID (''TUCASEID'') and counting the unique household members (''TULINENO'') for each household. We found which income range the from the ''HEFAMINC'' variable and further categorized them in ''low,'' ''mid,'' and ''high''. For the ATUS Well-Being (ATUS WB) dataset, we aligned the datasets on ''TUCASEID'' to get the right group of indivial's responses.

We categorized income levels into ``low,'' ``mid,'' and ``high'' based on thresholds set by the U.S. Census Bureau~\cite{census2021} and the Pew Research Center's analysis~\cite{pewResearch2021}. These thresholds depend on the number of household members. The upper boundary for the low-income category was taken from historical poverty thresholds provided by the U.S. Census Bureau~\cite{census2021}, while the lower boundary for the high-income category followed the Pew Research Center's analysis~\cite{pewResearch2021}. We calculated the number of household members by grouping the ATUS dataset by household ID (``TUCASEID'') and counting the unique household members (``TULINENO'') for each household. Then, we matched the income range from the ``HEFAMINC'' variable and further classified them into ``low,'' ``mid,'' and ``high.'' For the ATUS Well-Being dataset, we aligned the datasets on ``TUCASEID'' to ensure accurate alignment with the corresponding individuals' responses.

%the census and pew line is a copied line from main

In the BEHAVIOR-1K dataset, we classified responses by gender using the variable ``Answer.gender.gender-X,'' where X corresponds to female, male, other, or non-binary (nb). For income classification, we applied the same thresholds used in the ATUS dataset. Household size was determined from the ``Answer.household-members'' variable, and income ranges were derived from ``Answer.income.income-n,'' where n represents 1 to 14. These income brackets, specific to BEHAVIOR-1K, were then categorized into ``low,'' ``mid,'' and ``high'' income groups.

% \roberto{check exact question for Behavior, needs recheck to finish para.}

\subsection{Creating Rankings of the Activities} 
\label{sec:creating-rankings}

We ranked the activities based on the mean values of our dependent variables ---Desire for Automation, Time Spent, Happiness, Meaningfulness, Painfulness, Sadness, Stressfulness, and Tiredness--- in descending order. \textit{Rank 1} was assigned to the activity with the highest mean value, with subsequent ranks reflecting decreasing values. In instances where activities had identical values, they were assigned the same rank. As explained in the main text, there are two ways to compute the mean spent time on an activity and, thus, the ranking: considering all participants, where the survey participants who do not report performing the activity are equivalent to reporting 0 time, or considering only the participants that report performing the activity. We use the first option as it provides a better estimation of the importance of an activity over the entire population.
Since, for some activities, none of the members of a social subgroup performed the activity, the ranked lists do not include them. This explains the differences in number of bars in the plots in Fig.~\ref{fig:all-GP-M}, Fig.~\ref{fig:all-M-L}, and Fig.~\ref{fig:all-Mid-H}.

\begin{figure}[t!]
    \centering
    \includegraphics[width=0.9\linewidth]{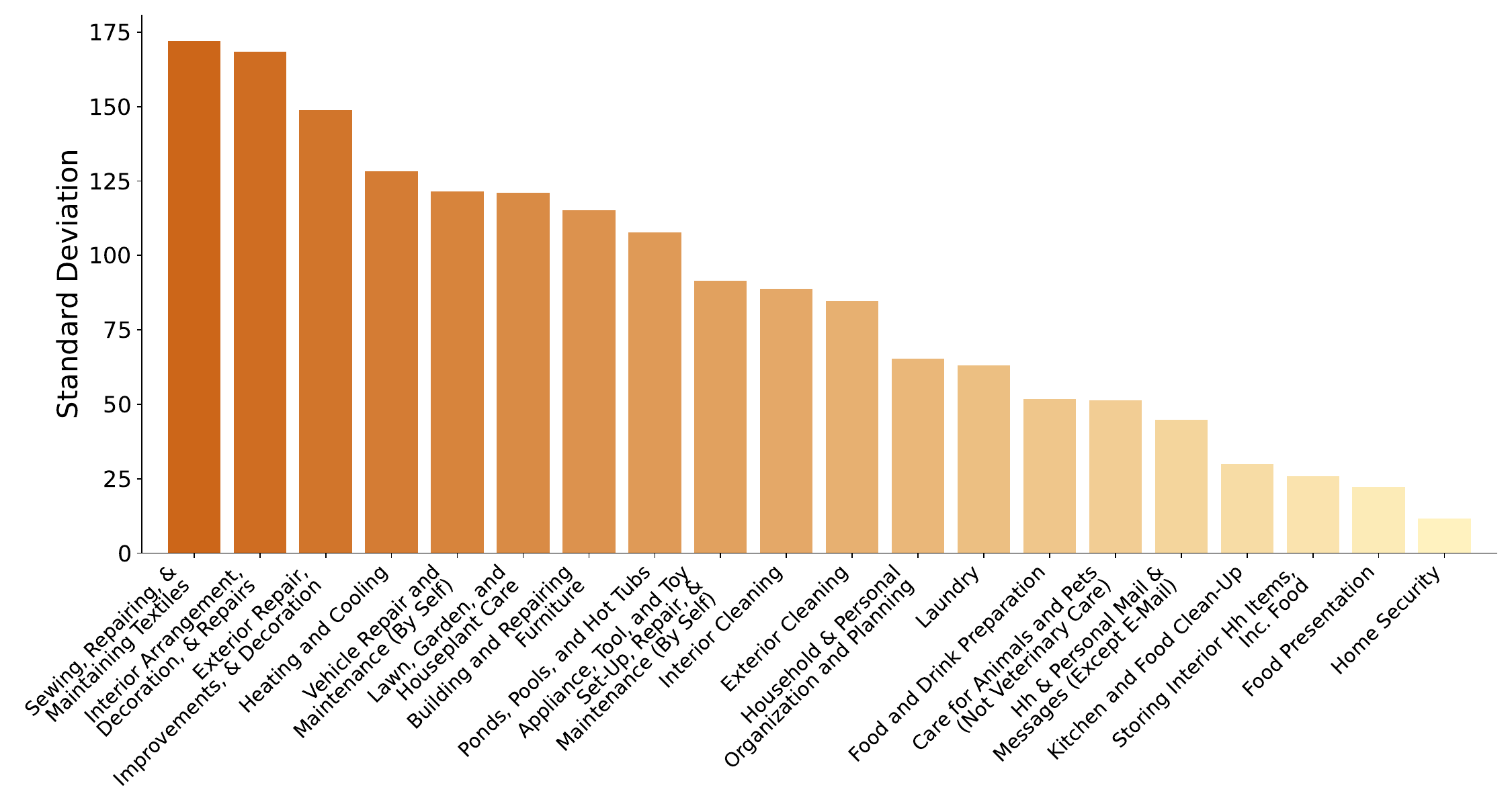}
    \caption{\textbf{Variability of time spent on activities (T) for the general population (GP)}. We measure the standard deviation of the survey responses around the mean value for each activity. Some activities (e.g., \textit{Home Security} and \textit{Food Preparation}) show low variability, indicating that most individuals report a similar time, while others (\textit{Sewing}, \textit{Repairs}) present a larger one, indicating very different time investment among individuals.}
    \label{fig:variability}
\end{figure}

\textbf{Variability in the dependant variables:} While our rankings are based on mean values for the dependant variables ---desire for automation, time spent, and the well-being metrics --- the variability (standard deviation) of those values for the population provides additional information about them. For example, if the mean time spent on an activity is $t$, does the entire surveyed population spend around $t$ time on the activity, or does each individual invest a very different time? Fig.~\ref{fig:variability} depicts the standard deviation of the time spent on activities for the general population. In future surveys we plan to collect responses from the same individuals about desire for automation, time and well-being variables to explore correlations between the individuals that report over/under mean time/well-being values and their desire for automation.

\subsection{The Household Activities Subset}
% Fill in here:
% - where does it come from:
Throughout the paper, our plots visualize activities that are categorized by ATUS as Household Activities (Category Code 20000). This includes the regular "activities done by individuals to maintain their households," including activities like \textit{interior cleaning}, \textit{laundry},\textit{ food and drink preparation}, as well as \textit{yard care}, \textit{pet care}, \textit{vehicle repair}, etc. We identified a total of 20 activities from the 34 that are common to both the ATUS and the B1K datasets.

We used this subset of household activities given the potential for robotic automation in these areas. This subset also contained relevant activities to automate, while the full list of activities include some that are clearly not an automation target such as playing sports or leisure activities. The Household Activity selection allows us to highlight key insights while keeping the visualizations manageable in the main body of the paper. Nevertheless, the full subset, consisting of over 50 activities, is included in this Appendix to provide a comprehensive overview of the data and ensure completeness in our analysis.

\subsection{On the \textit{Gender} Variable}
\label{sec:gendervar}

Current philosophies on gender measurement in statistical methods emphasize the need for inclusivity, accuracy, and respect for gender diversity in survey response options -- criteria that B1K, ATUS and ATUS-WB don't meet sufficiently. Guidelines have been developed to ensure that gender questions are respectful and accurately capture respondents' identities, including offering multiple response options in demographic questions~\cite{scheuerman2020hci}, using gender-neutral language in question design~\cite{cameron2019gender}, and adapting questions to local cultural norms around gender~\cite{bauer2017transgender}.

BEHAVIOR-1K uses a more inclusive gender measurement, however, the non-binary and other gender identities are underrepresented based on population-based survey research that includes trans, nonbinary, and genderqueer identities (for instance, research estimates that 5\% of young adults identify as nonbinary or transgender~\cite{brown2022experiences}). 
% This indicates an avenue for further studies centered on the needs and perception of robots and automation for underrepresented (or unclearly identified) populations.
Datasets like the ones used in our study that do not accurately capture or adequately include non-binary and gender-diverse individuals contribute to the problem of ``data invisibility'' -- erasing or marginalizing queer identities in research~\cite{subramaniam2024queer}. Recent efforts within the robotics community try to shed more light on the specific trends among those groups, e.g., the work of the recently created \textit{QueerInRobotics}~\cite{korpan2024launching} affinity group\footnote{\url{https://sites.google.com/view/queerinrobotics/}}.

% \todo{do we include some analysis of the variability intra-activity? yes. that requires explanation, so it will need also text}

\subsection{Additional Limitations and Directions for Future Research}
The theoretical constructs around the gendered division of labor that informed our analysis assume heteronormative household structures and a binary classification of gender. Time use studies show that gender disparities in household tasks are less pronounced in gay and lesbian couples. Straight women spend more time on cleaning and maintenance than lesbian women, though both spend more time than gay men. Less than 3\% of HRI studies published between 2006–2022 report on family configurations~\cite{seaborn2023not}. There is a clear need to understand how diverse family structures shape domestic labor. 
Further, given B1K, ATUS, and ATUS-WB were collected from different samples (and in different years), our results reflect activity-level associations rather than individual-level relationships. Thus, a unified survey and a study of individual-level relationships are important future work. Further, in future work, we plan to further explore the trends identified in our results and discussion by conducting more targeted analyses.

\clearpage

%\subsection{Figure 1 Credits}
%Fig. 1 was designed by Helen Shewolfe Tseng and makes-use of two images: ``Robotic Vacuum Cleaner'' by Olga Miltsova~\cite{miltsova2018robotic} and ``Knox County, Tennessee. Mrs. Wiegel Uses Electric Vacuum Cleaner'' by Arthur Rothstein~\cite{rothstein1942wiegel}.

% \subsection{}

\newcommand{\highoffignew}{0.6}
\begin{figure*}[t!]
\centering
\includegraphics[height=\highoffignew\columnwidth]{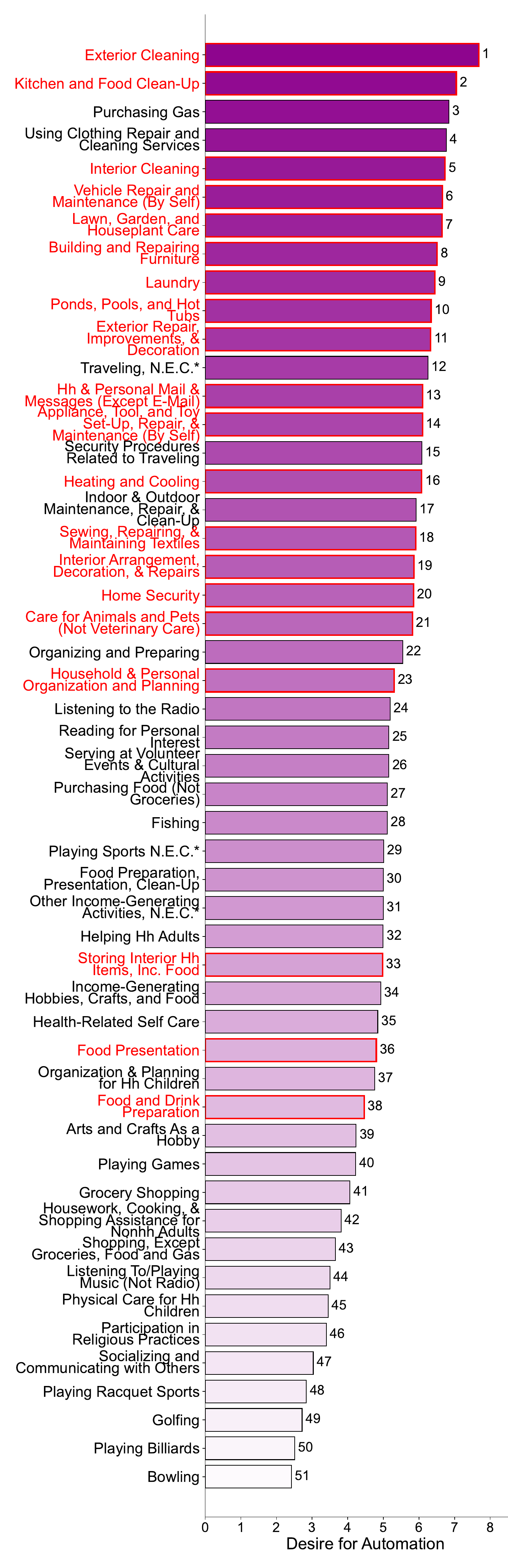}
\includegraphics[height=\highoffignew\columnwidth]{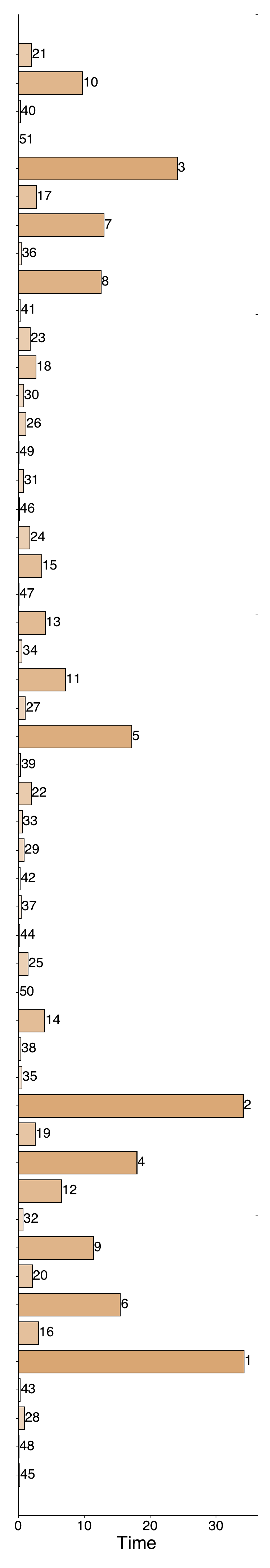}
\includegraphics[height=\highoffignew\columnwidth]{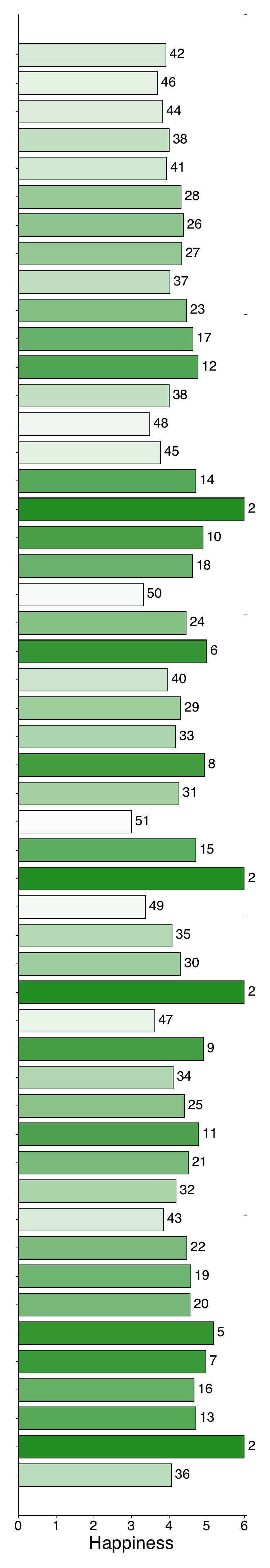}
\includegraphics[height=\highoffignew\columnwidth]{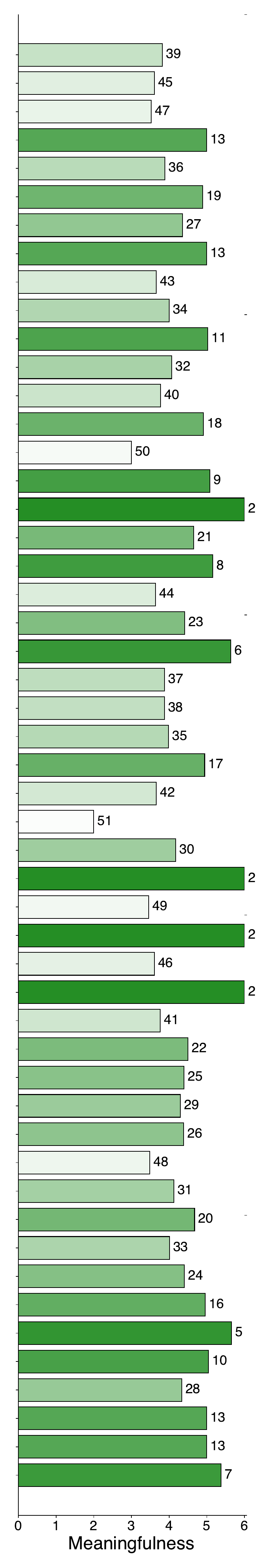}
\includegraphics[height=\highoffignew\columnwidth]{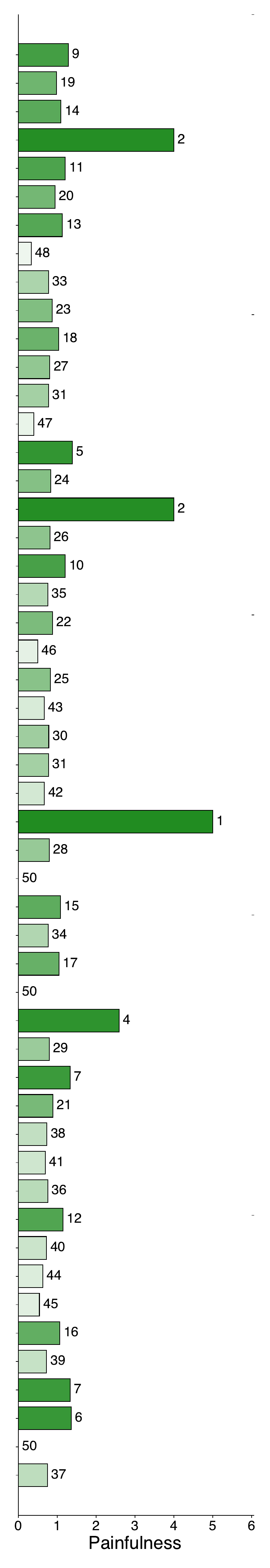}
\includegraphics[height=\highoffignew\columnwidth]{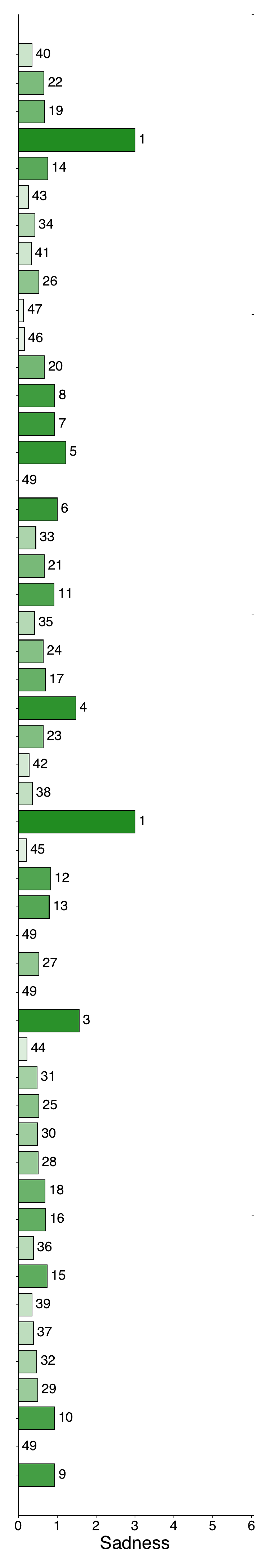}
\includegraphics[height=\highoffignew\columnwidth]{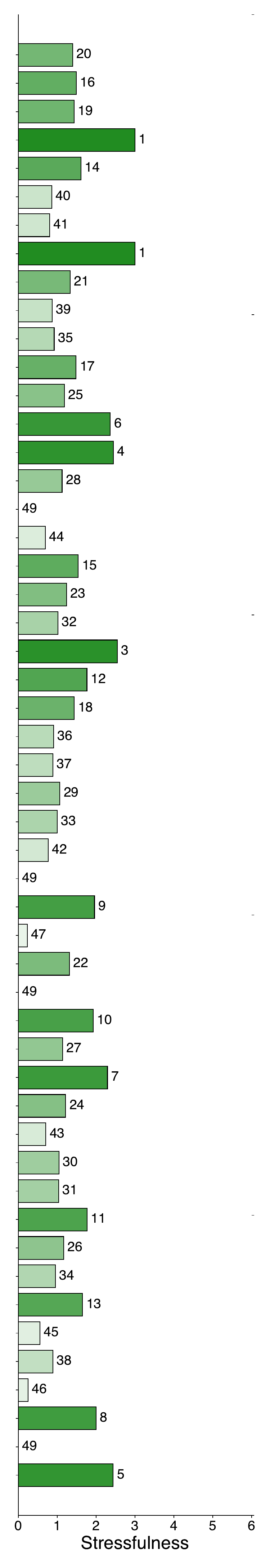}
\includegraphics[height=\highoffignew\columnwidth]{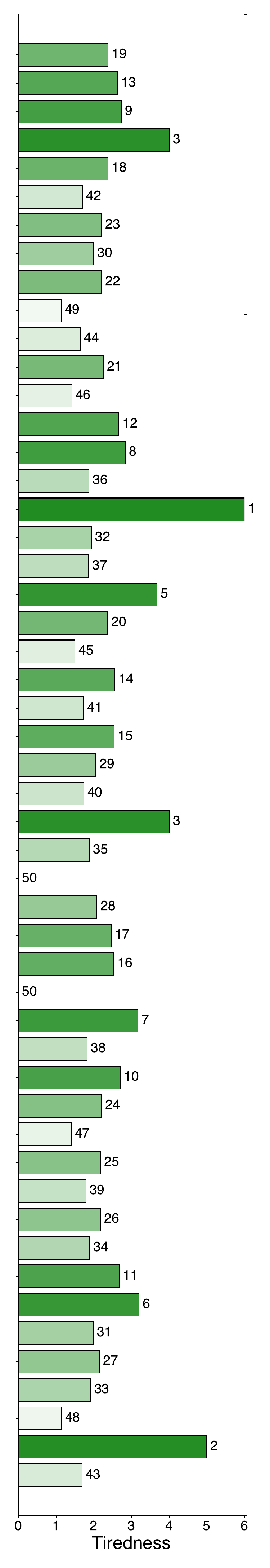}\\
\includegraphics[height=\highoffignew\columnwidth]{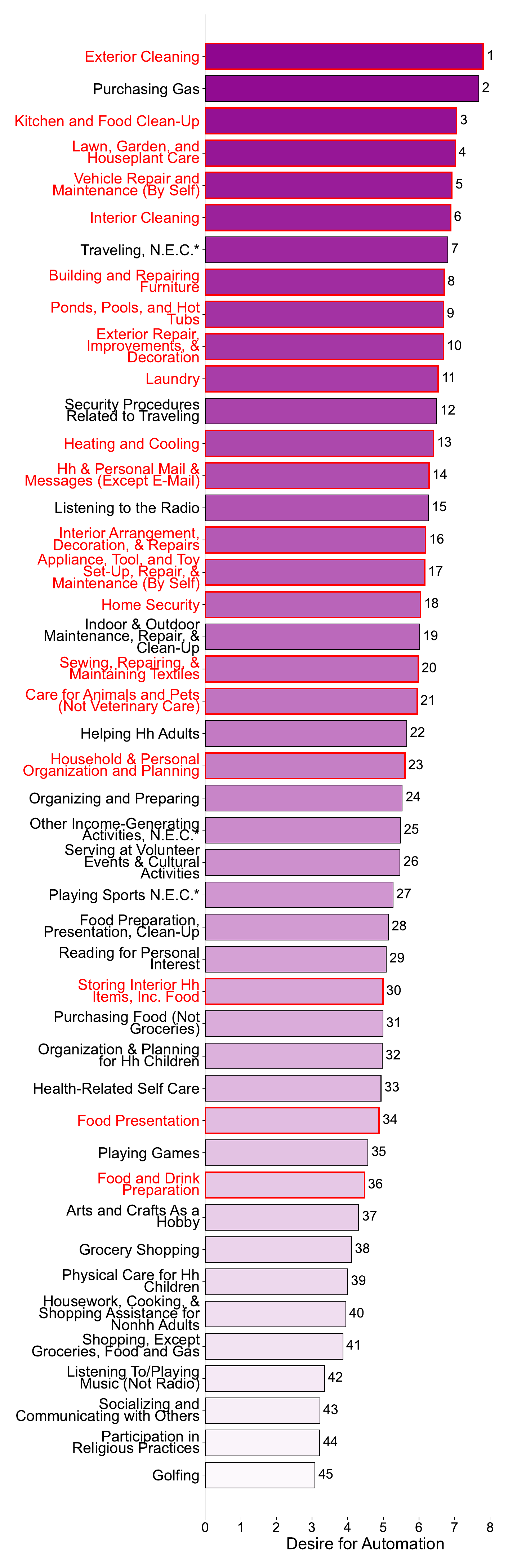}
\includegraphics[height=\highoffignew\columnwidth]{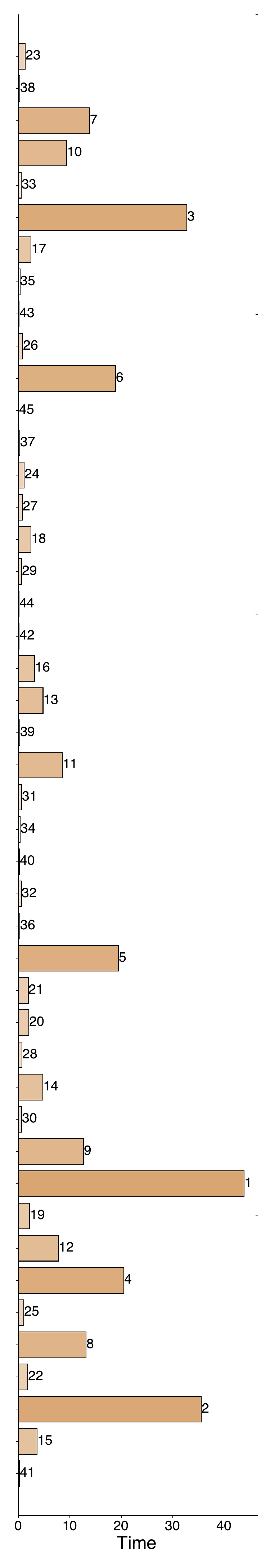}
\includegraphics[height=\highoffignew\columnwidth]{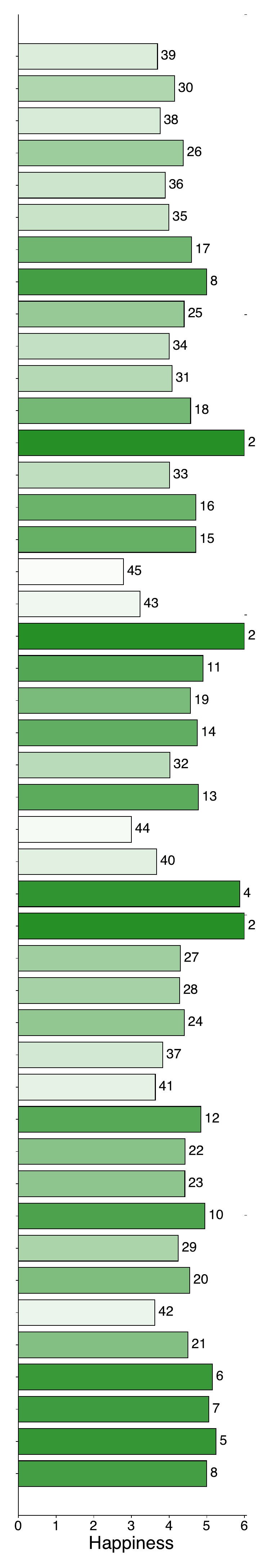}
\includegraphics[height=\highoffignew\columnwidth]{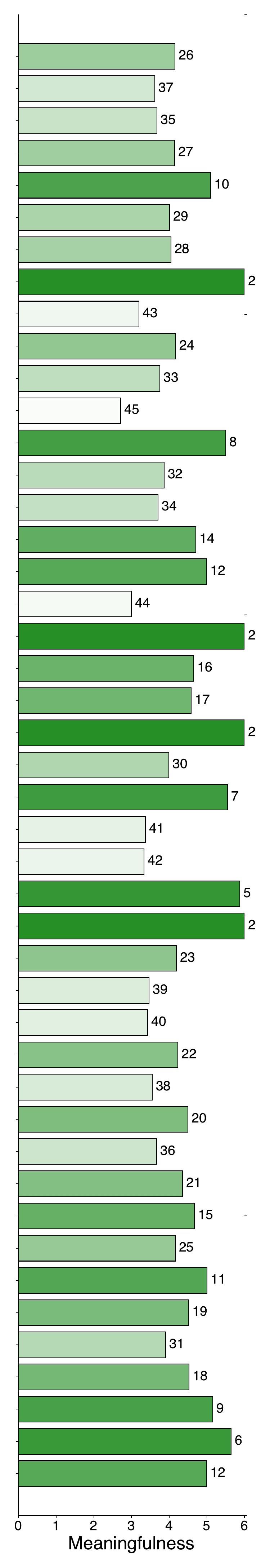}
\includegraphics[height=\highoffignew\columnwidth]{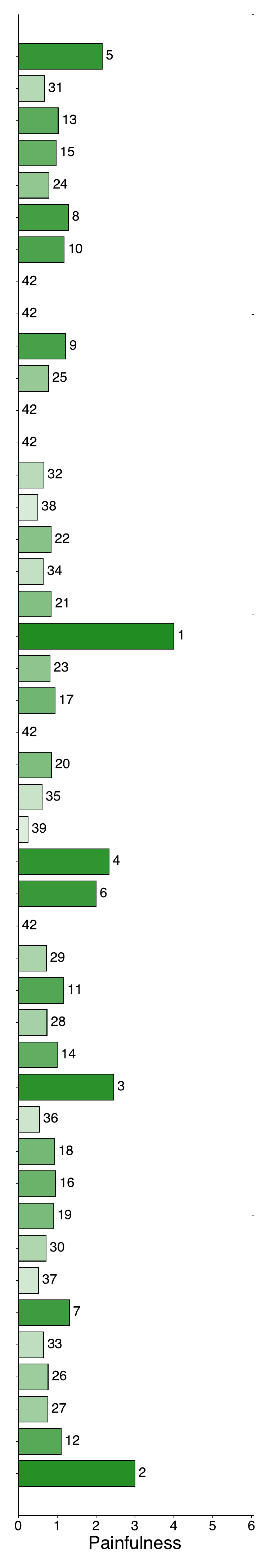}
\includegraphics[height=\highoffignew\columnwidth]{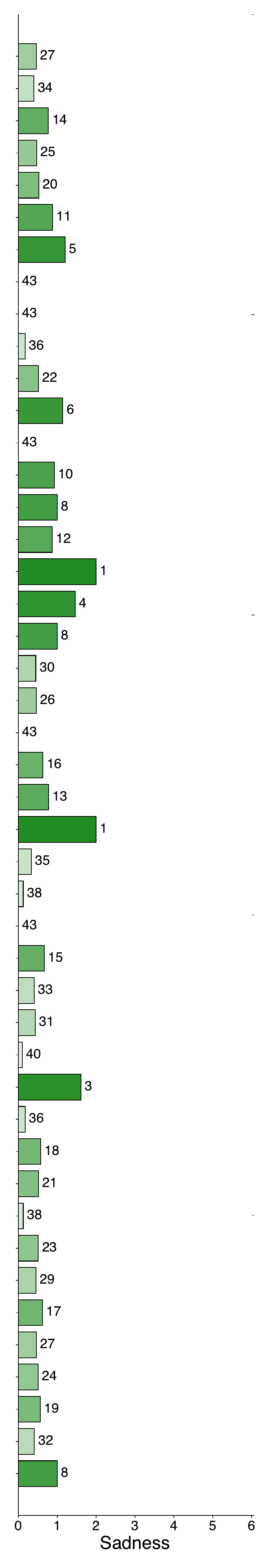}
\includegraphics[height=\highoffignew\columnwidth]{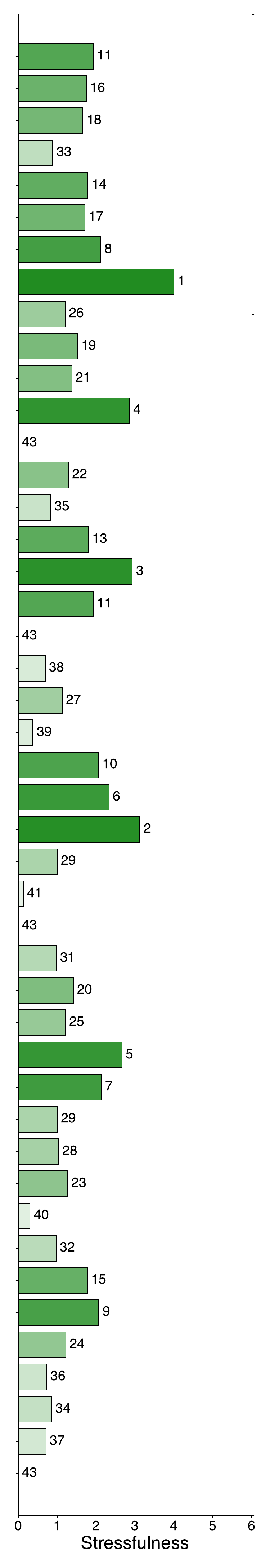}
\includegraphics[height=\highoffignew\columnwidth]{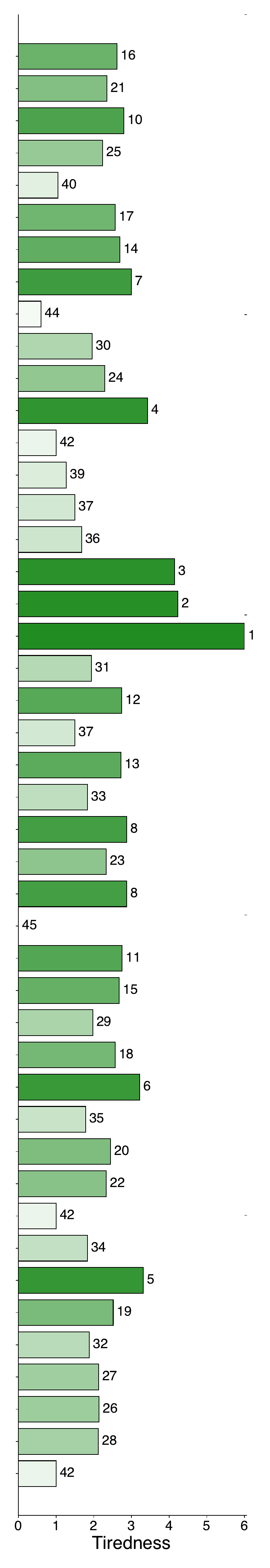}
\caption{\textbf{Absolute values and ranked activities} for the \textbf{general population (top row)} and \textbf{women (bottom row)} based on Desire for Automation (1st from left), Time spent (2nd from left), Happiness (3rd from left), Meaningfulness (4th from left), Painfulness (5th from left), Sadness (6th from left), Stressfulness (7th from left) and Tiredness (most right); Darker color tone indicates higher rank, numbers next to the bars indicate ranking positions; Red labels and bar lines indicate activities of the \textit{Household Activities} subset.}
\label{fig:all-GP-M}
\end{figure*}

\begin{figure*}[t!]
\centering
\includegraphics[height=\highoffignew\columnwidth]{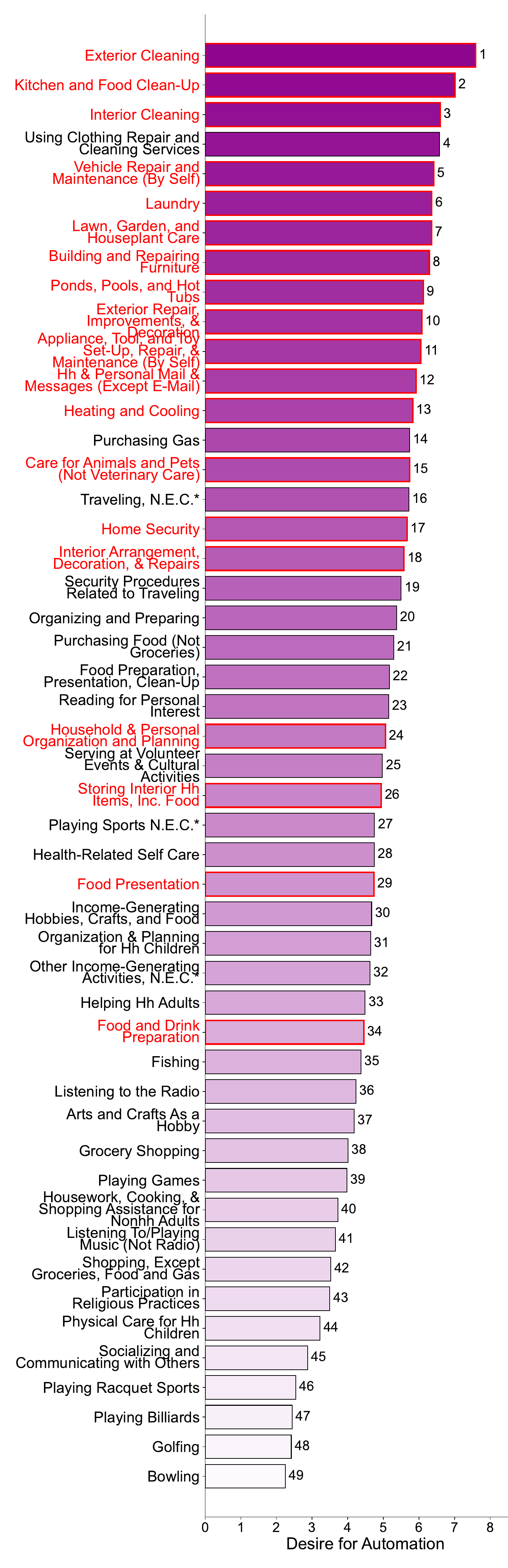}
\includegraphics[height=\highoffignew\columnwidth]{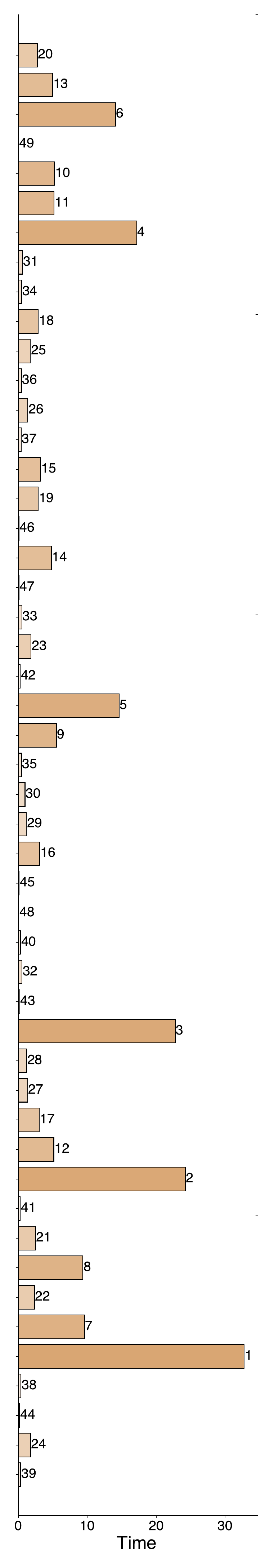}
\includegraphics[height=\highoffignew\columnwidth]{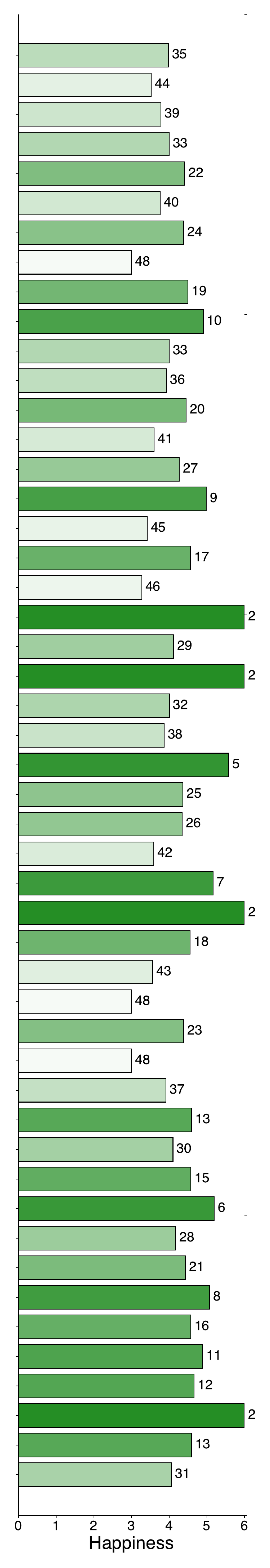}
\includegraphics[height=\highoffignew\columnwidth]{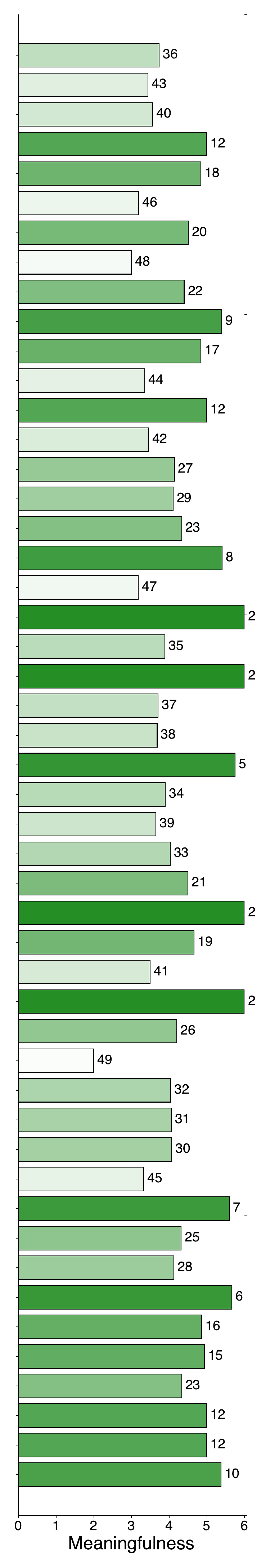}
\includegraphics[height=\highoffignew\columnwidth]{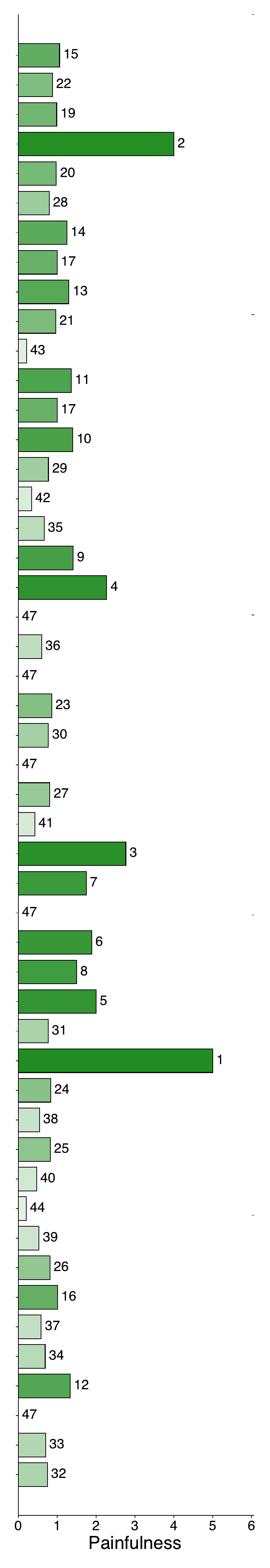}
\includegraphics[height=\highoffignew\columnwidth]{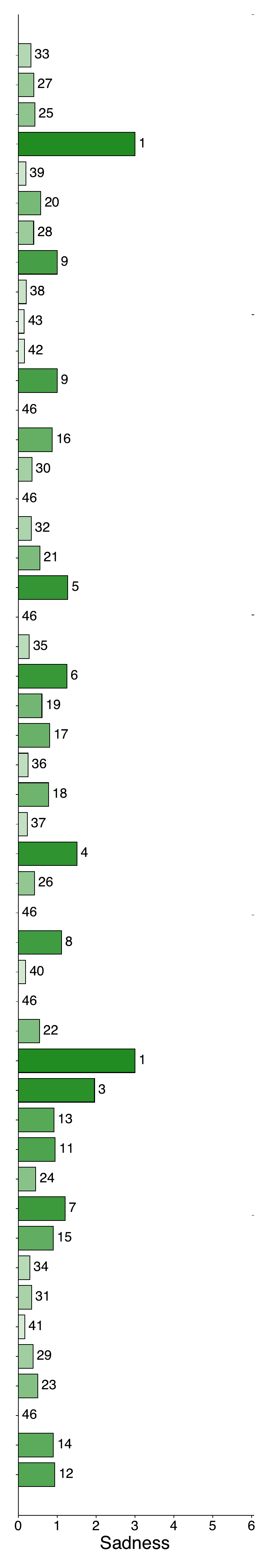}
\includegraphics[height=\highoffignew\columnwidth]{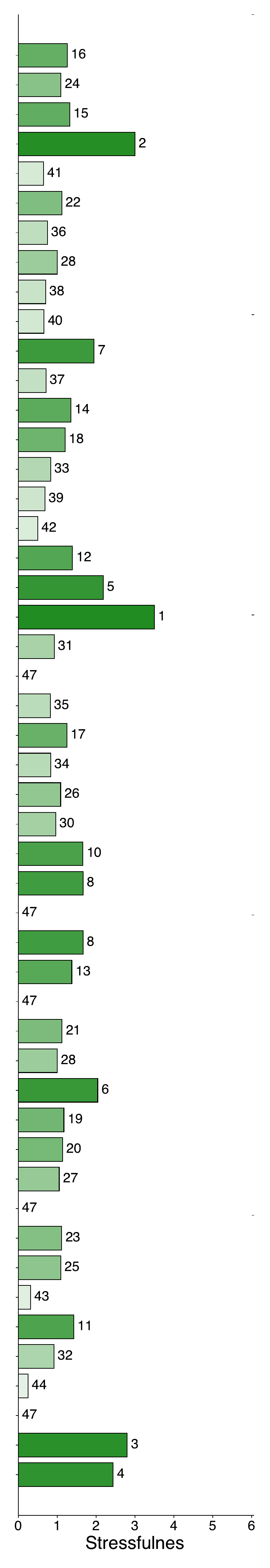}
\includegraphics[height=\highoffignew\columnwidth]{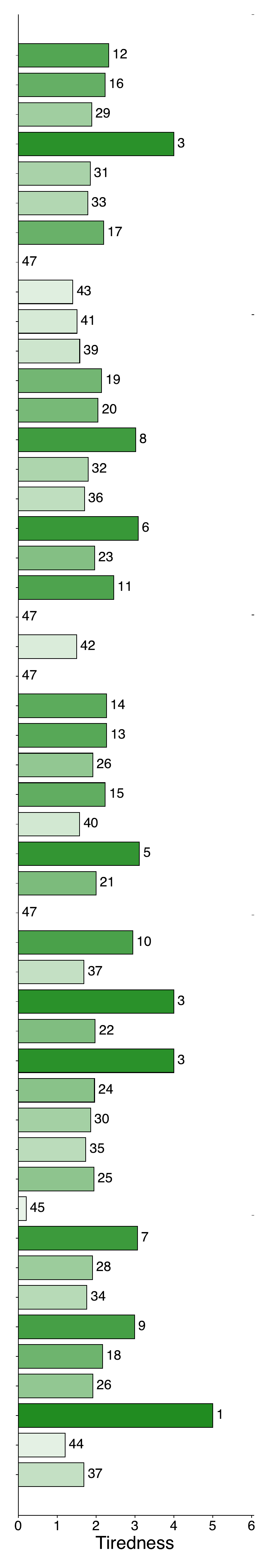}\\
\includegraphics[height=\highoffignew\columnwidth]{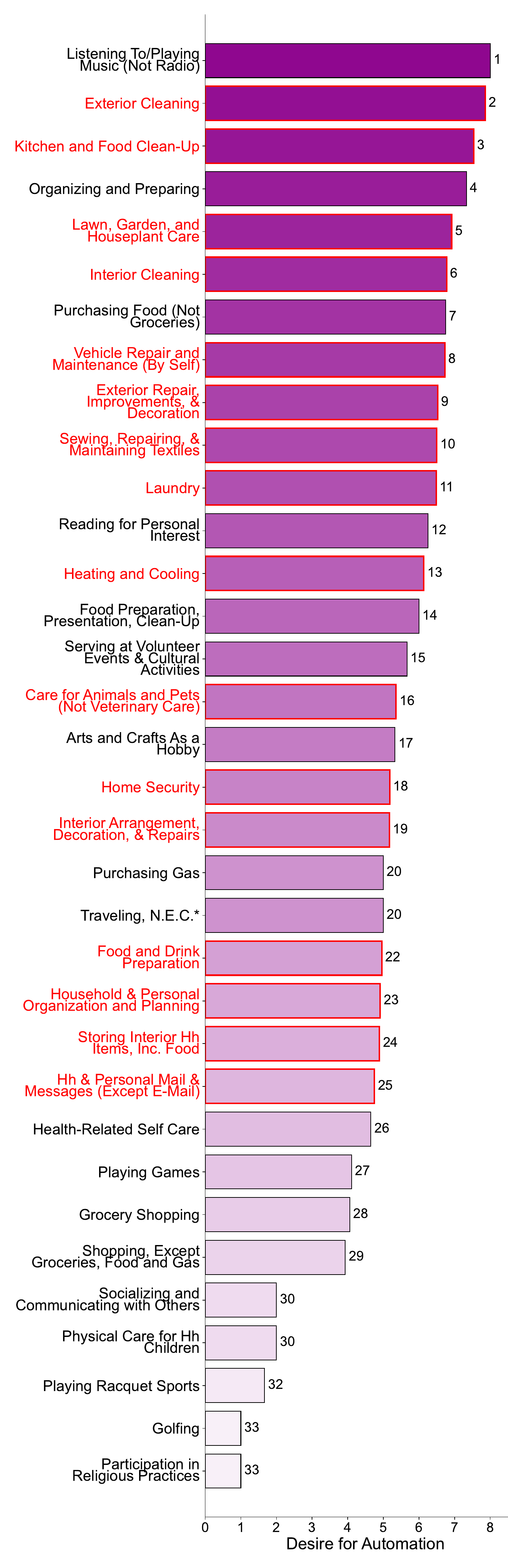}
\includegraphics[height=\highoffignew\columnwidth]{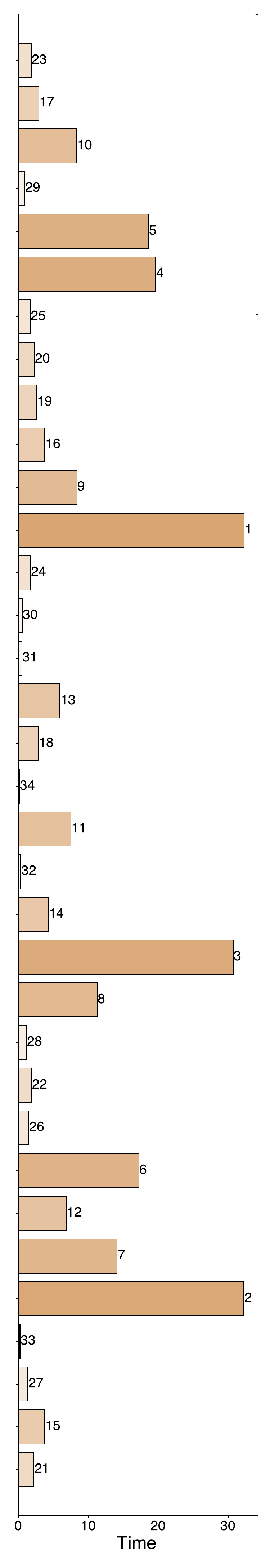}
\includegraphics[height=\highoffignew\columnwidth]{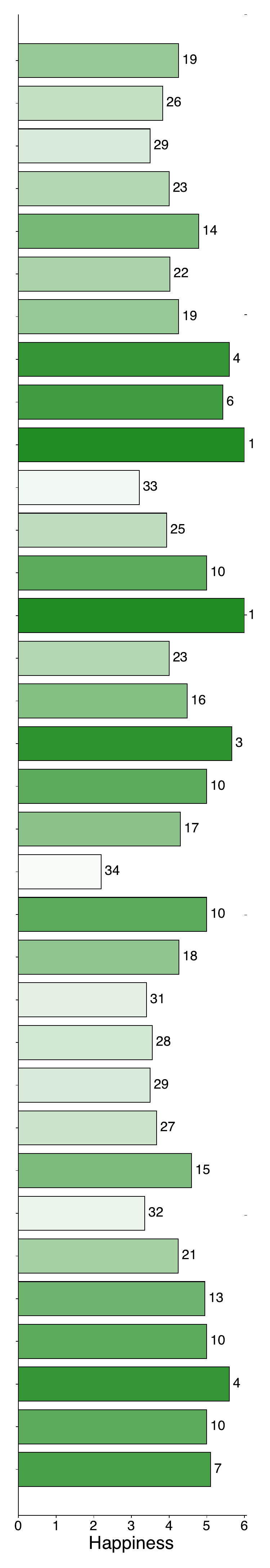}
\includegraphics[height=\highoffignew\columnwidth]{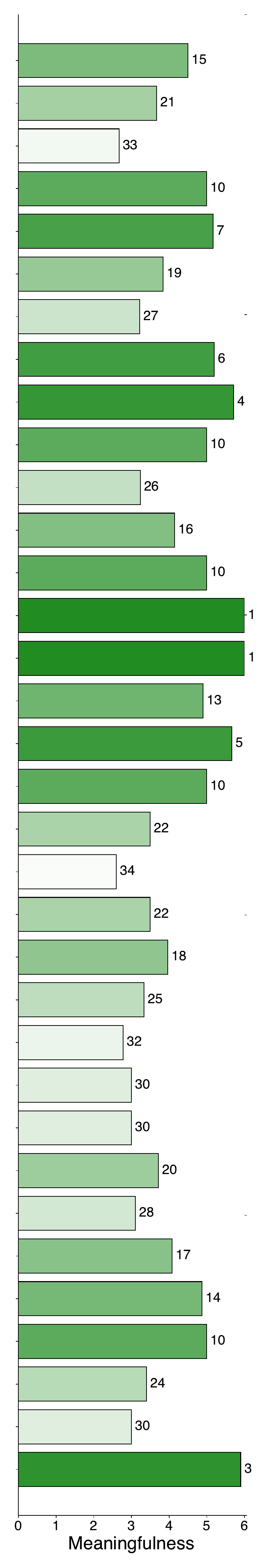}
\includegraphics[height=\highoffignew\columnwidth]{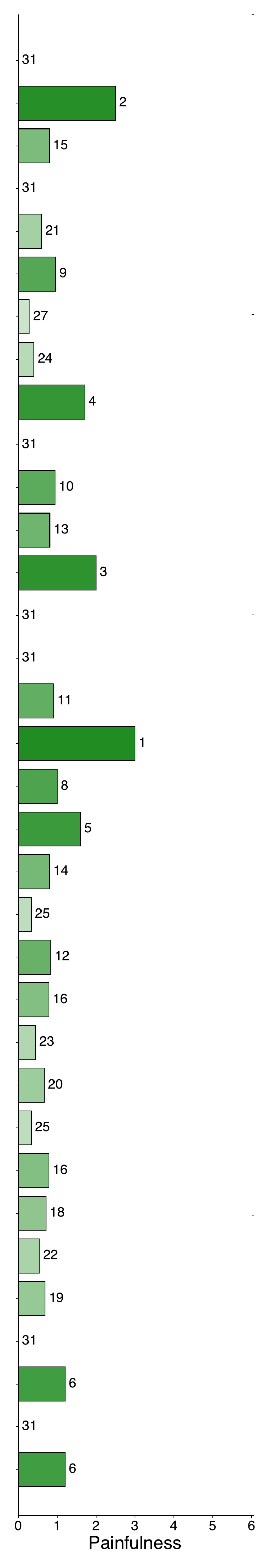}
\includegraphics[height=\highoffignew\columnwidth]{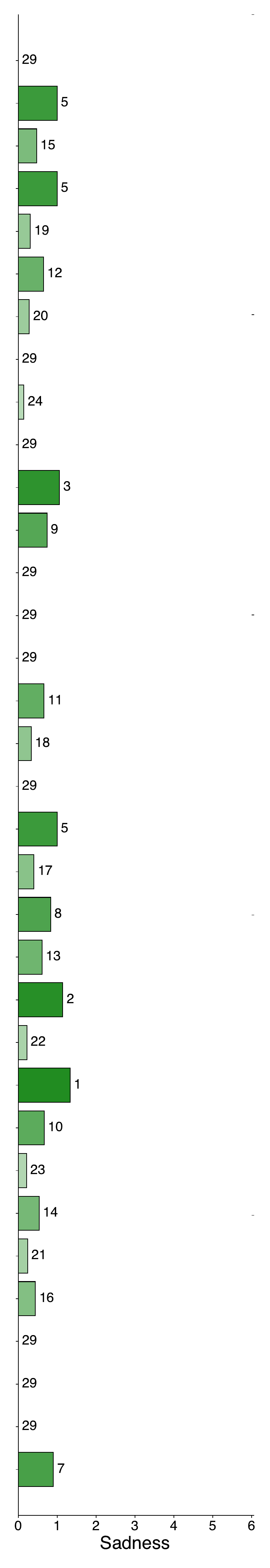}
\includegraphics[height=\highoffignew\columnwidth]{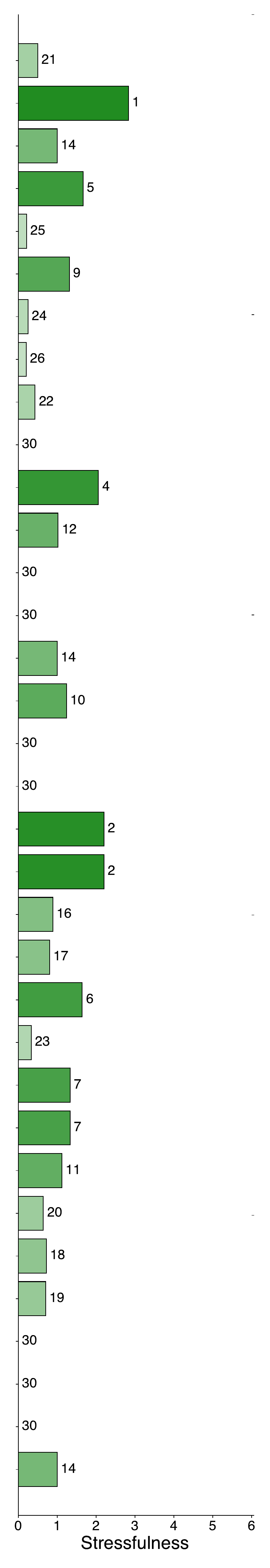}
\includegraphics[height=\highoffignew\columnwidth]{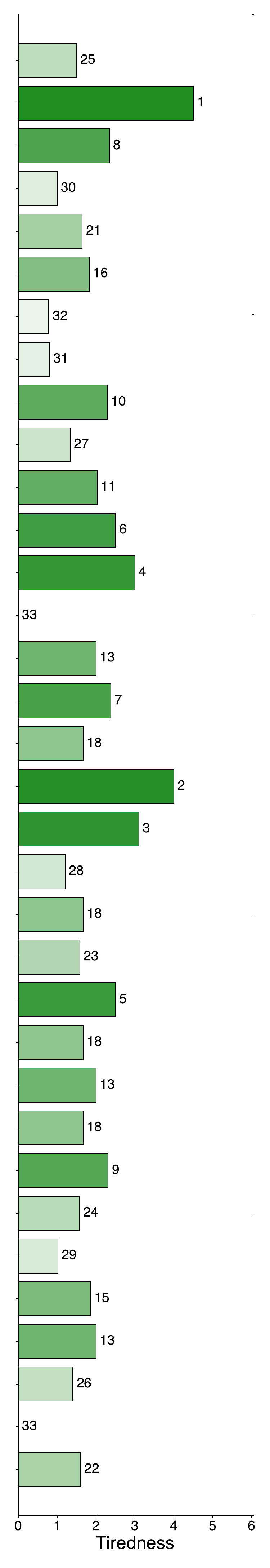}
\caption{\textbf{Absolute values and ranked activities} for \textbf{men (top row)} and \textbf{high income (bottom row)} participants based on Desire for Automation (1st from left), Time spent (2nd from left), Happiness (3rd from left), Meaningfulness (4th from left), Painfulness (5th from left), Sadness (6th from left), Stressfulness (7th from left) and Tiredness (most right); Darker color tone indicates higher rank, numbers next to the bars indicate ranking positions; Red labels and bar lines indicate activities of the \textit{Household Activities} subset}
\label{fig:all-M-L}
\end{figure*}

\begin{figure*}[t!]
\centering
\includegraphics[height=\highoffignew\columnwidth]{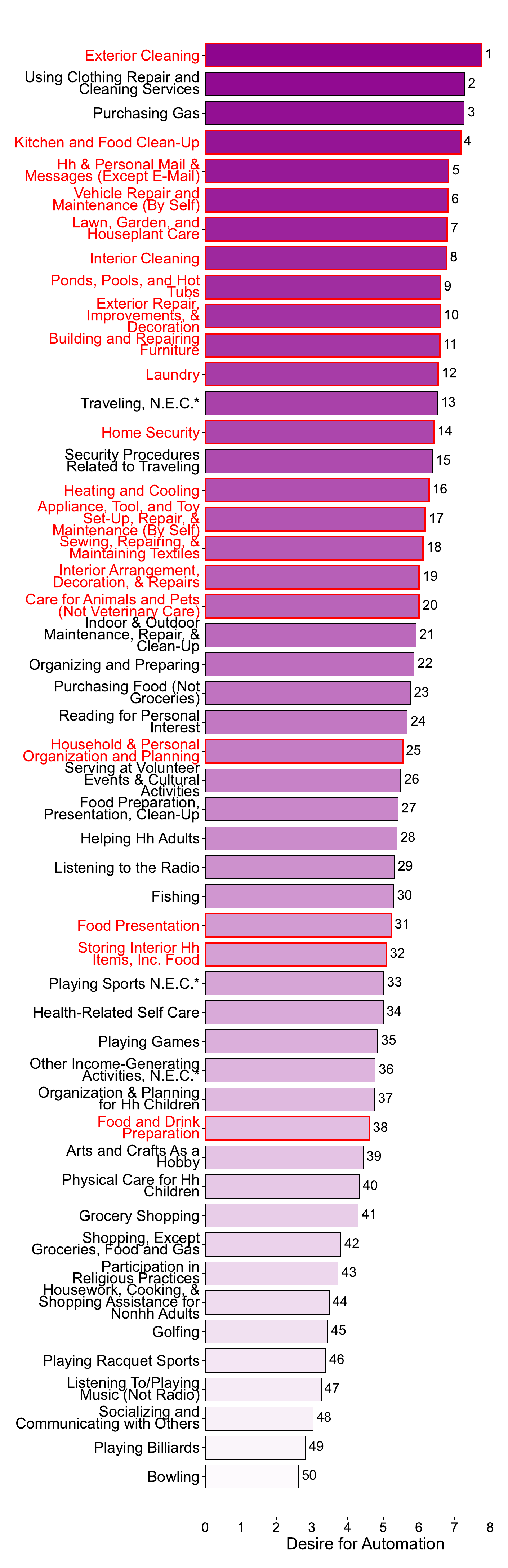}
\includegraphics[height=\highoffignew\columnwidth]{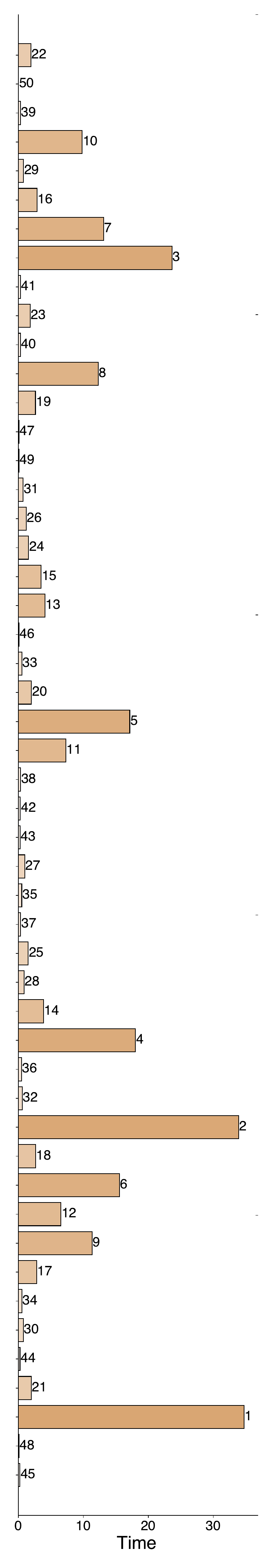}
\includegraphics[height=\highoffignew\columnwidth]{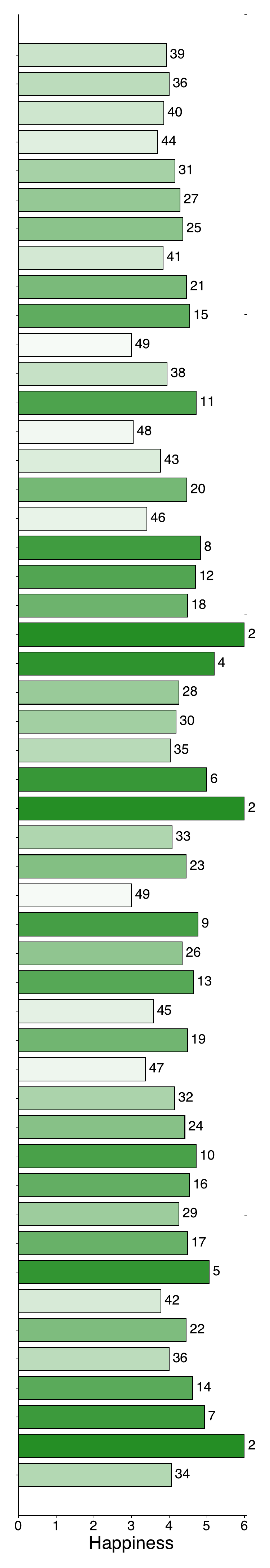}
\includegraphics[height=\highoffignew\columnwidth]{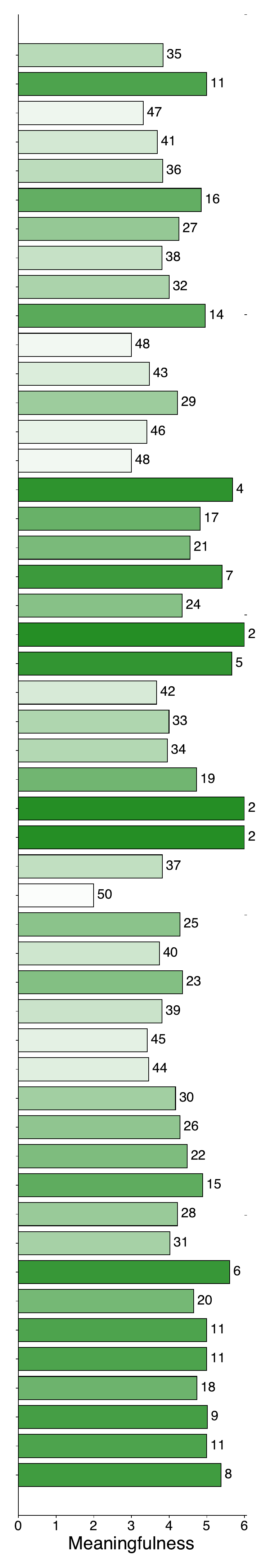}
\includegraphics[height=\highoffignew\columnwidth]{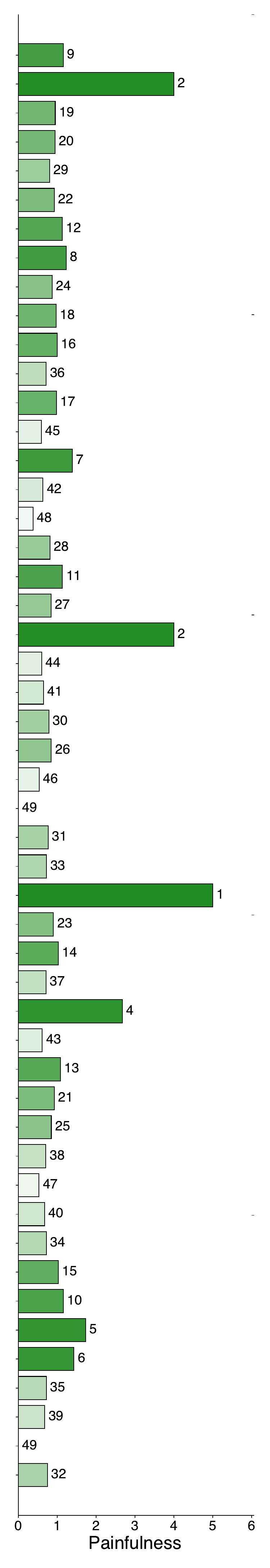}
\includegraphics[height=\highoffignew\columnwidth]{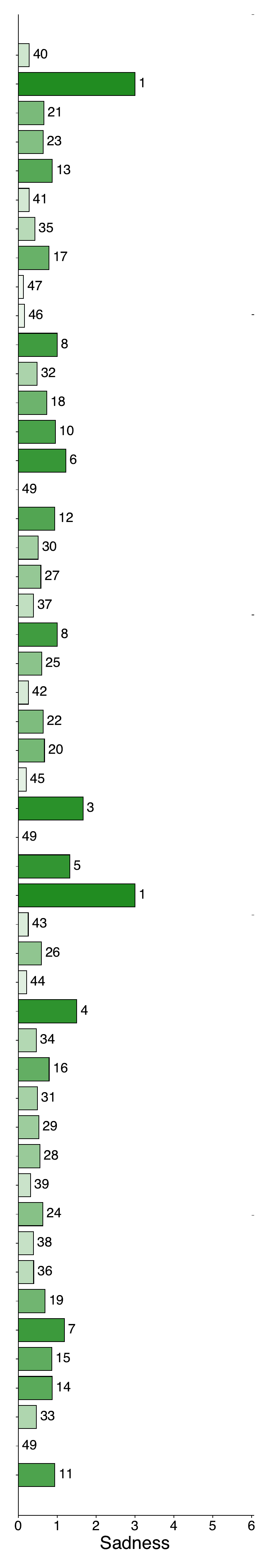}
\includegraphics[height=\highoffignew\columnwidth]{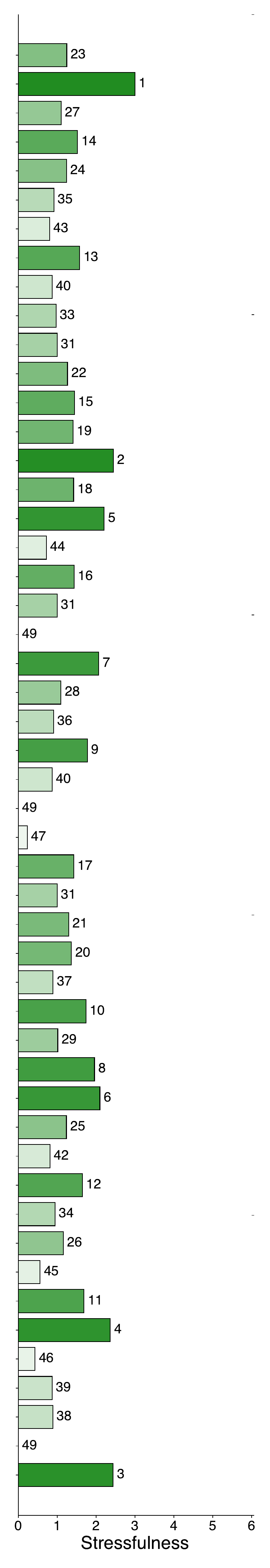}
\includegraphics[height=\highoffignew\columnwidth]{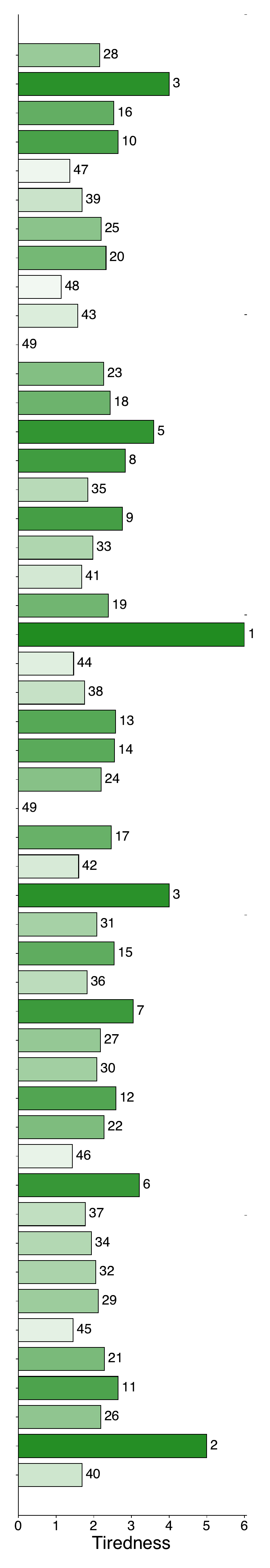}\\
\includegraphics[height=\highoffignew\columnwidth]{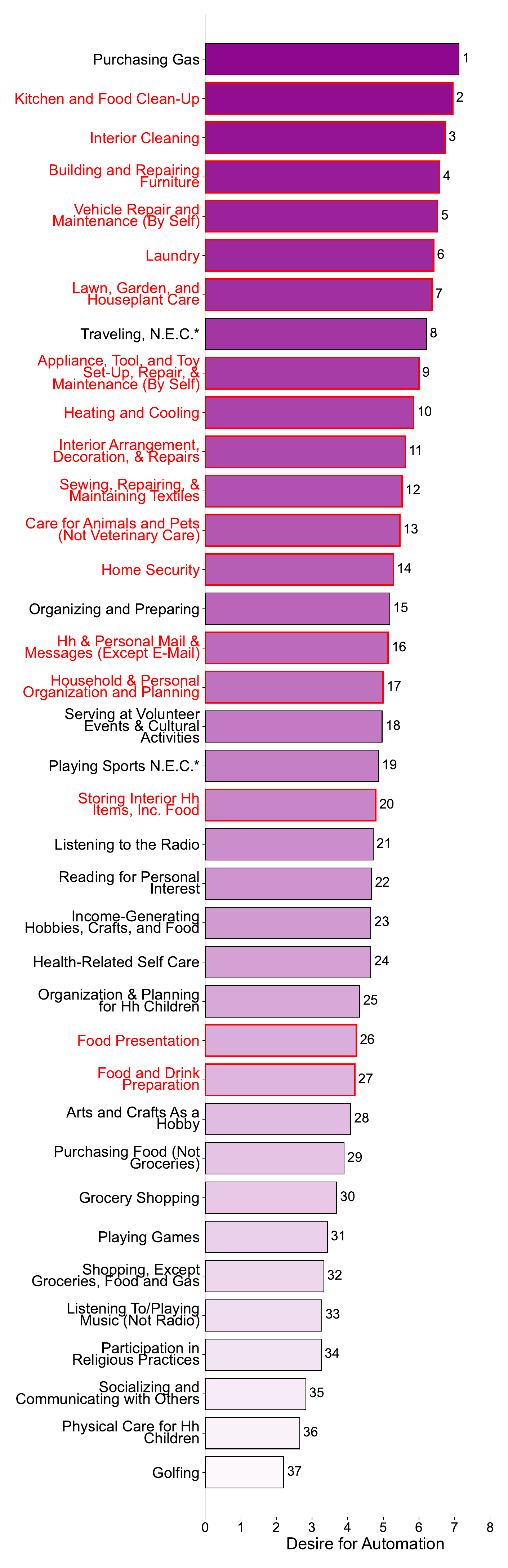}
\includegraphics[height=\highoffignew\columnwidth]{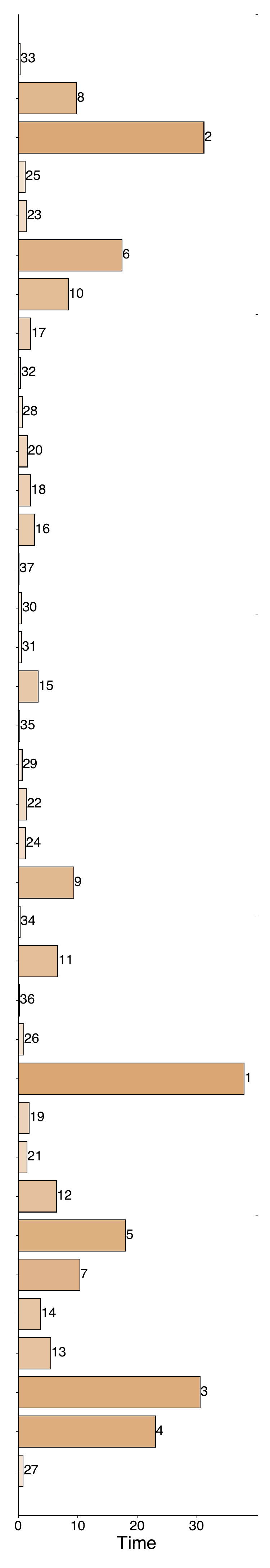}
\includegraphics[height=\highoffignew\columnwidth]{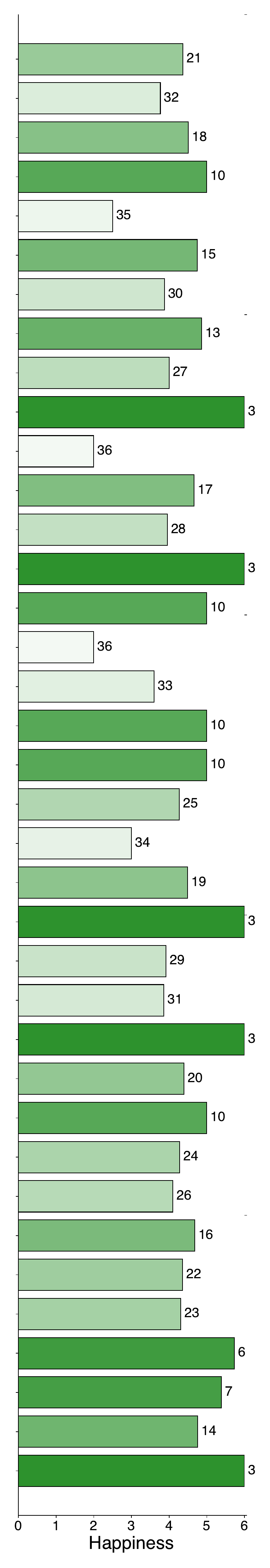}
\includegraphics[height=\highoffignew\columnwidth]{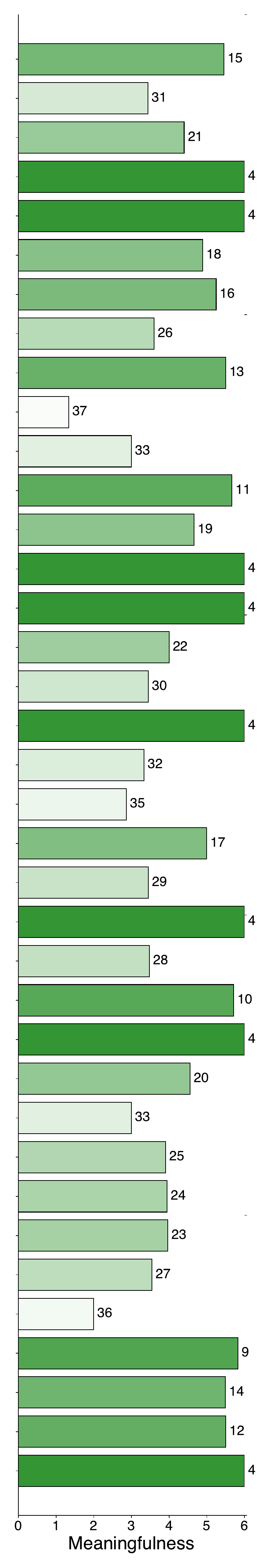}
\includegraphics[height=\highoffignew\columnwidth]{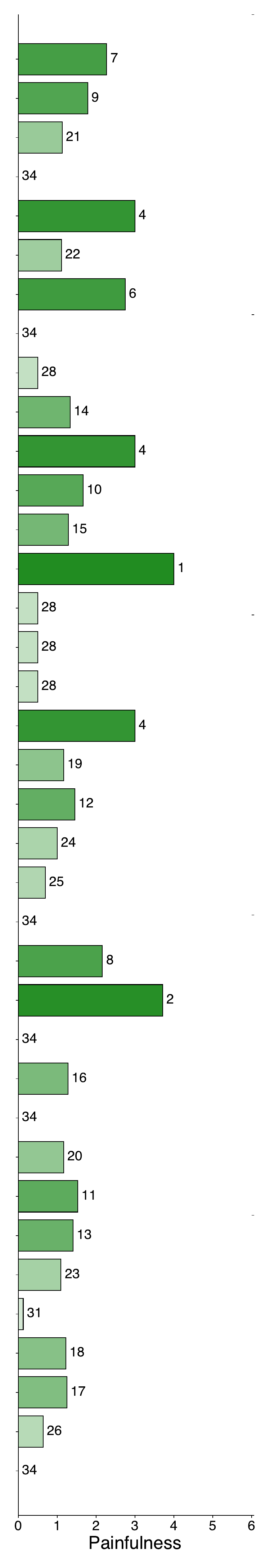}
\includegraphics[height=\highoffignew\columnwidth]{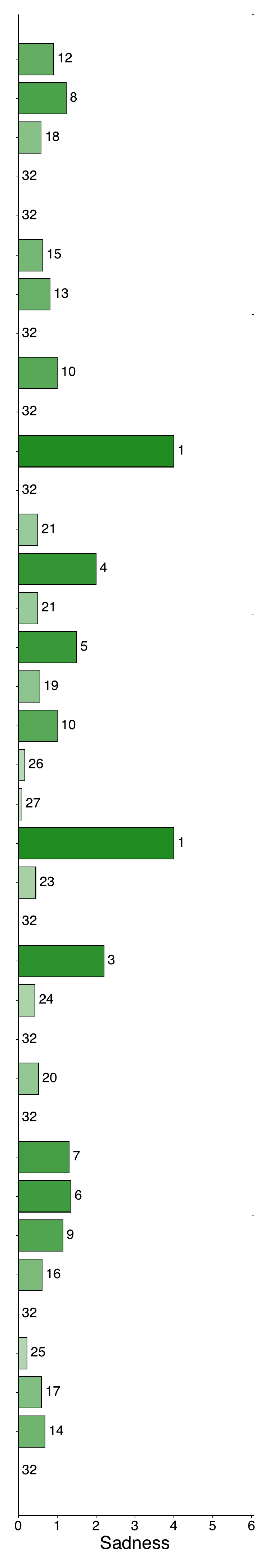}
\includegraphics[height=\highoffignew\columnwidth]{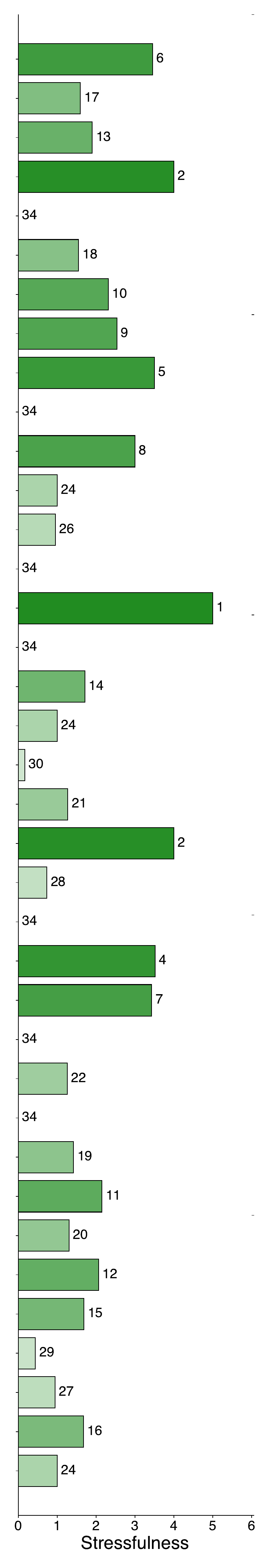}
\includegraphics[height=\highoffignew\columnwidth]{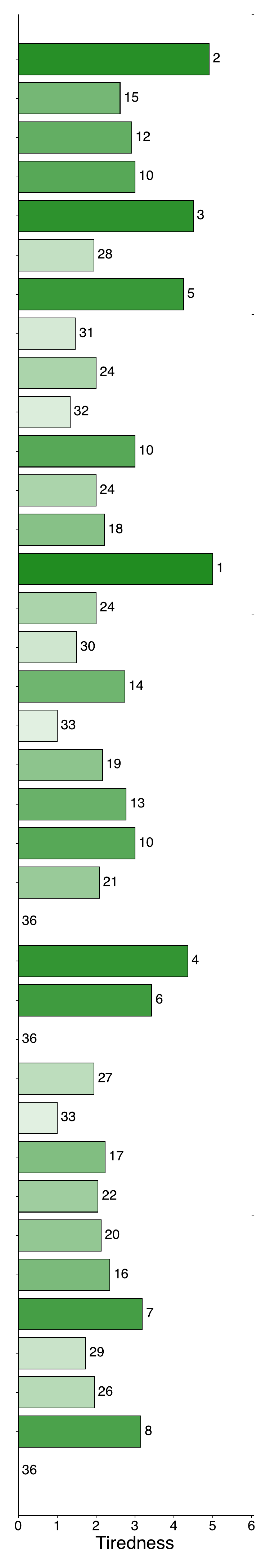}
\caption{\textbf{Absolute values and ranked activities} for \textbf{middle income (top row)} and \textbf{low income (bottom row)} participants based on Desire for Automation (1st from left), Time spent (2nd from left), Happiness (3rd from left), Meaningfulness (4th from left), Painfulness (5th from left), Sadness (6th from left), Stressfulness (7th from left) and Tiredness (most right); Darker color tone indicates higher rank, numbers next to the bars indicate ranking positions; Red labels and bar lines indicate activities of the \textit{Household Activities} subset.}
\label{fig:all-Mid-H}
\end{figure*}

\newcommand{\rankdiff}{0.195}

% female

\newcommand{\rankdiffr}{0.24}

\begin{figure*}[t!]
\centering
\includegraphics[width=\rankdiffr\textwidth]{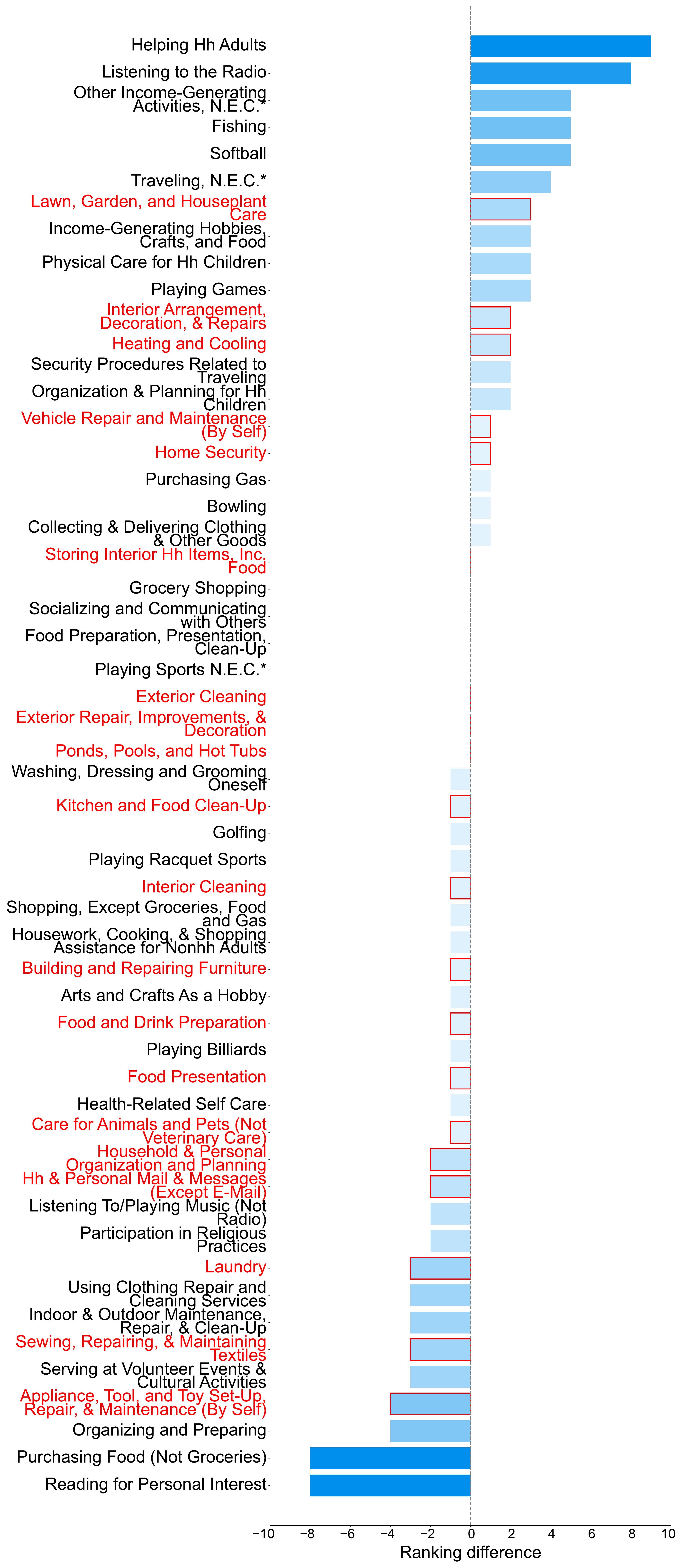}
\includegraphics[width=\rankdiffr\textwidth]{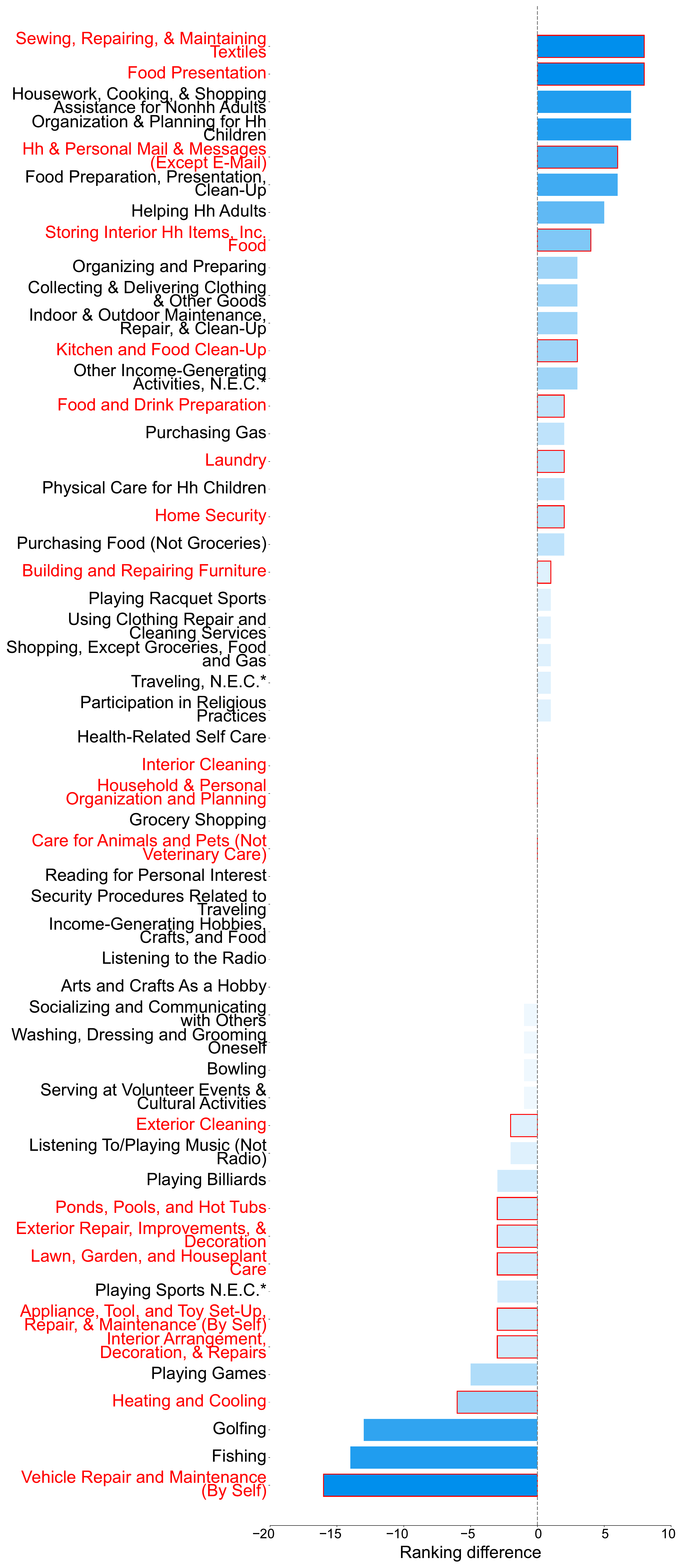}
\includegraphics[width=\rankdiffr\textwidth]{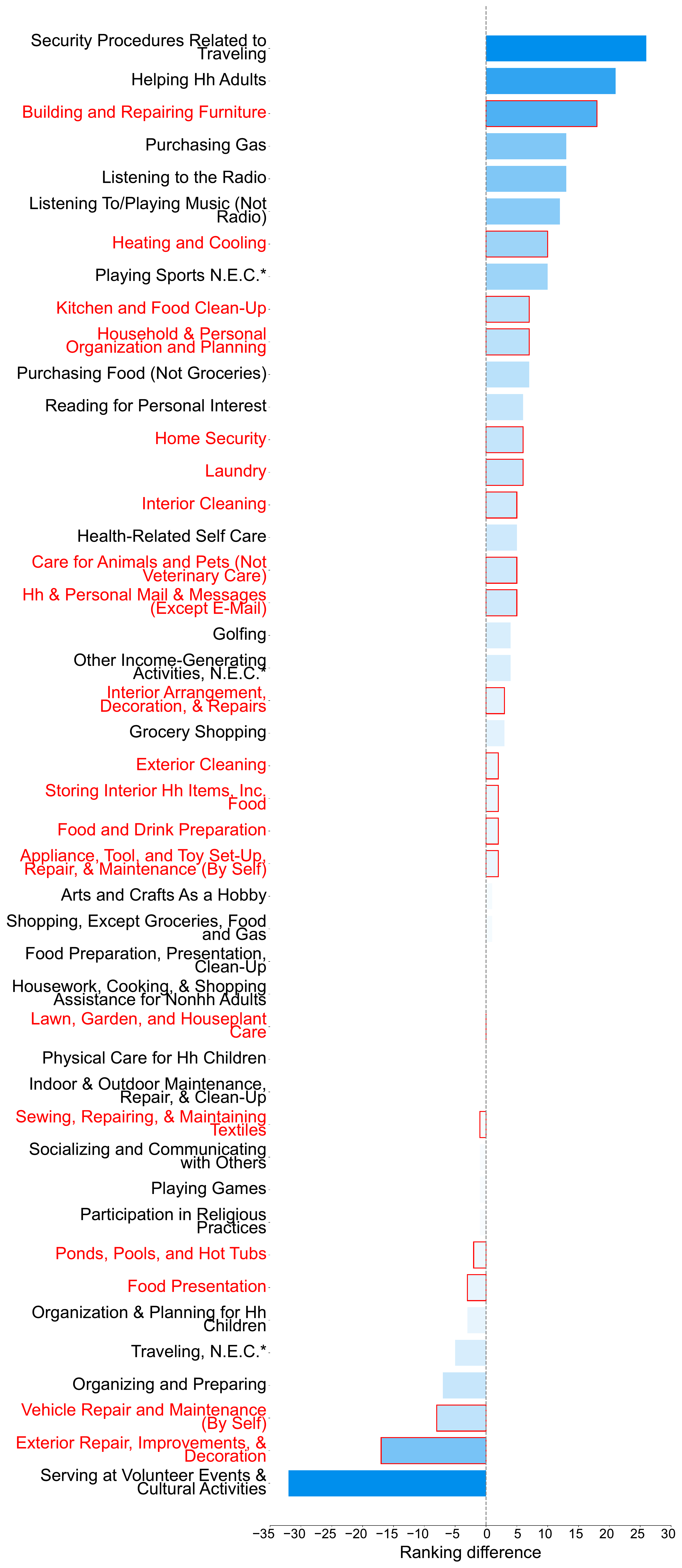}
\includegraphics[width=\rankdiffr\textwidth]{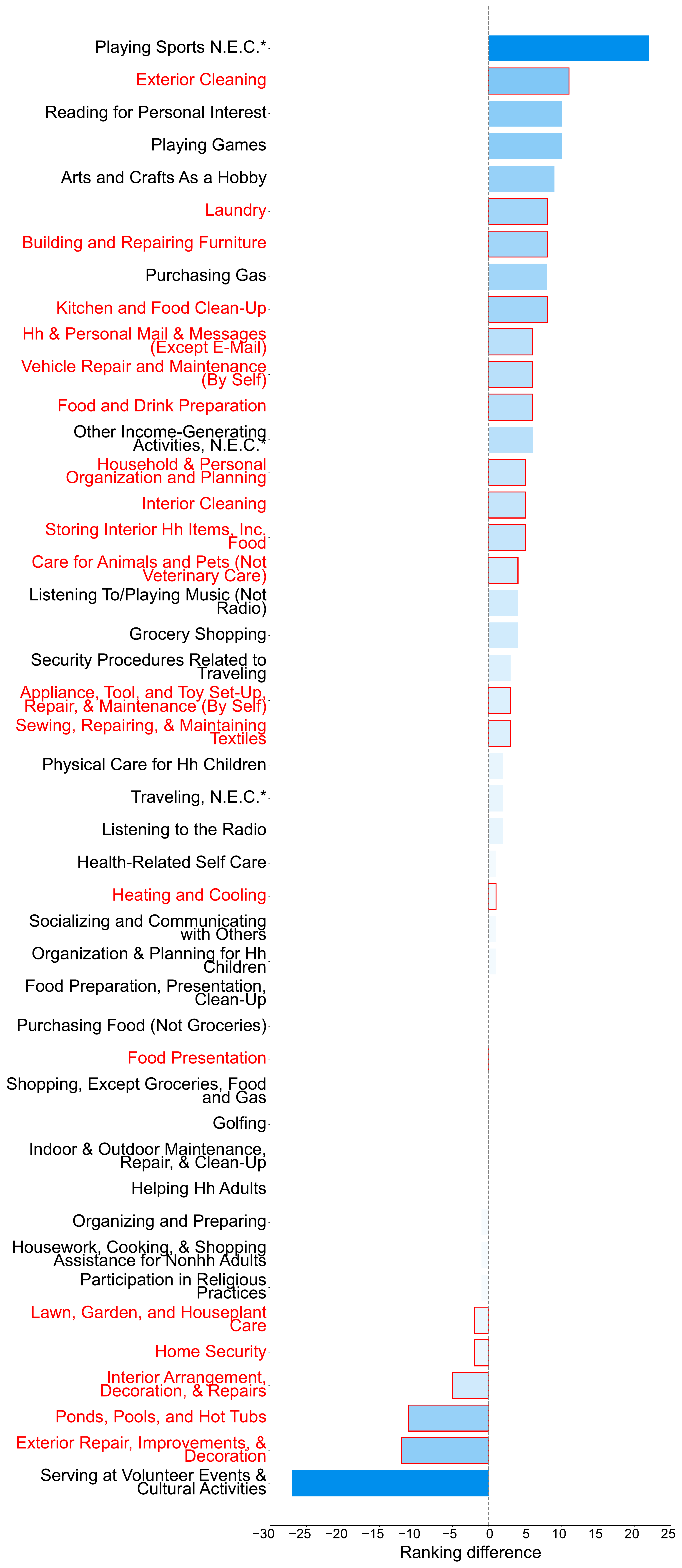}\\
\vspace{1em}
\includegraphics[width=\rankdiffr\textwidth]{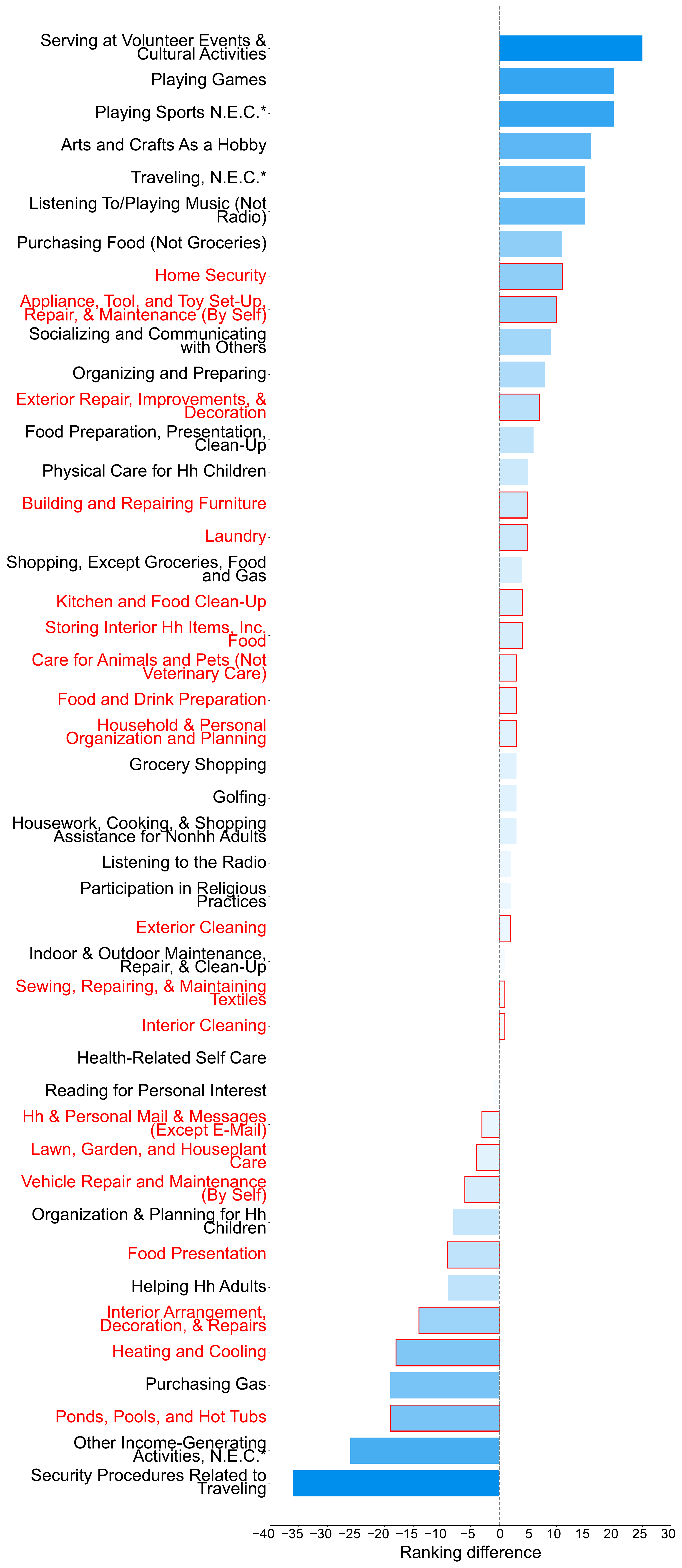}
\includegraphics[width=\rankdiffr\textwidth]{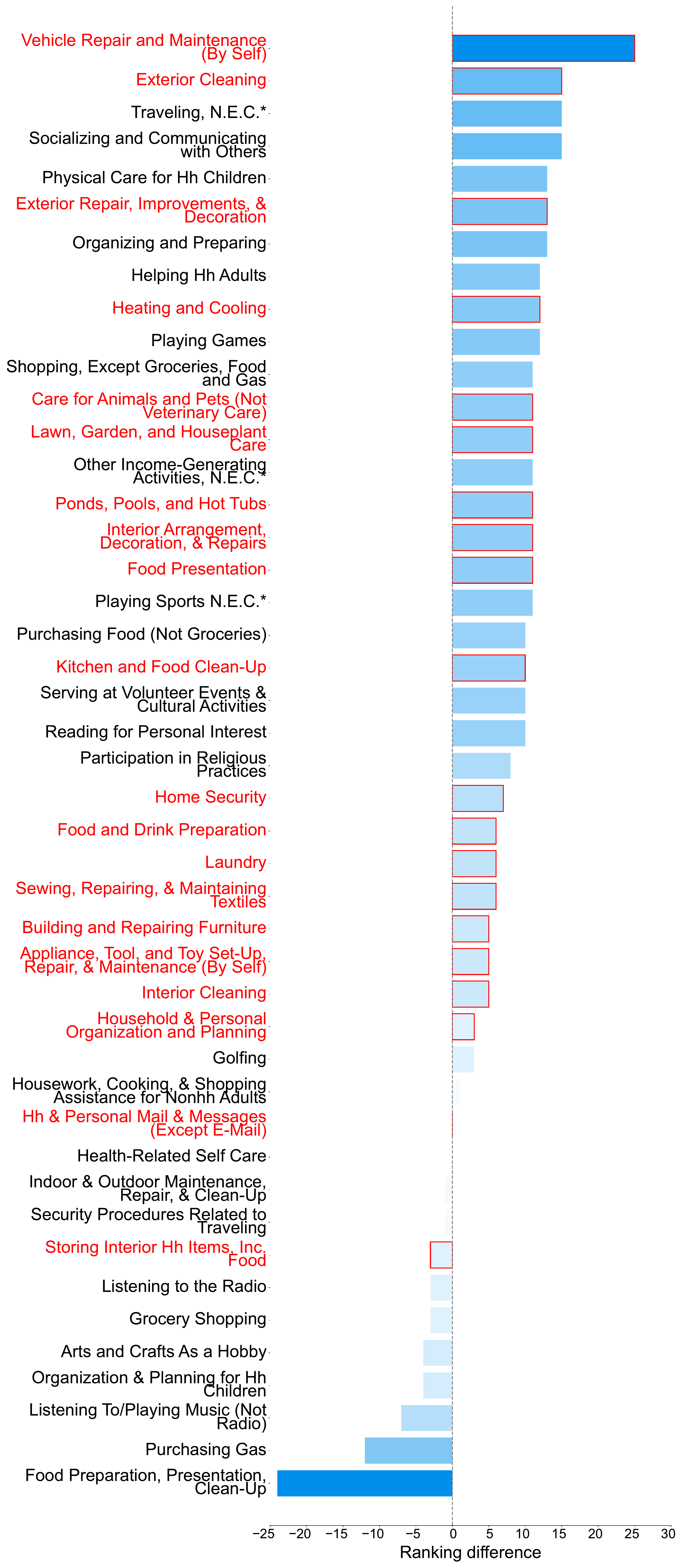}
\includegraphics[width=\rankdiffr\textwidth]{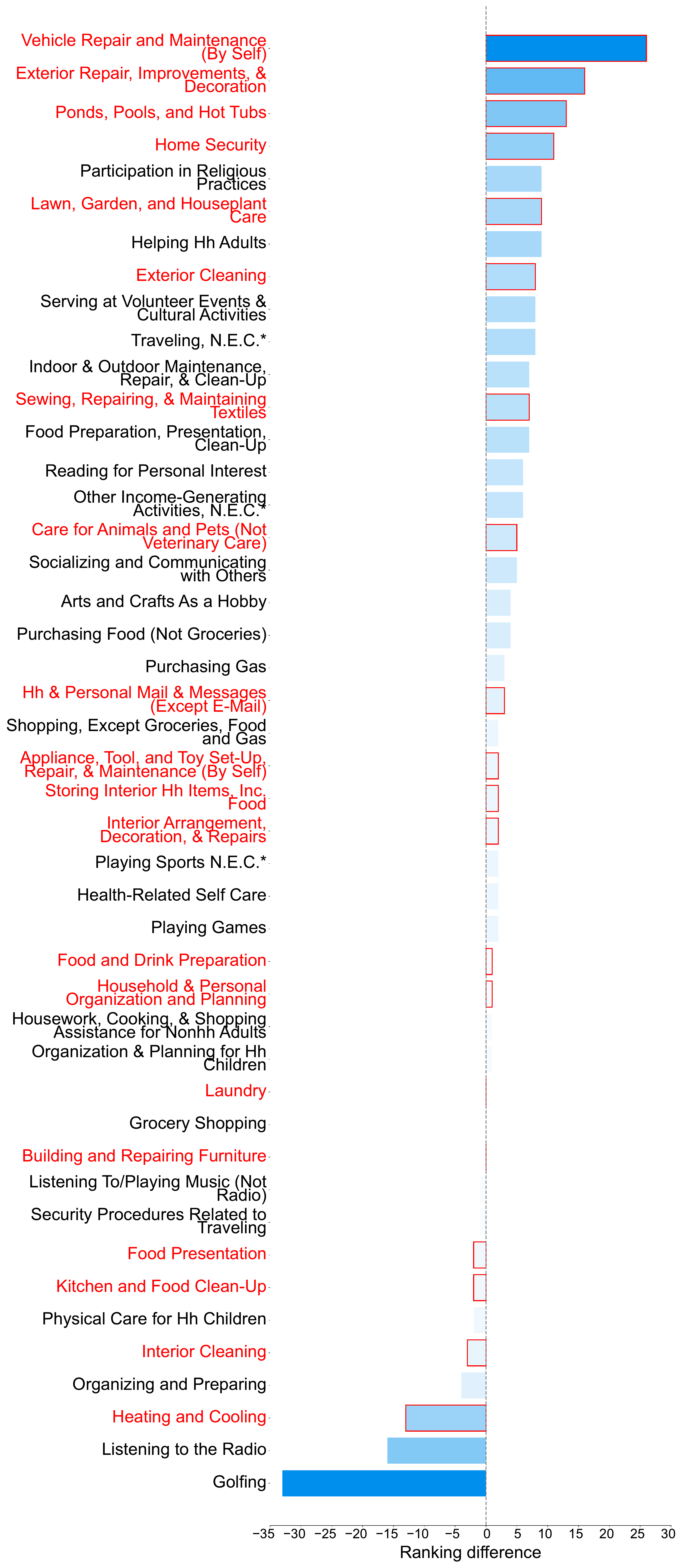}
\includegraphics[width=\rankdiffr\textwidth]{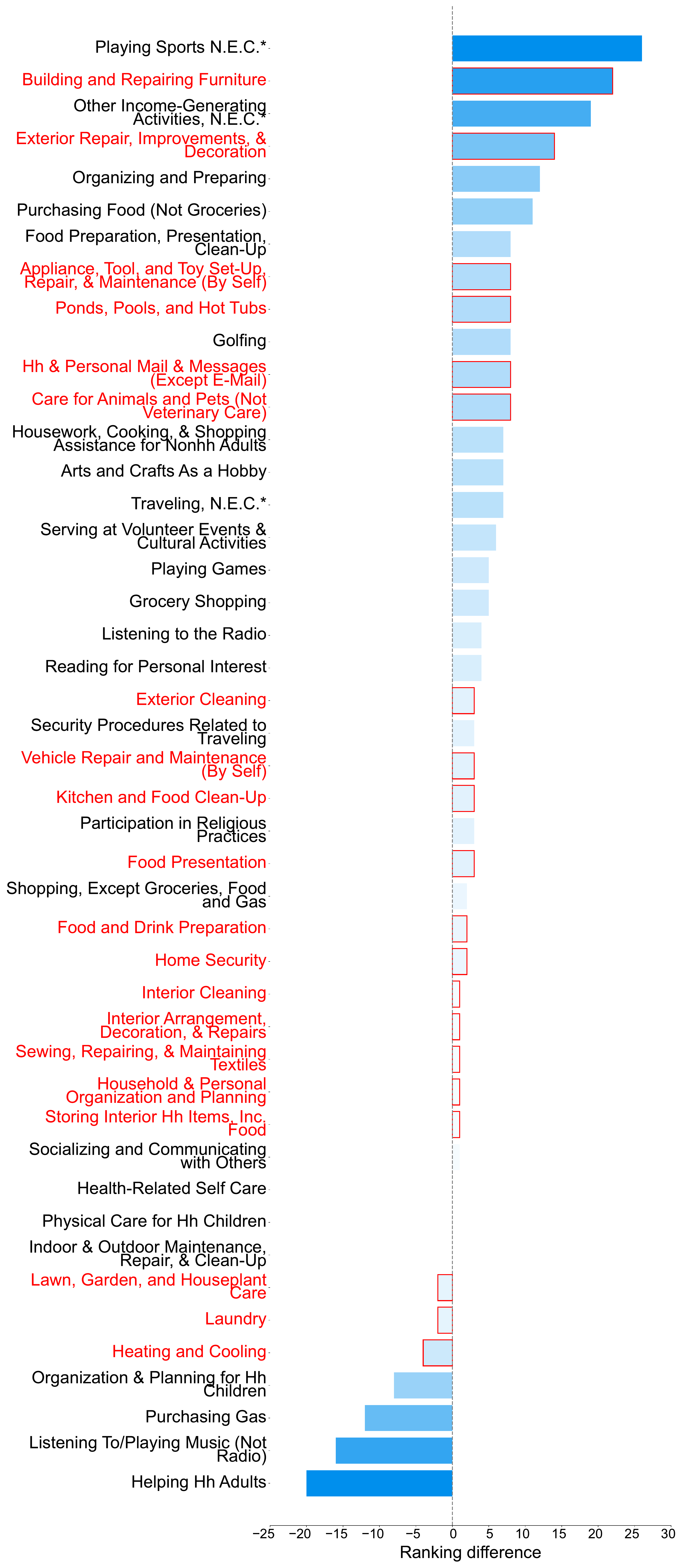}

\caption{\textbf{Difference in ranking} for all activities between the \textbf{general population and women}; Darker color indicates higher rank differences; Differences for Desire for Automation (top row, 1st from left), Time spent (top row, 2nd from left), Happiness (top row, 3rd from left), Meaningfulness (top row, most right), Painfulness (bottom row, 1st from left), Sadness (bottom row, 2nd from left), Stressfulness (bottom row, 3rd from left) and Tiredness (bottom row, most right)}

\end{figure*}

% male

\begin{figure*}[t!]
\centering
\includegraphics[width=\rankdiffr\textwidth]{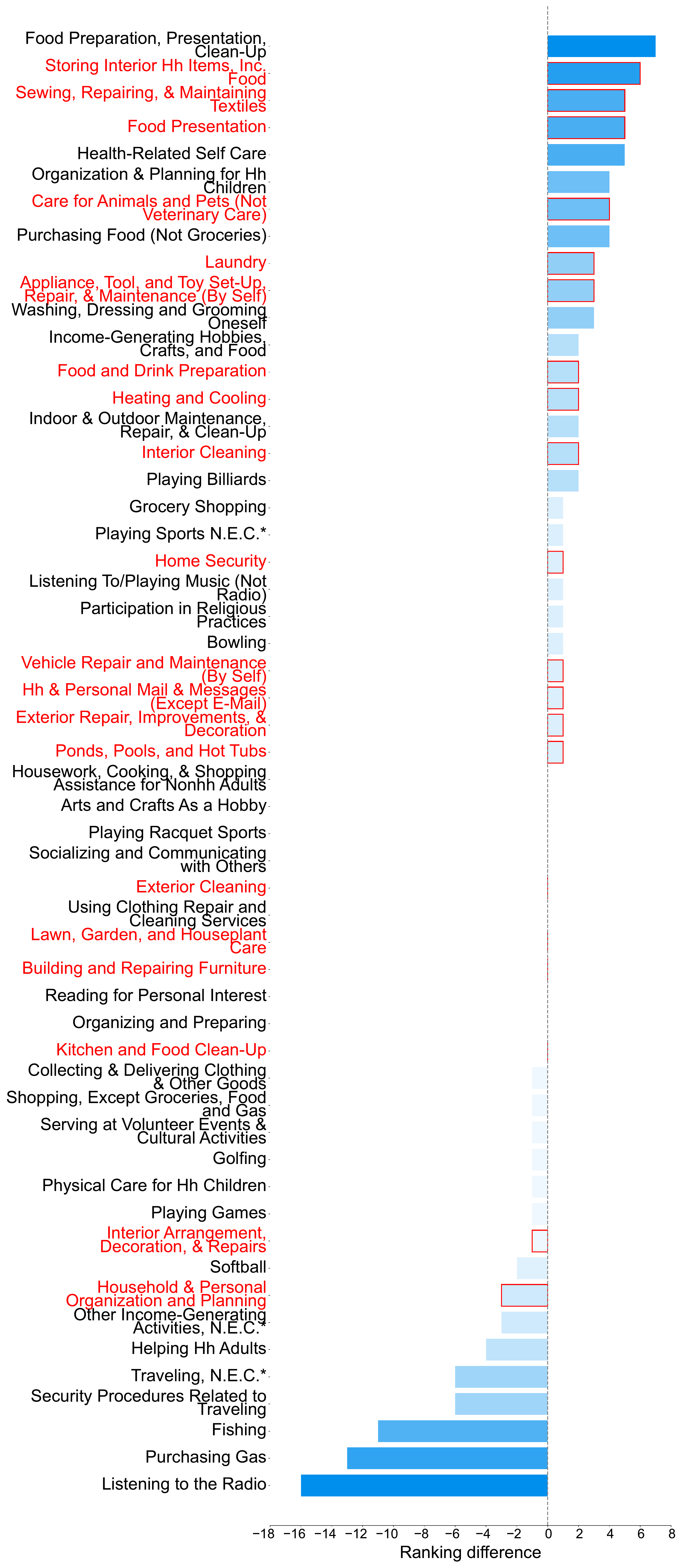}
\includegraphics[width=\rankdiffr\textwidth]{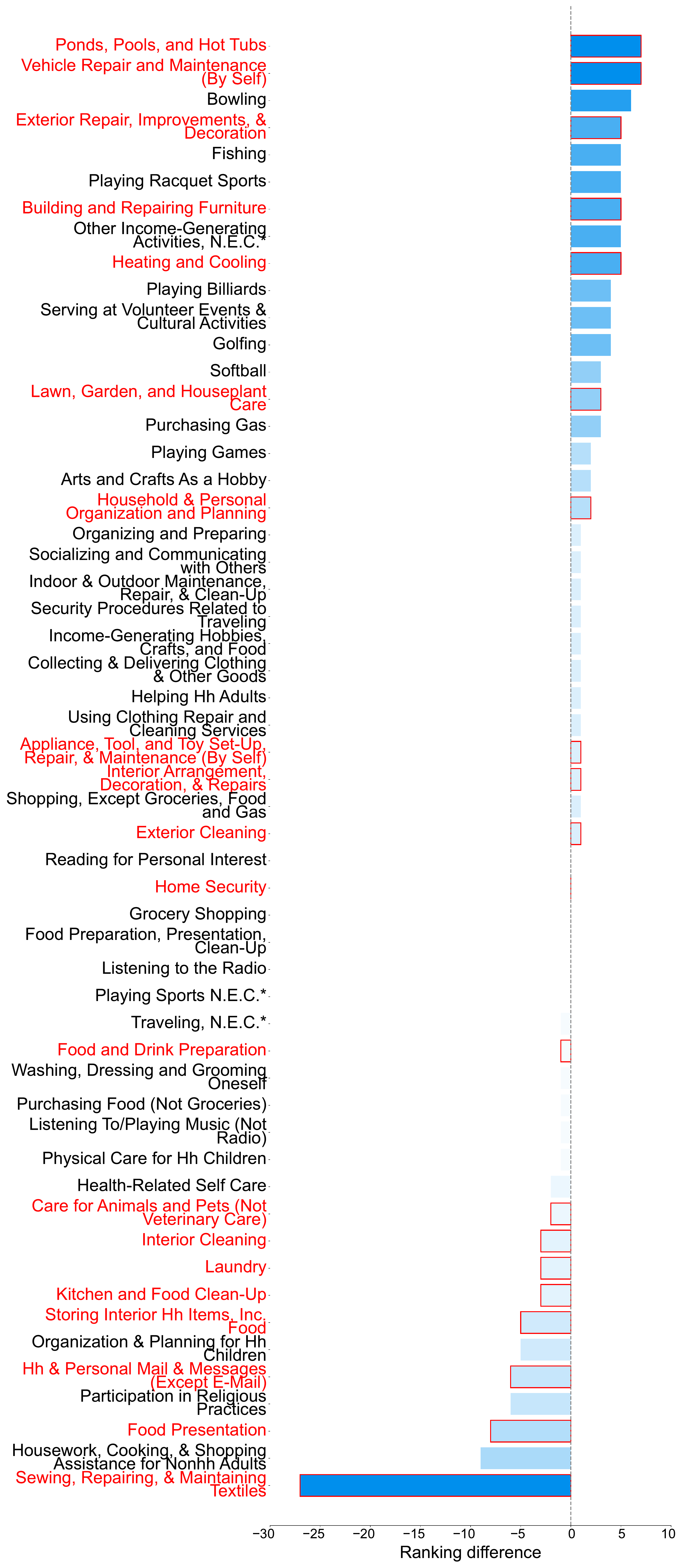}
\includegraphics[width=\rankdiffr\textwidth]{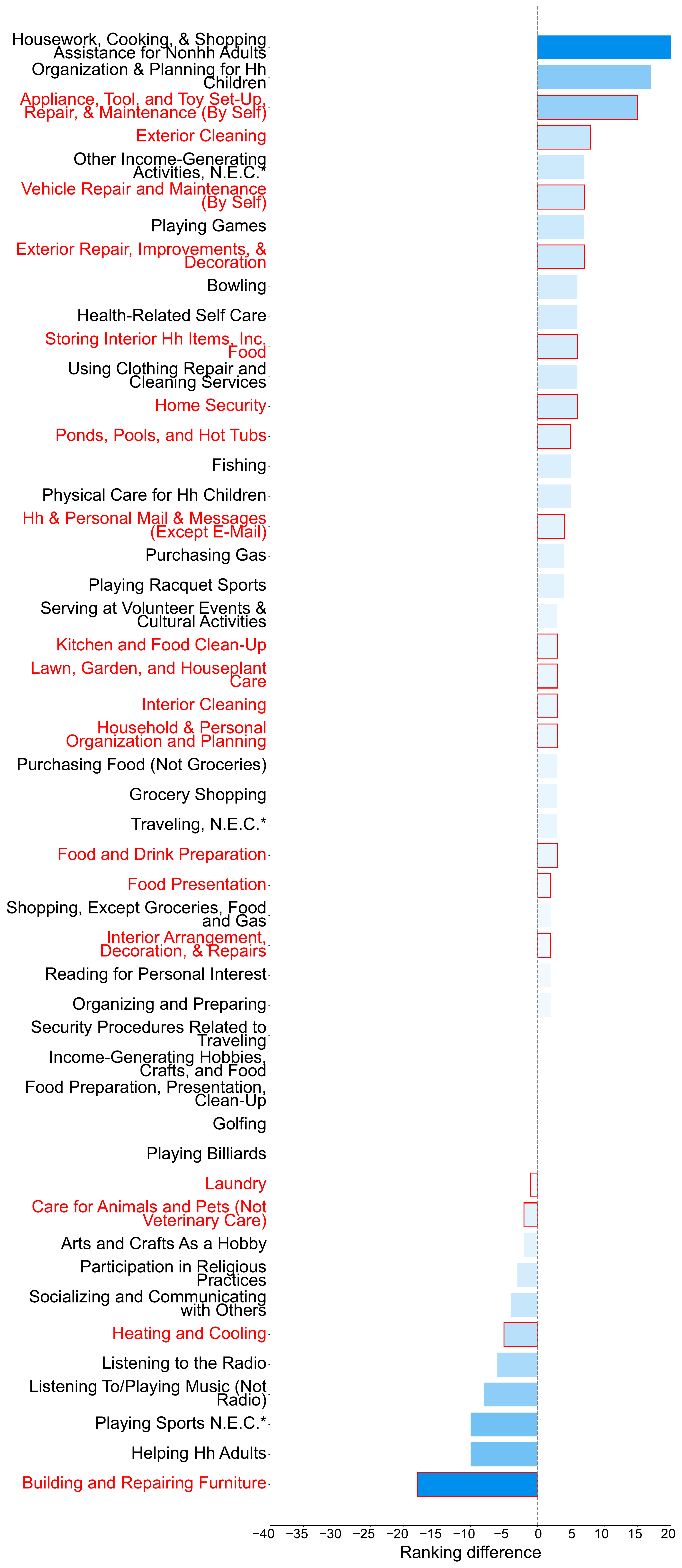}
\includegraphics[width=\rankdiffr\textwidth]{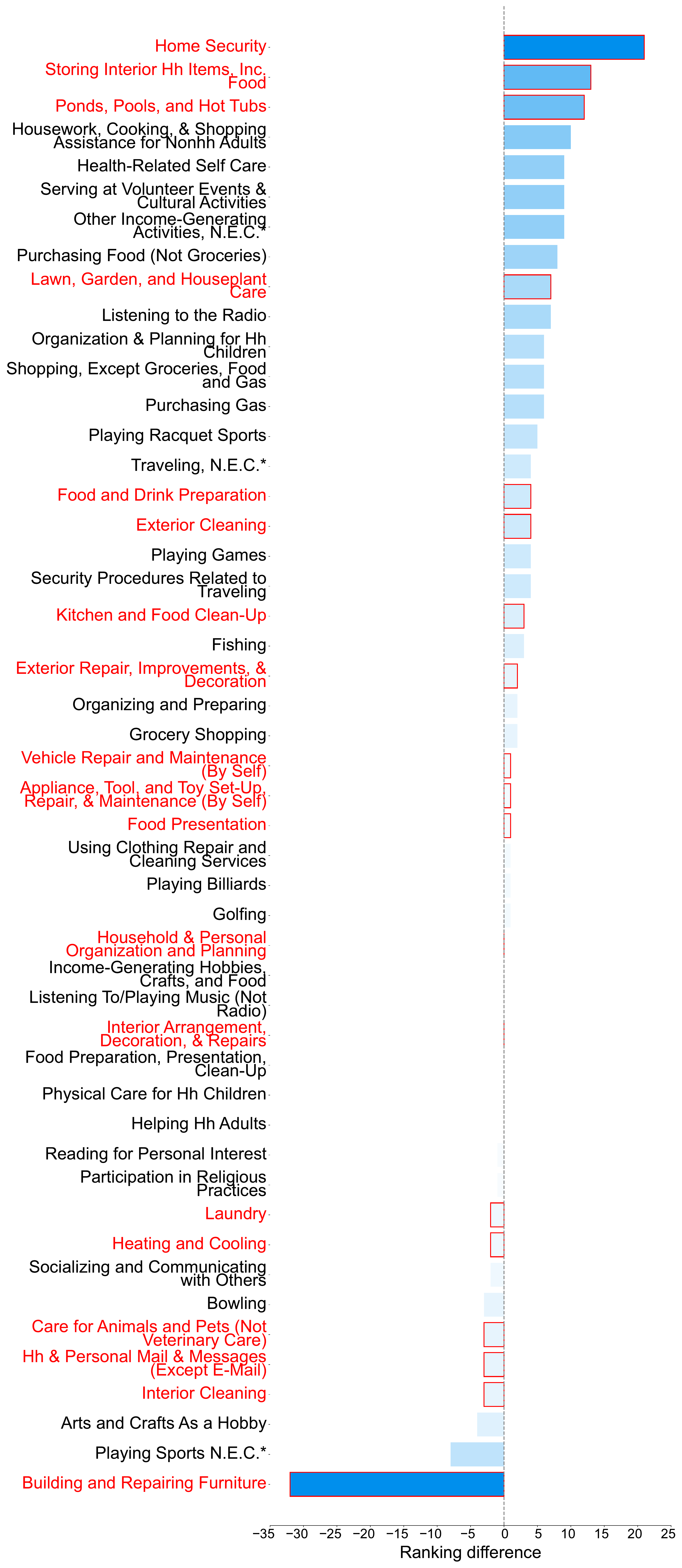}\\
\vspace{1em}
\includegraphics[width=\rankdiffr\textwidth]{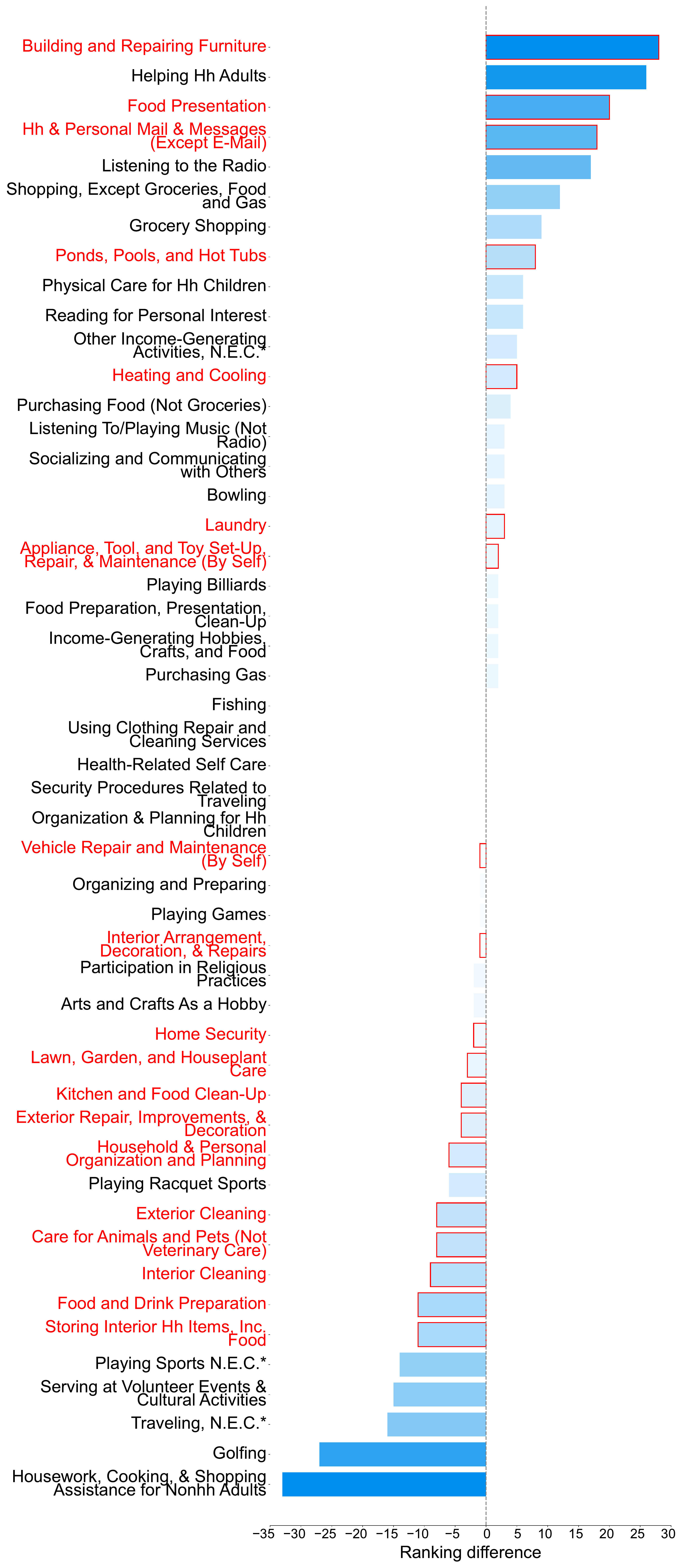}
\includegraphics[width=\rankdiffr\textwidth]{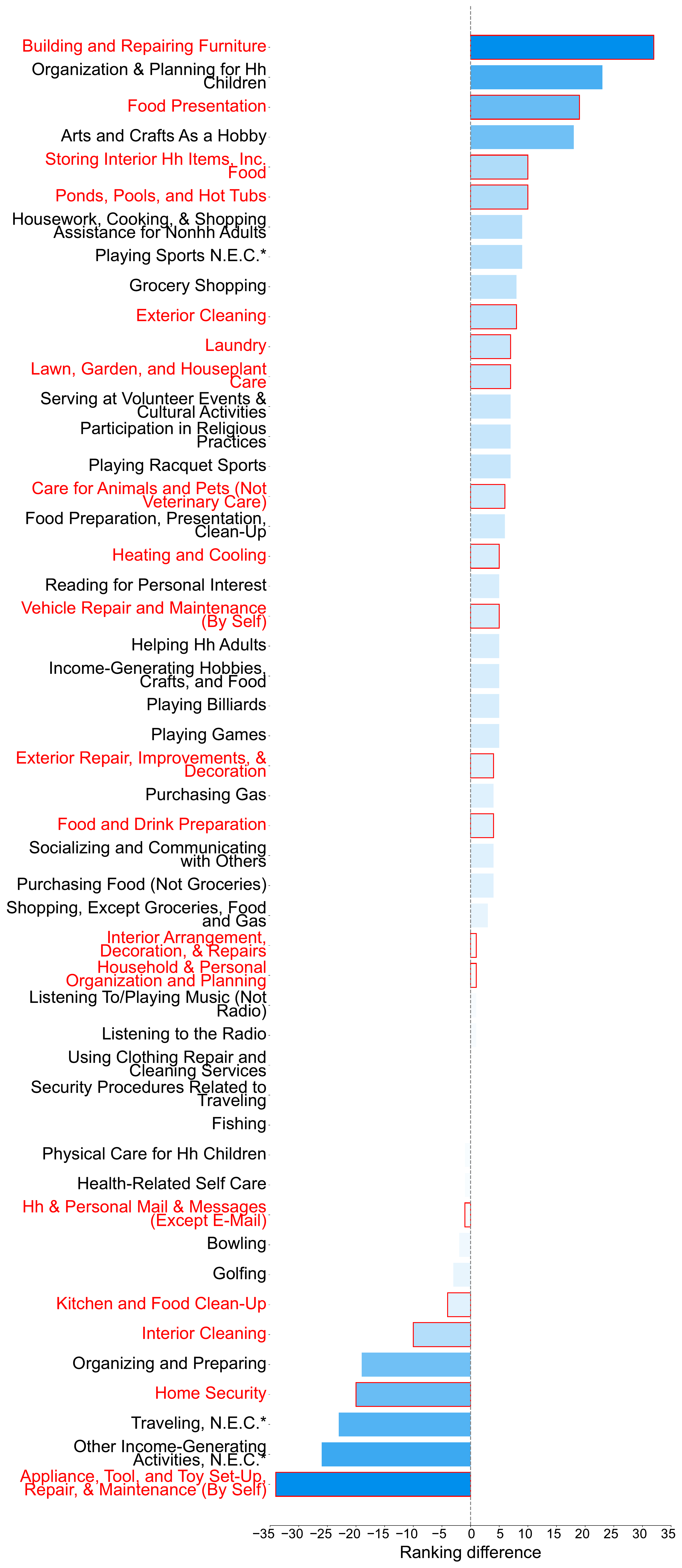}
\includegraphics[width=\rankdiffr\textwidth]{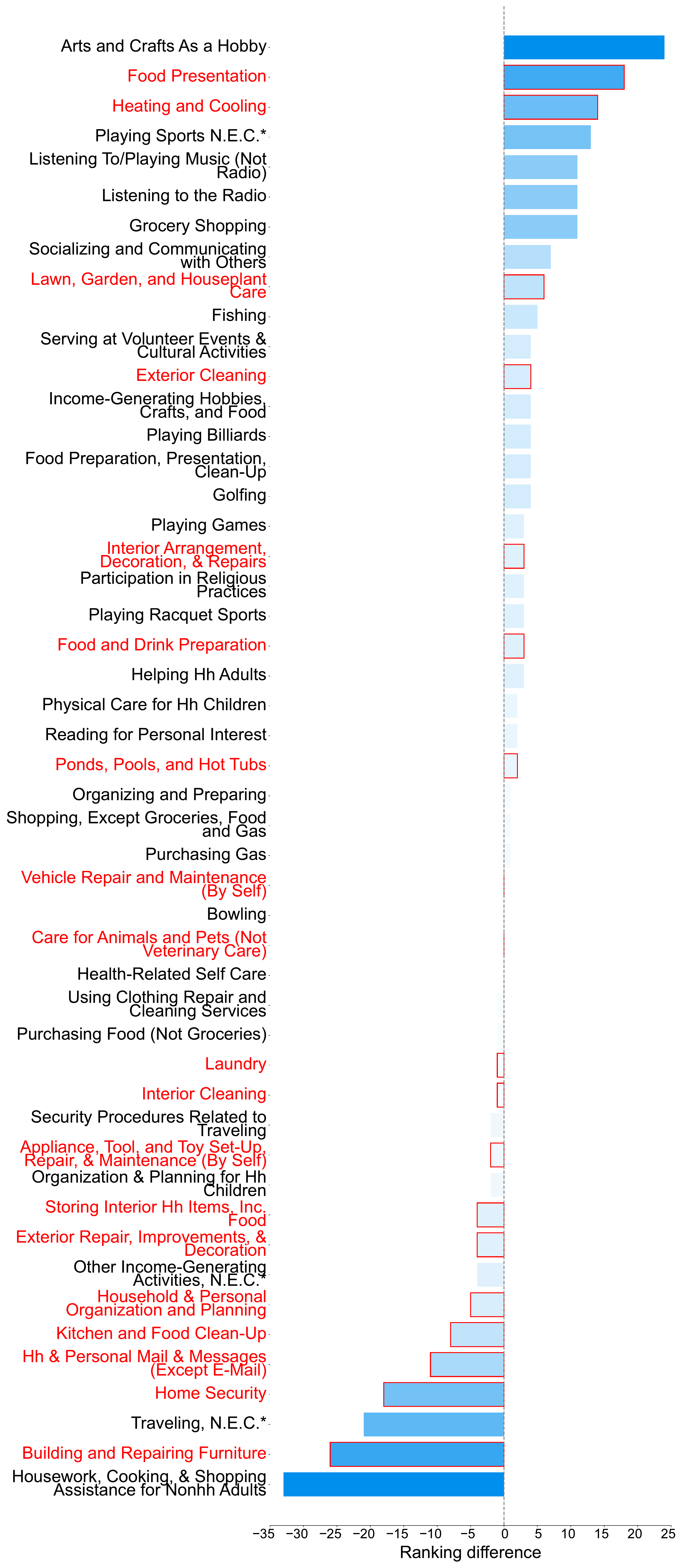}
\includegraphics[width=\rankdiffr\textwidth]{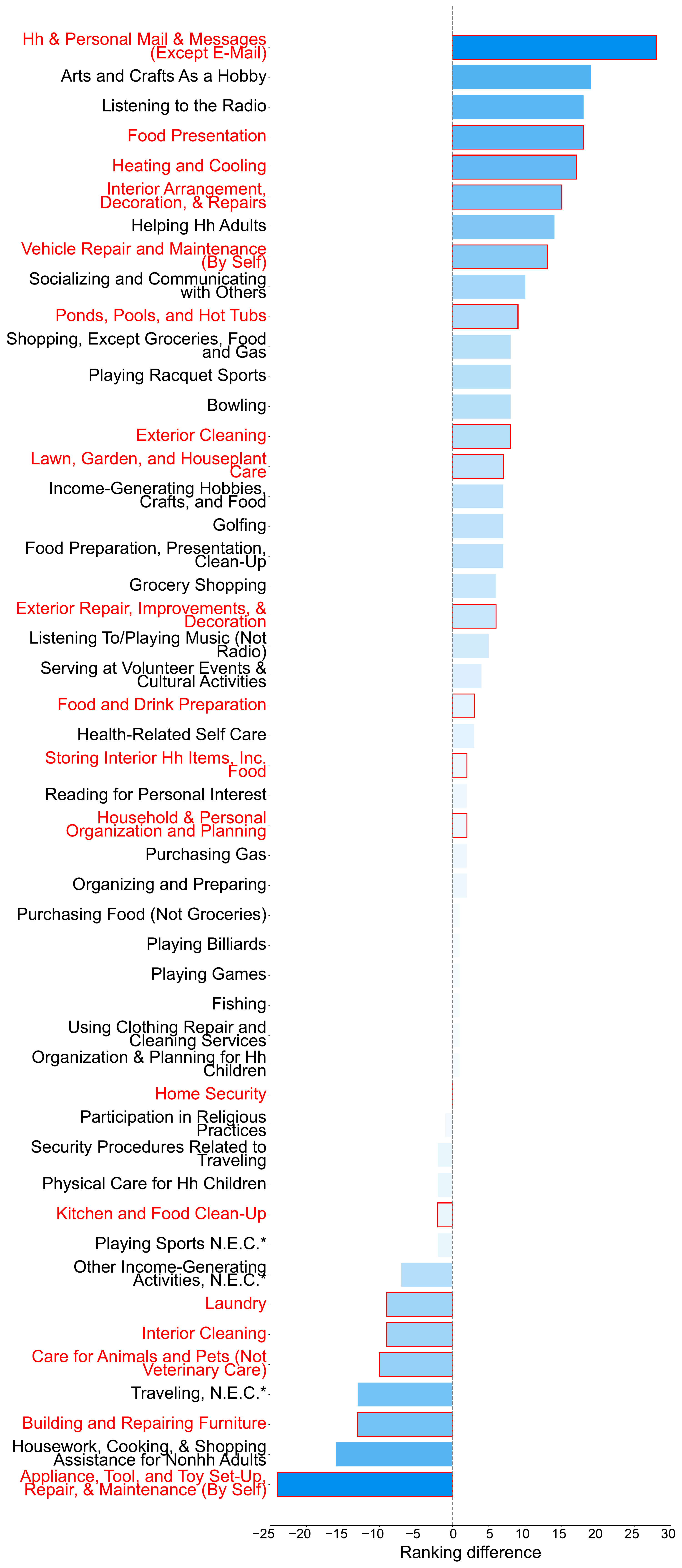}

\caption{\textbf{Difference in ranking} for all activities between the \textbf{general population and men}; Darker color indicates higher rank differences; Red labels and bar lines indicate activities of the \textit{Household Activities} subset; Differences for Desire for Automation (top row, 1st from left), Time spent (top row, 2nd from left), Happiness (top row, 3rd from left), Meaningfulness (top row, most right), Painfulness (bottom row, 1st from left), Sadness (bottom row, 2nd from left), Stressfulness (bottom row, 3rd from left) and Tiredness (bottom row, most right).}

\end{figure*}

% high

\begin{figure*}[t!]
\centering
\includegraphics[width=\rankdiffr\textwidth]{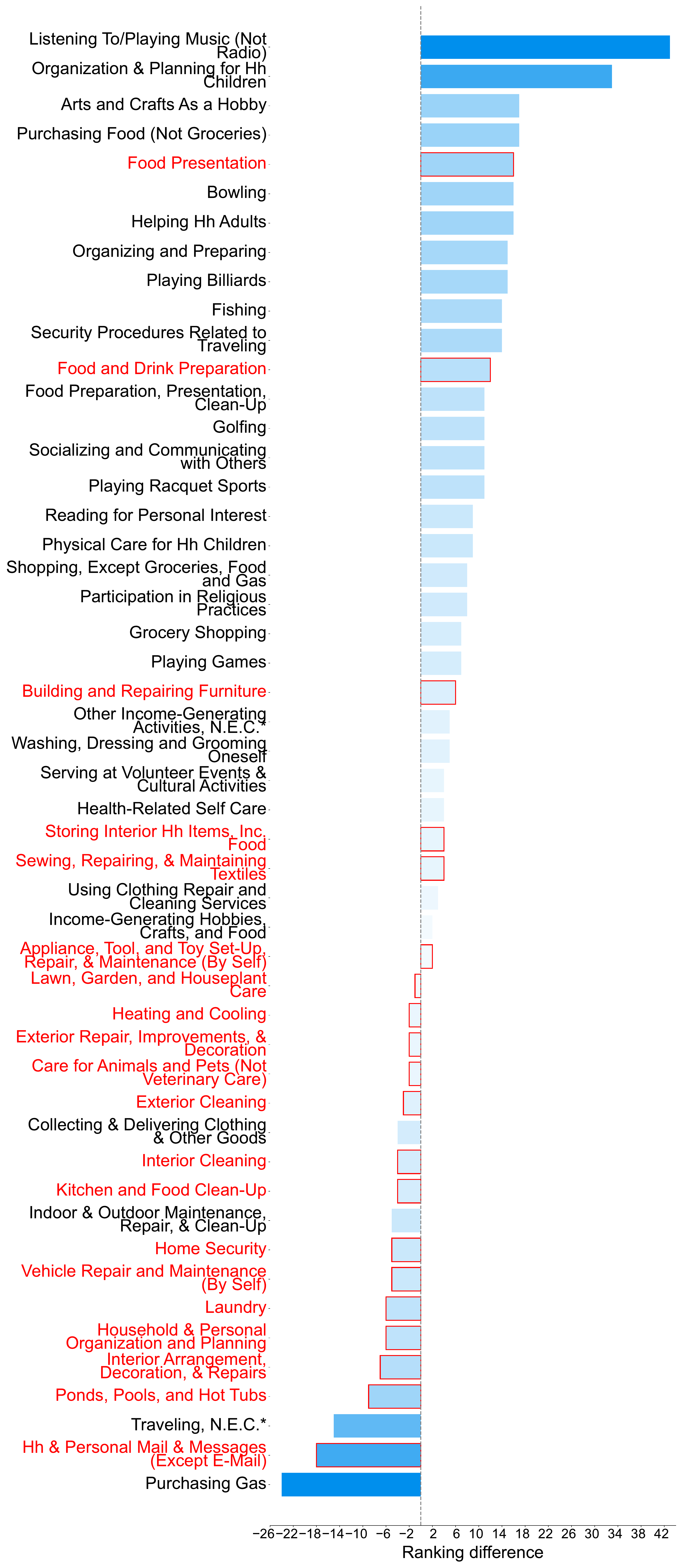}
\includegraphics[width=\rankdiffr\textwidth]{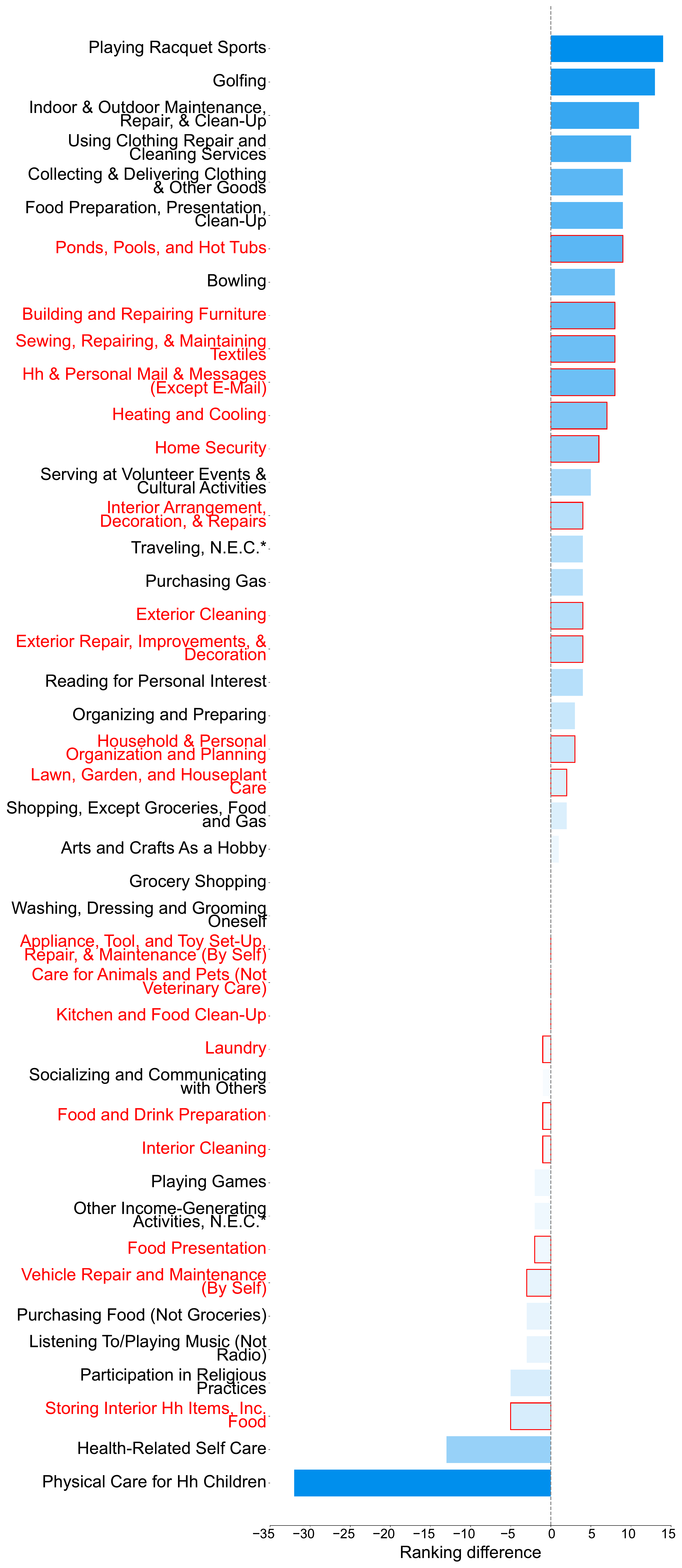}
\includegraphics[width=\rankdiffr\textwidth]{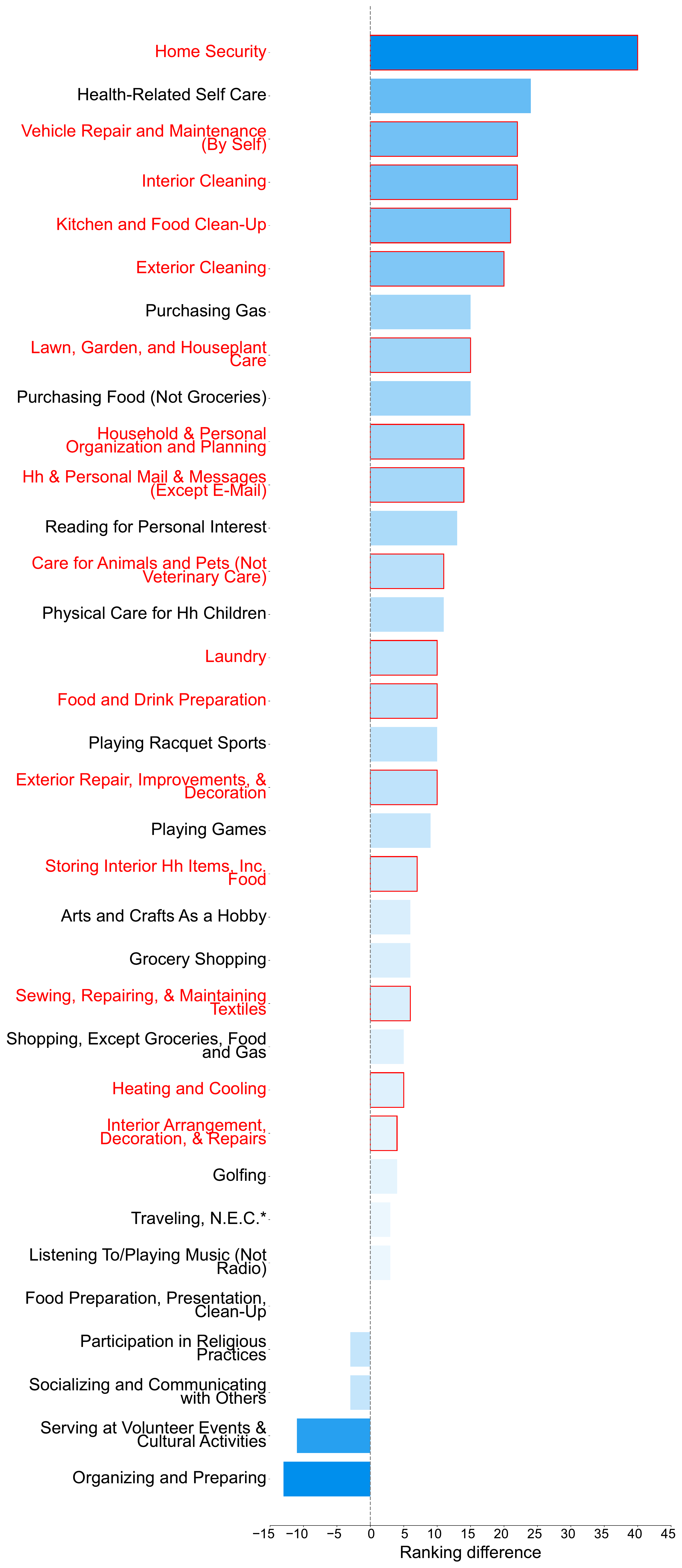}
\includegraphics[width=\rankdiffr\textwidth]{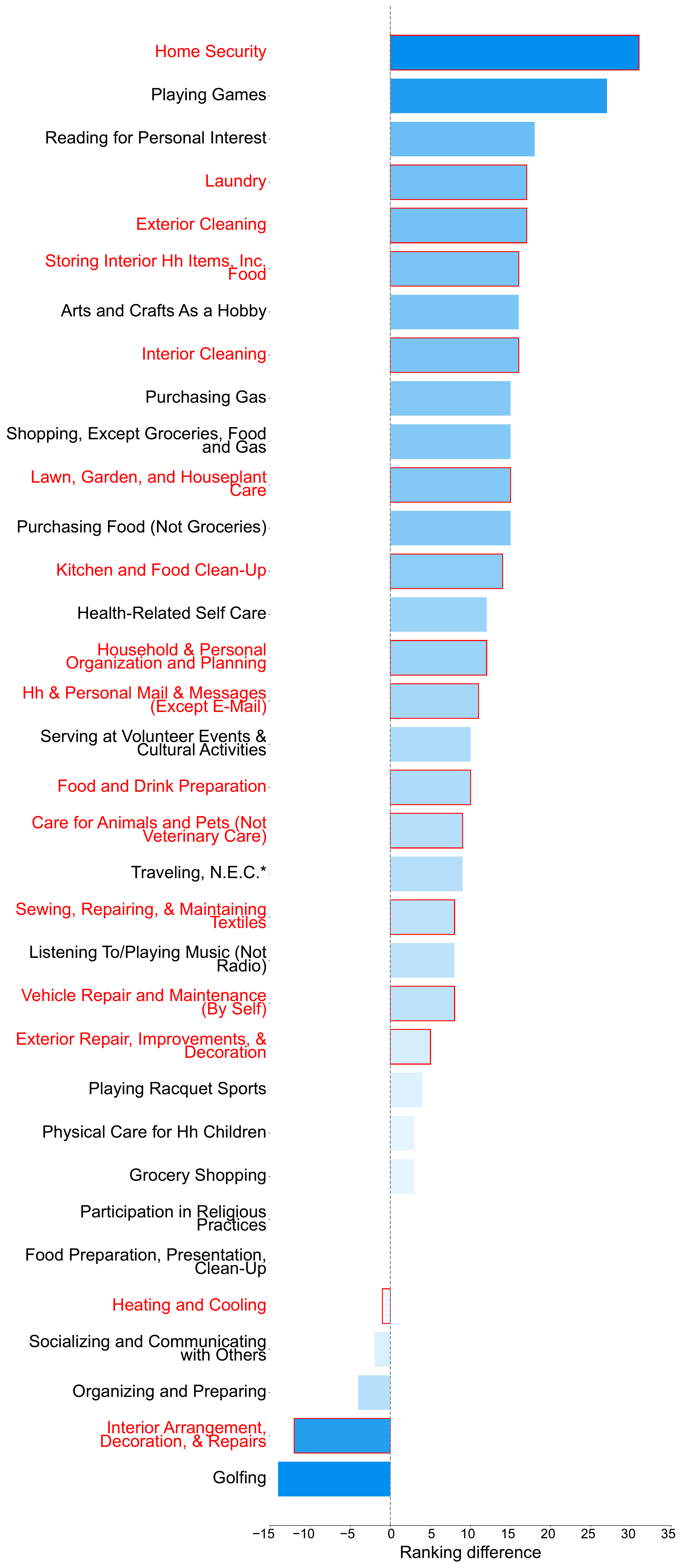}\\
\vspace{1em}
\includegraphics[width=\rankdiffr\textwidth]{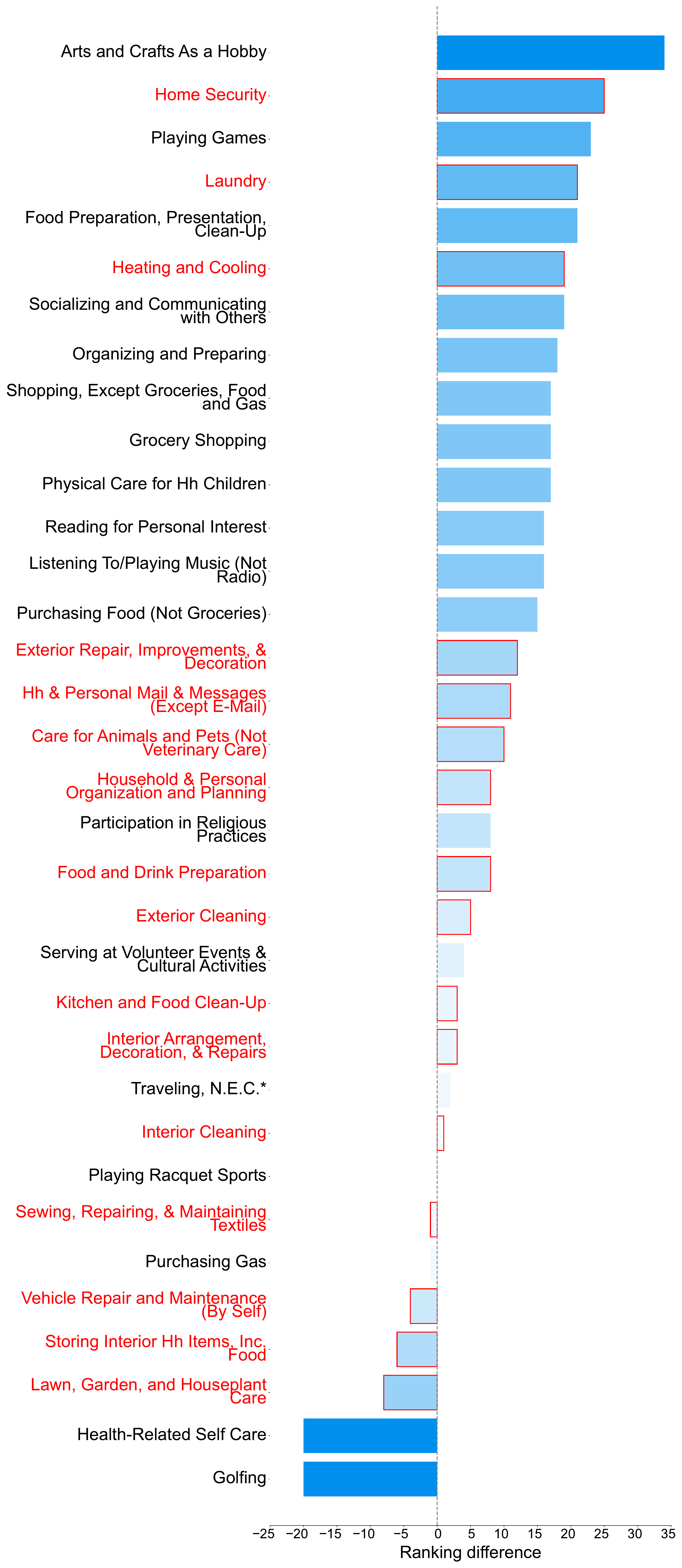}
\includegraphics[width=\rankdiffr\textwidth]{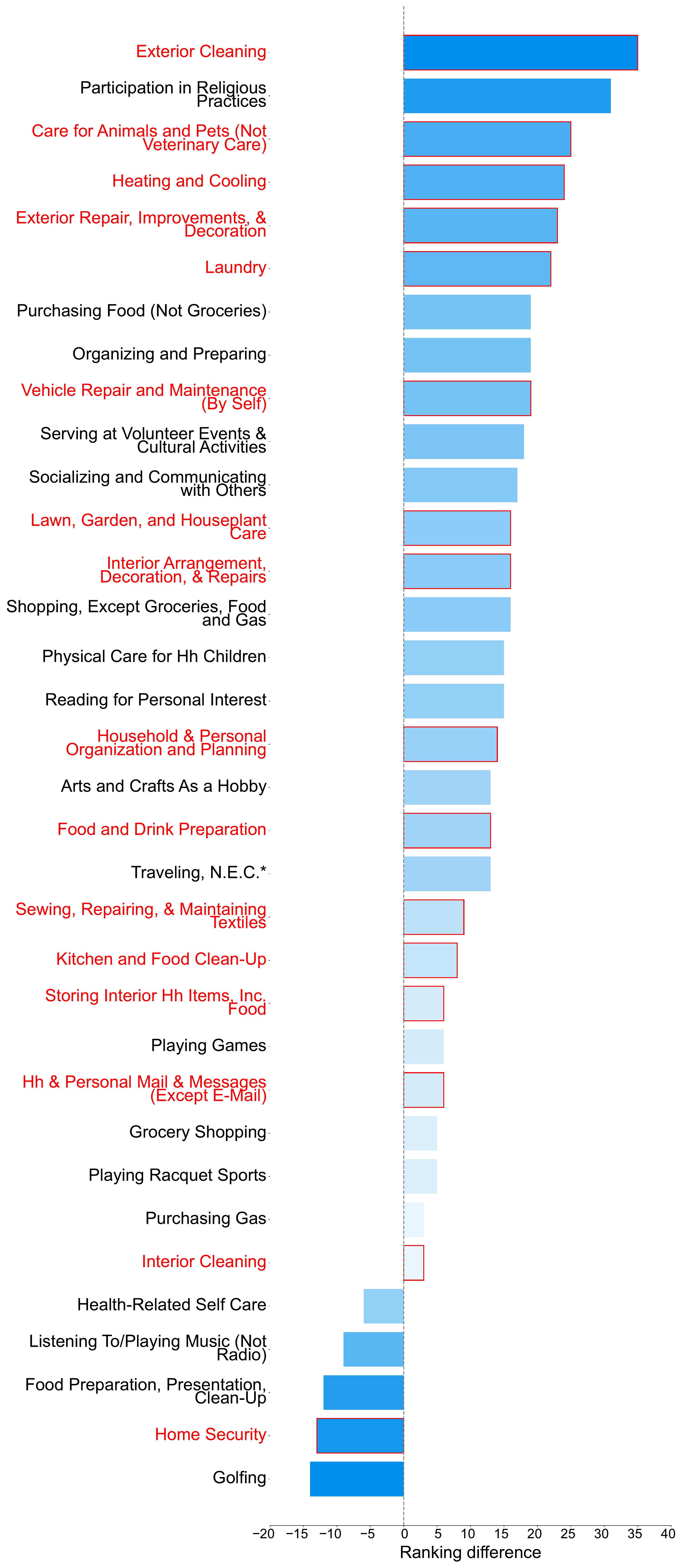}
\includegraphics[width=\rankdiffr\textwidth]{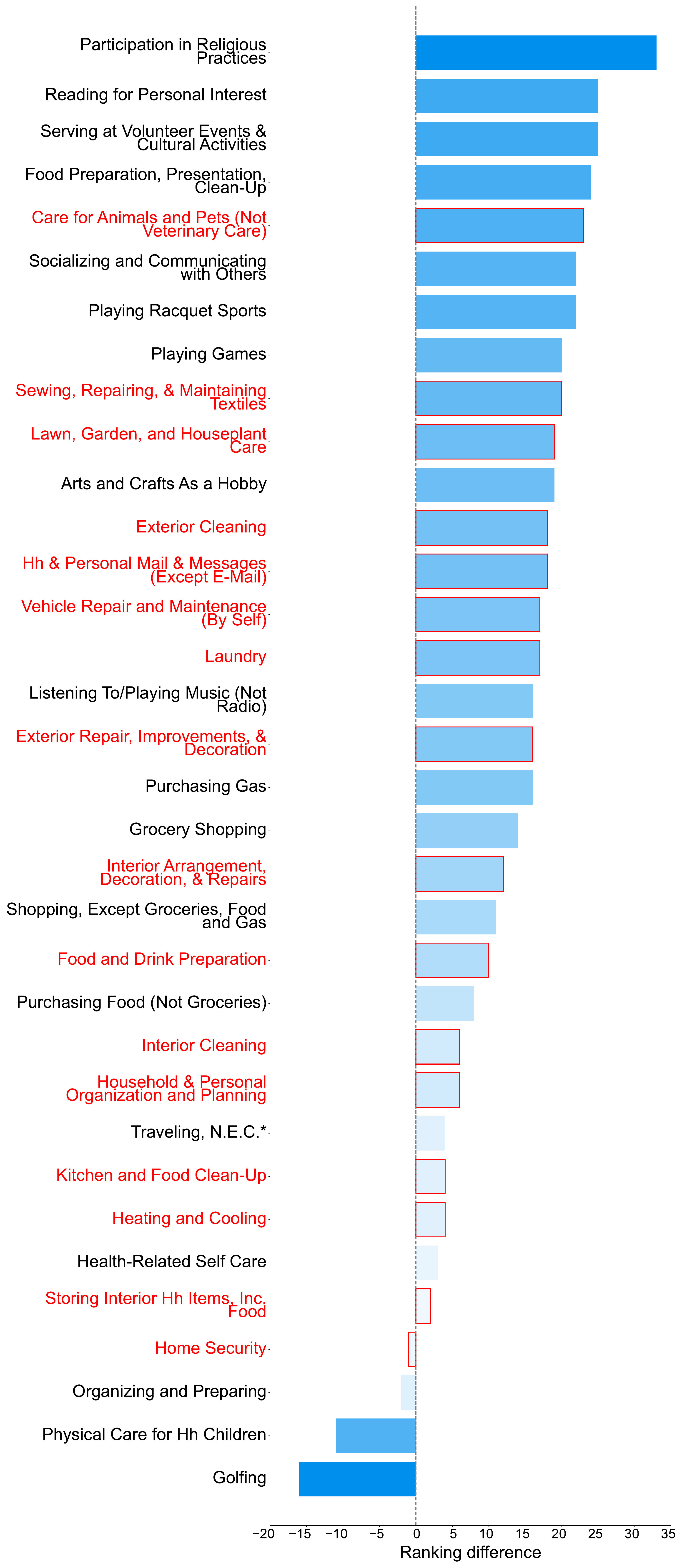}
\includegraphics[width=\rankdiffr\textwidth]{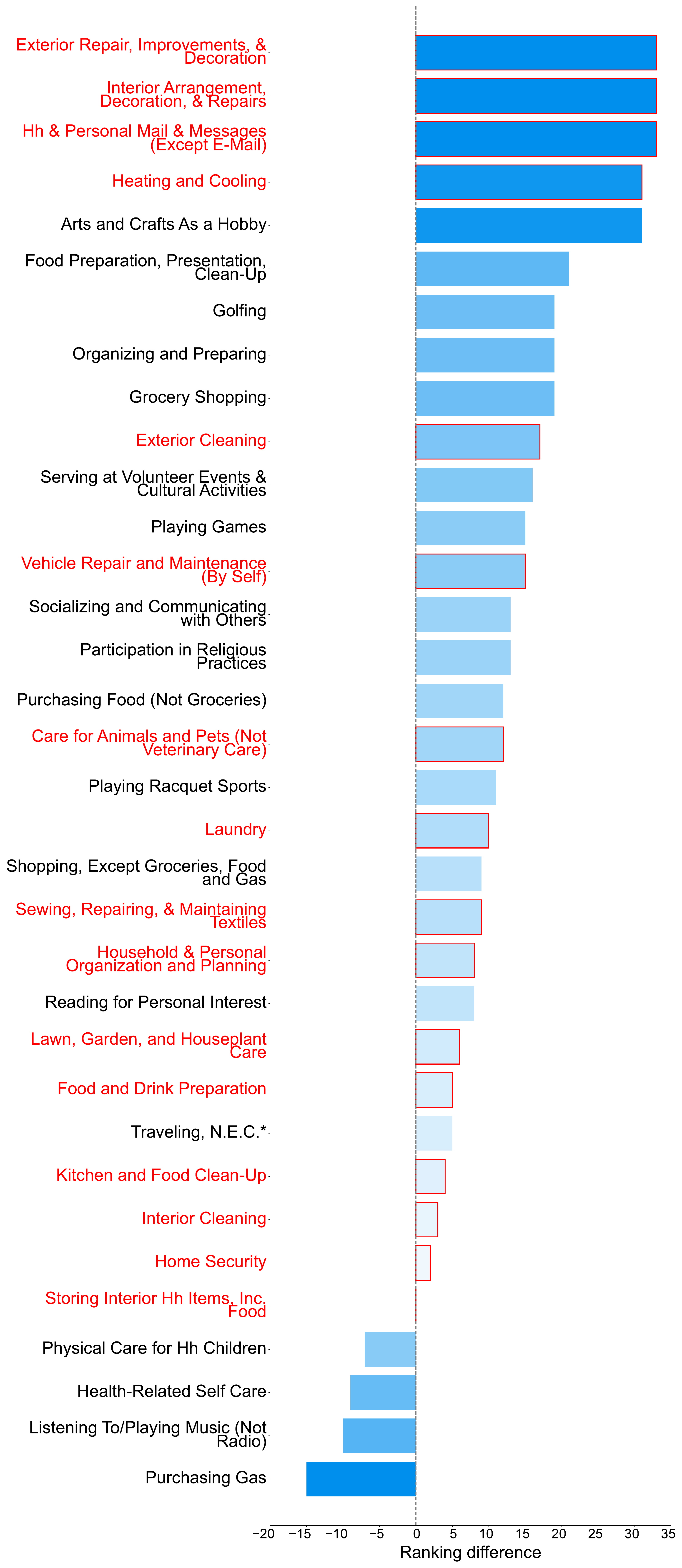}

\caption{\textbf{Difference in ranking} for all activities between the \textbf{general population and high-income} subset; Darker color indicates higher rank differences; Red labels and bar lines indicate activities of the \textit{Household Activities} subset; Differences for Desire for Automation (top row, 1st from left), Time spent (top row, 2nd from left), Happiness (top row, 3rd from left), Meaningfulness (top row, most right), Painfulness (bottom row, 1st from left), Sadness (bottom row, 2nd from left), Stressfulness (bottom row, 3rd from left) and Tiredness (bottom row, most right).}

\end{figure*}

% mid

\begin{figure*}[t!]
\centering
\includegraphics[width=\rankdiffr\textwidth]{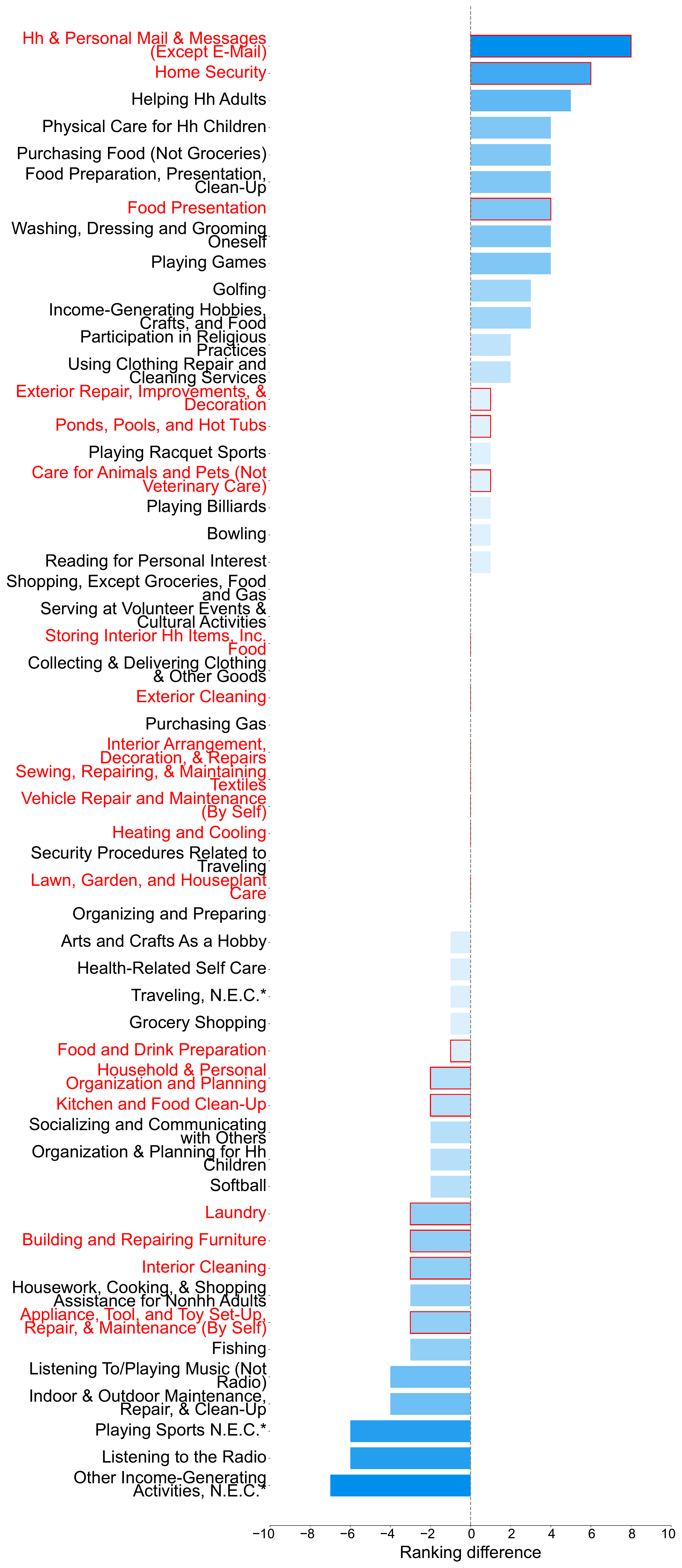}
\includegraphics[width=\rankdiffr\textwidth]{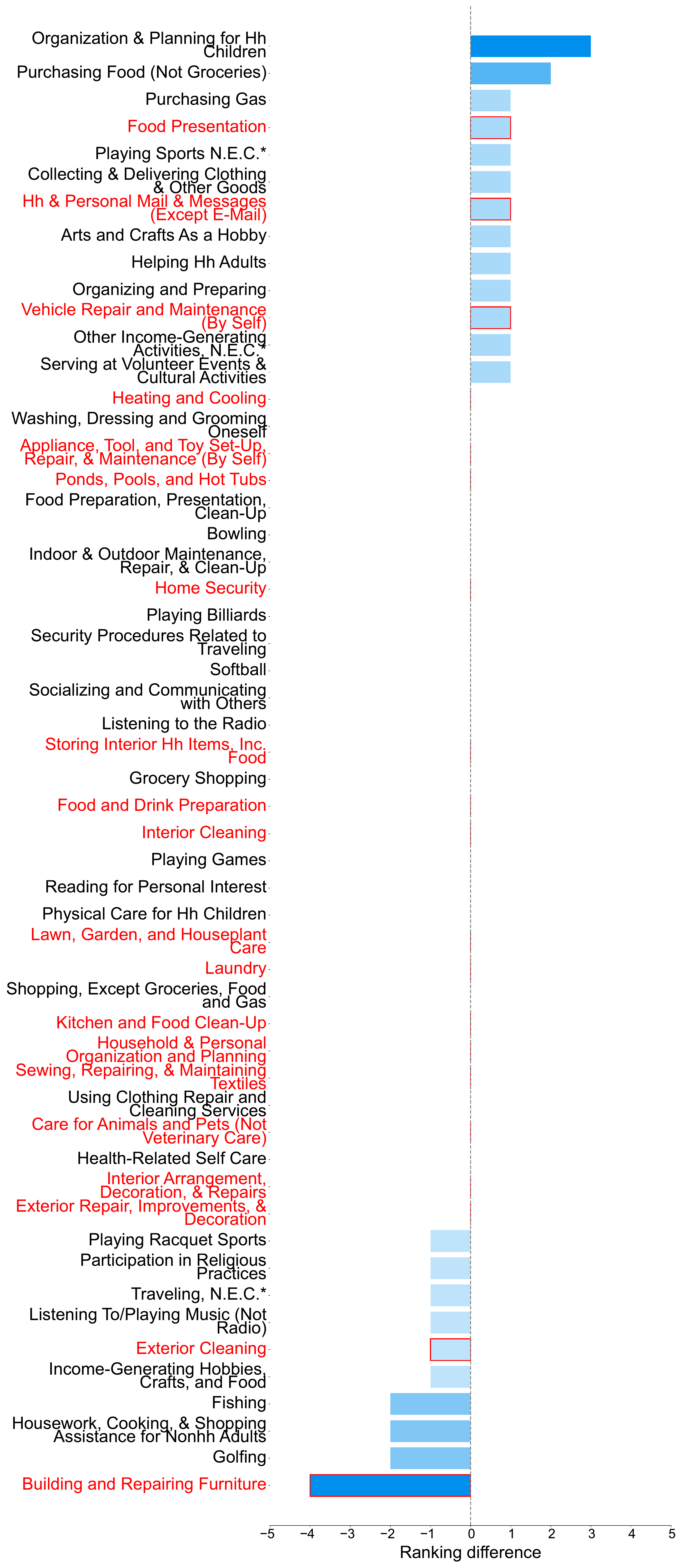}
\includegraphics[width=\rankdiffr\textwidth]{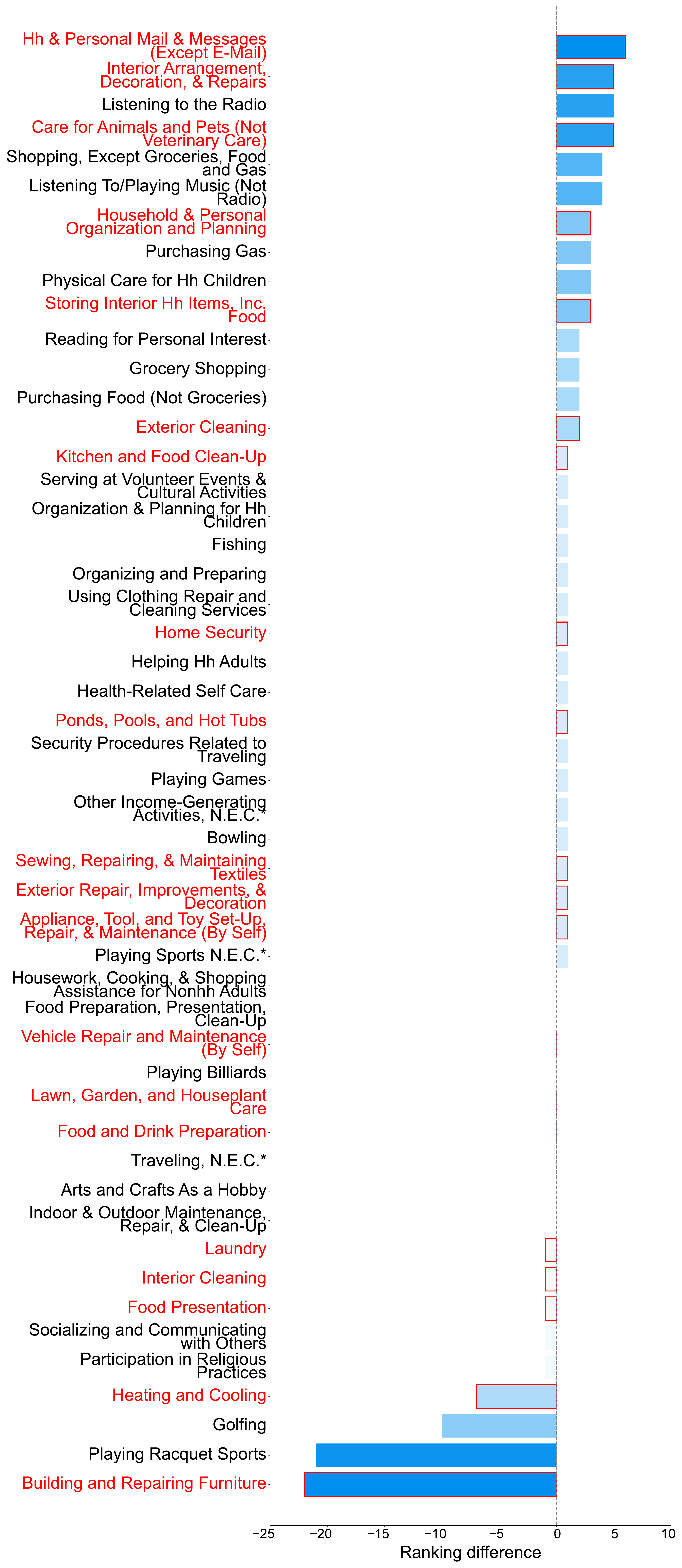}
\includegraphics[width=\rankdiffr\textwidth]{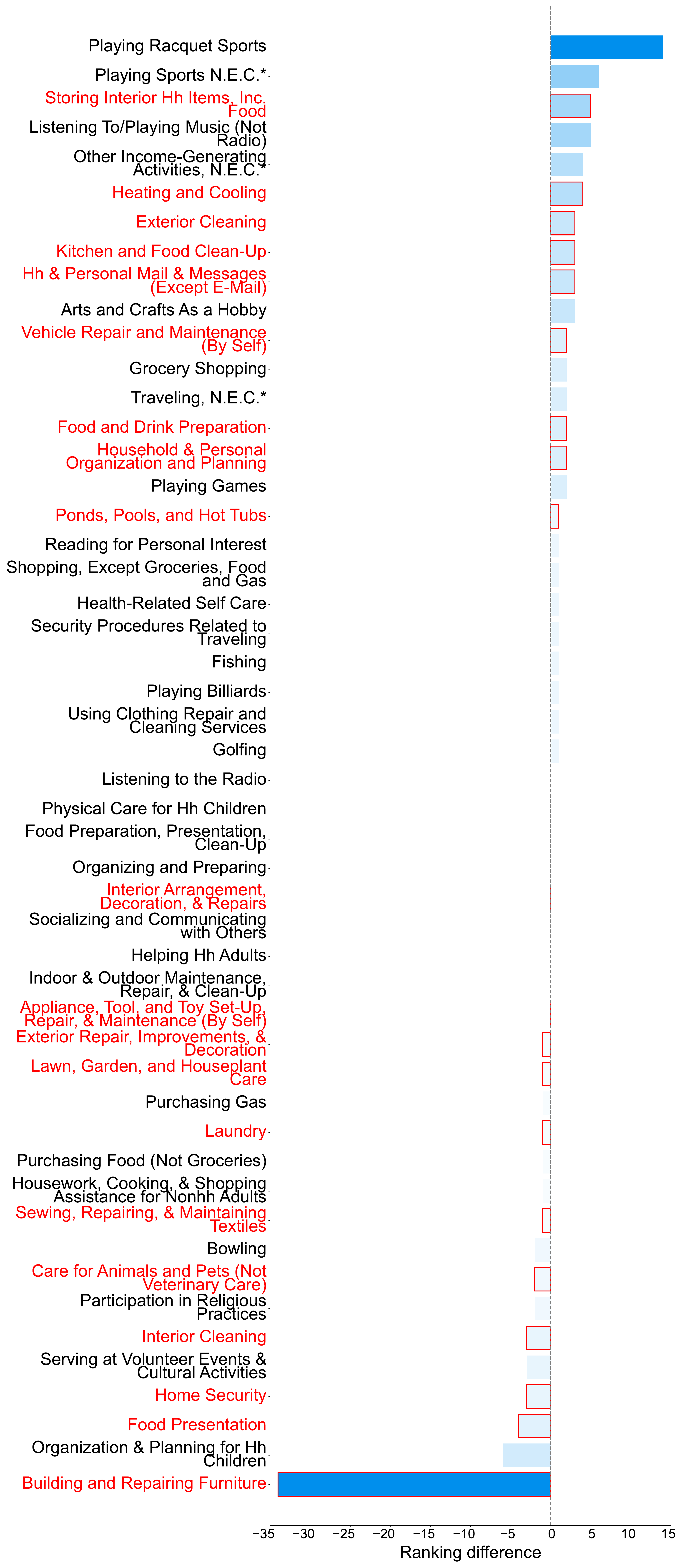}\\
\vspace{1em}
\includegraphics[width=\rankdiffr\textwidth]{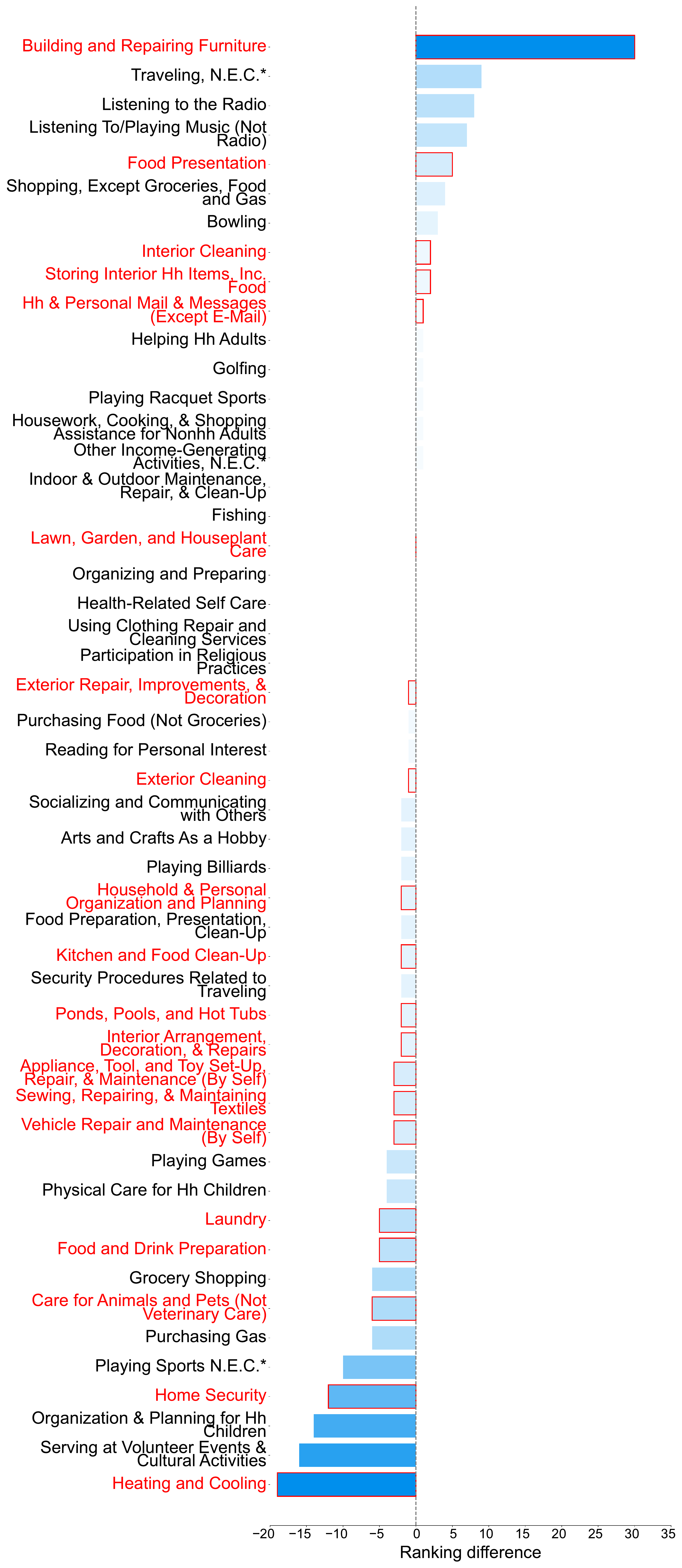}
\includegraphics[width=\rankdiffr\textwidth]{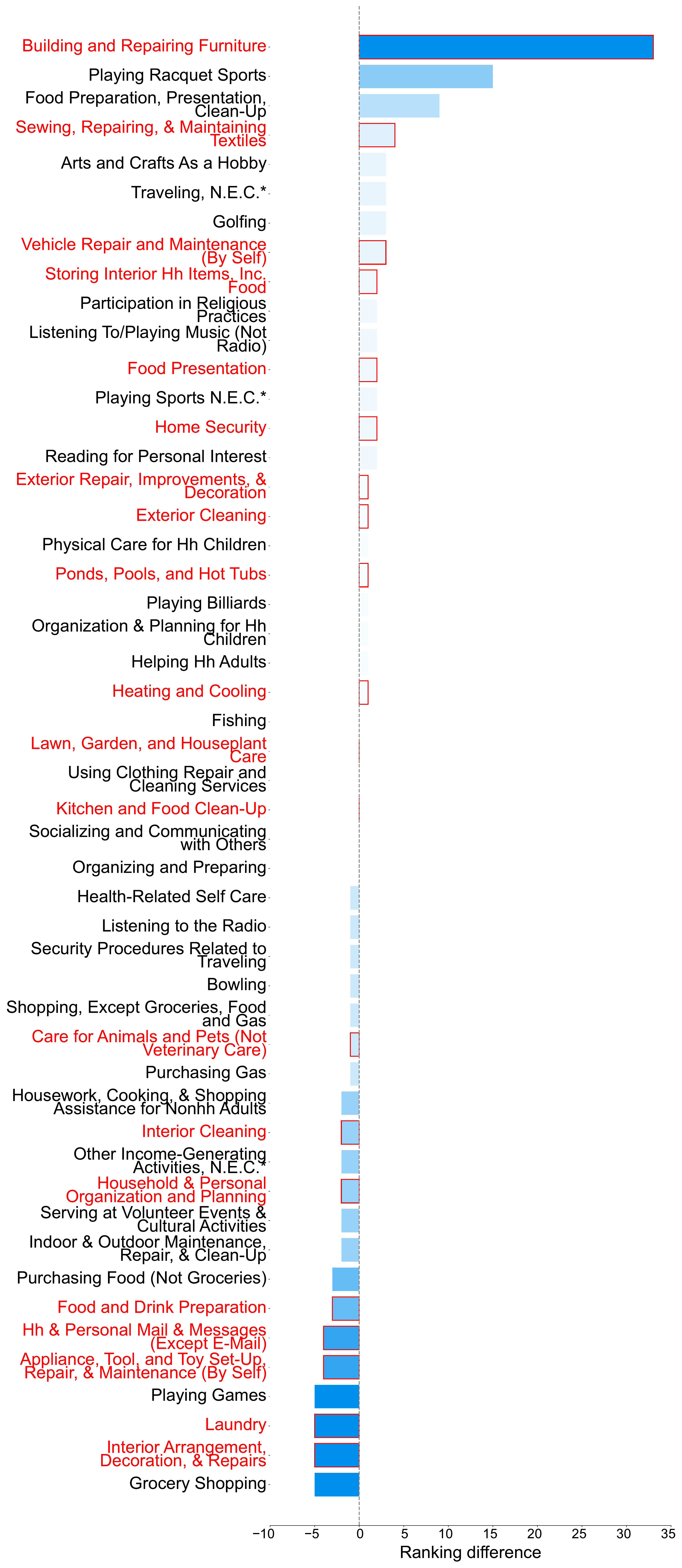}
\includegraphics[width=\rankdiffr\textwidth]{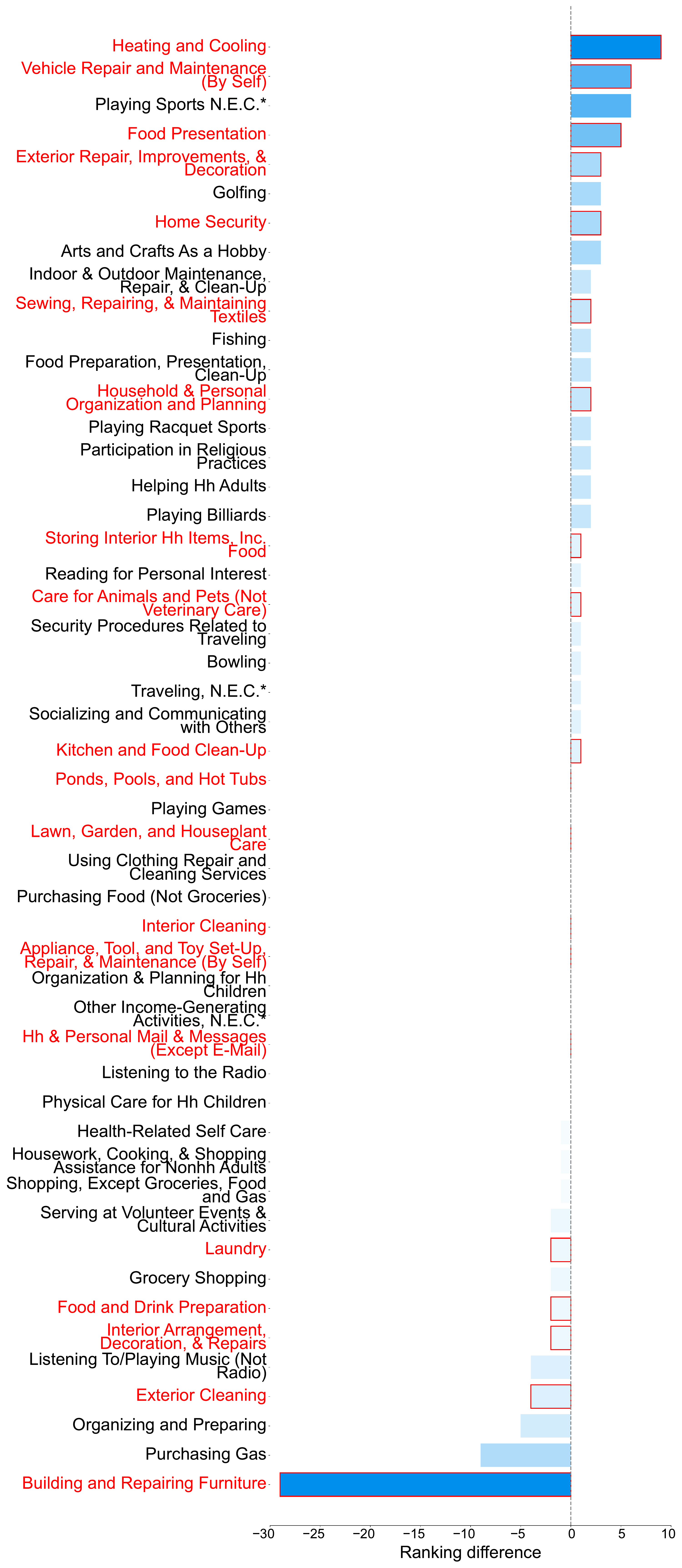}
\includegraphics[width=\rankdiffr\textwidth]{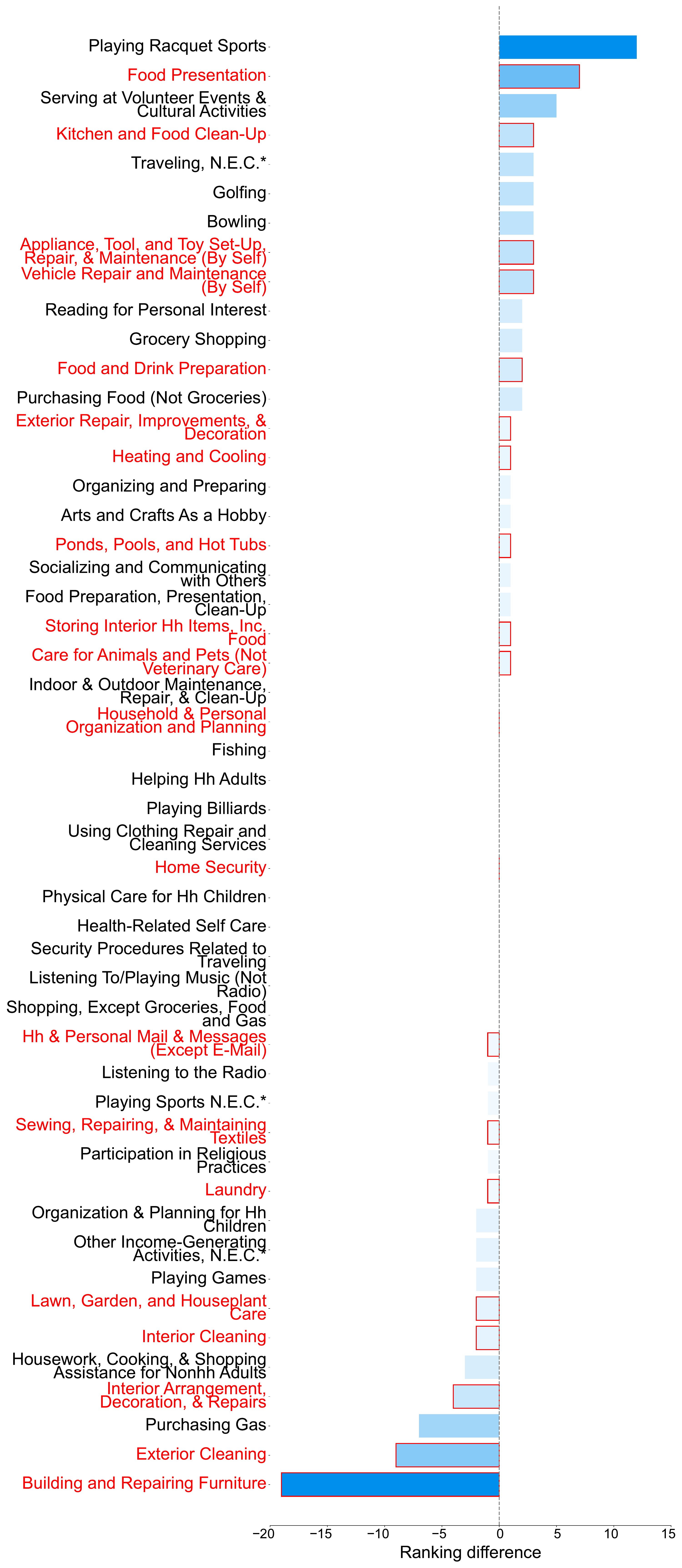}

\caption{\textbf{Difference in ranking} for all activities between the \textbf{general population and mid-income} subset; Darker color indicates higher rank differences; Red labels and bar lines indicate activities of the \textit{Household Activities} subset; Differences for Desire for Automation (top row, 1st from left), Time spent (top row, 2nd from left), Happiness (top row, 3rd from left), Meaningfulness (top row, most right), Painfulness (bottom row, 1st from left), Sadness (bottom row, 2nd from left), Stressfulness (bottom row, 3rd from left) and Tiredness (bottom row, most right).}

\end{figure*}

% low

\begin{figure*}[t!]
\centering
\includegraphics[width=\rankdiffr\textwidth]{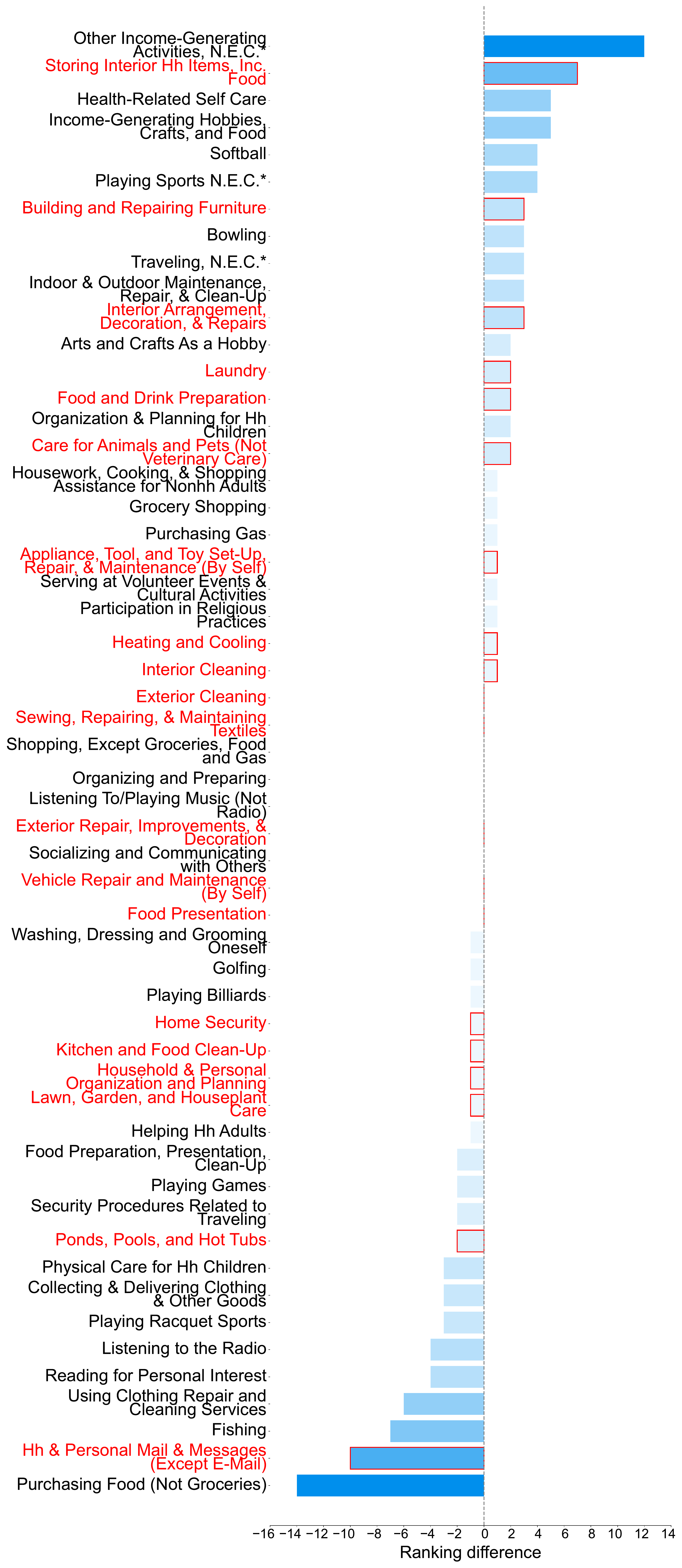}
\includegraphics[width=\rankdiffr\textwidth]{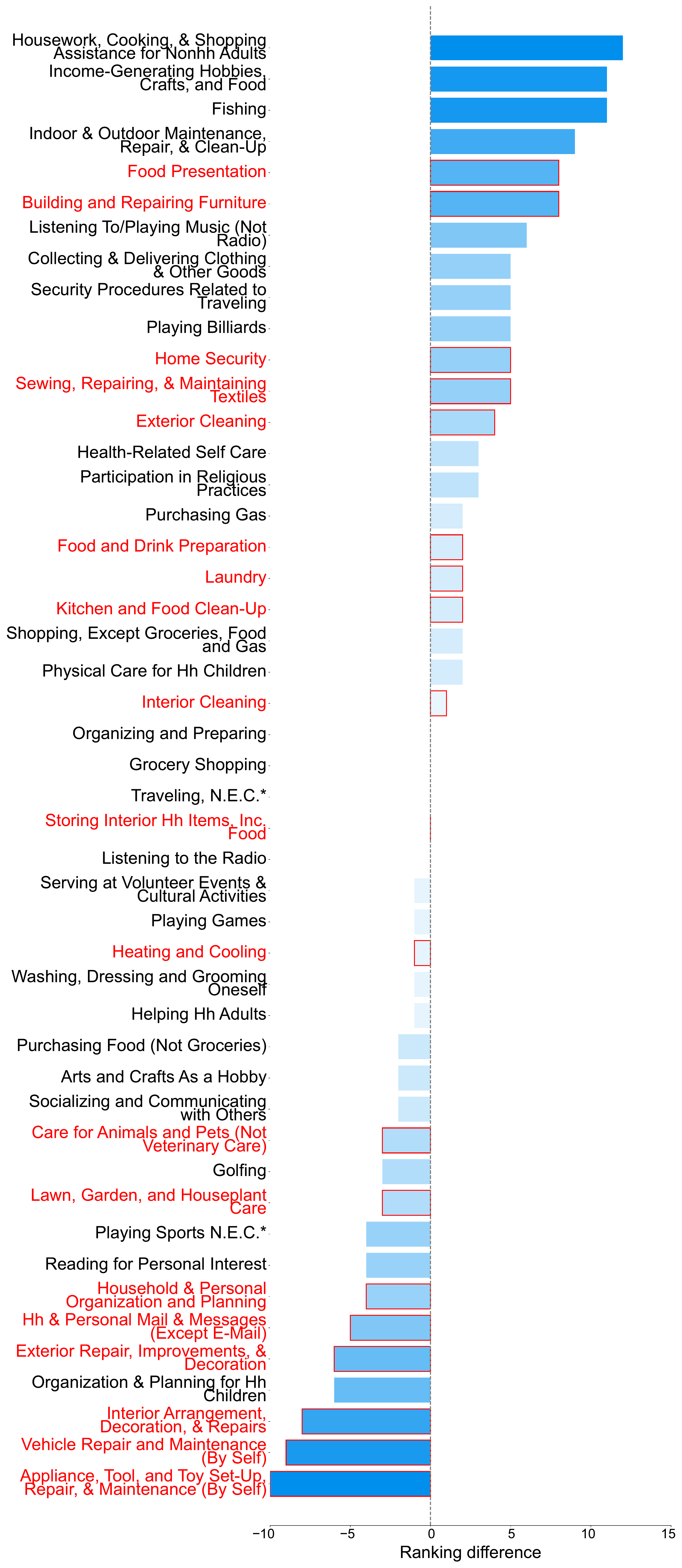}
\includegraphics[width=\rankdiffr\textwidth]{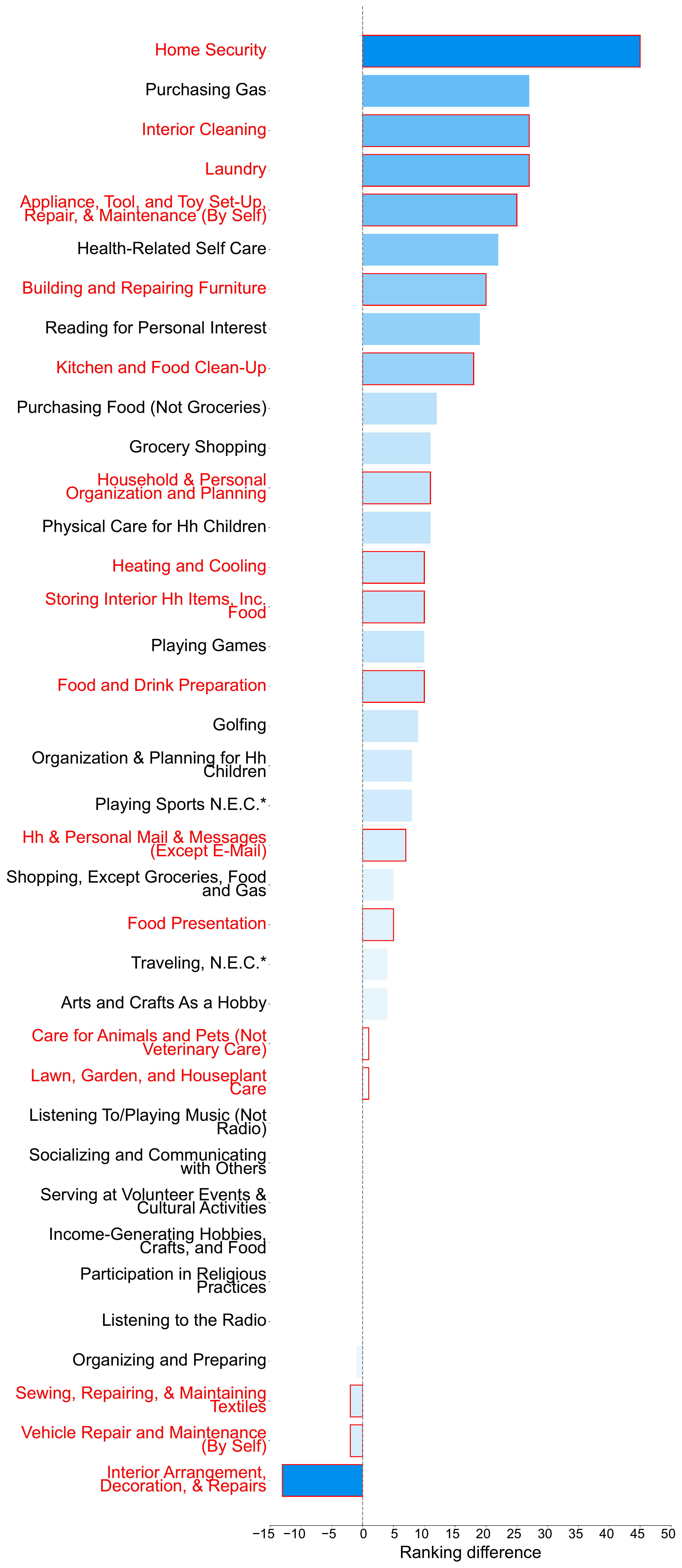}
\includegraphics[width=\rankdiffr\textwidth]{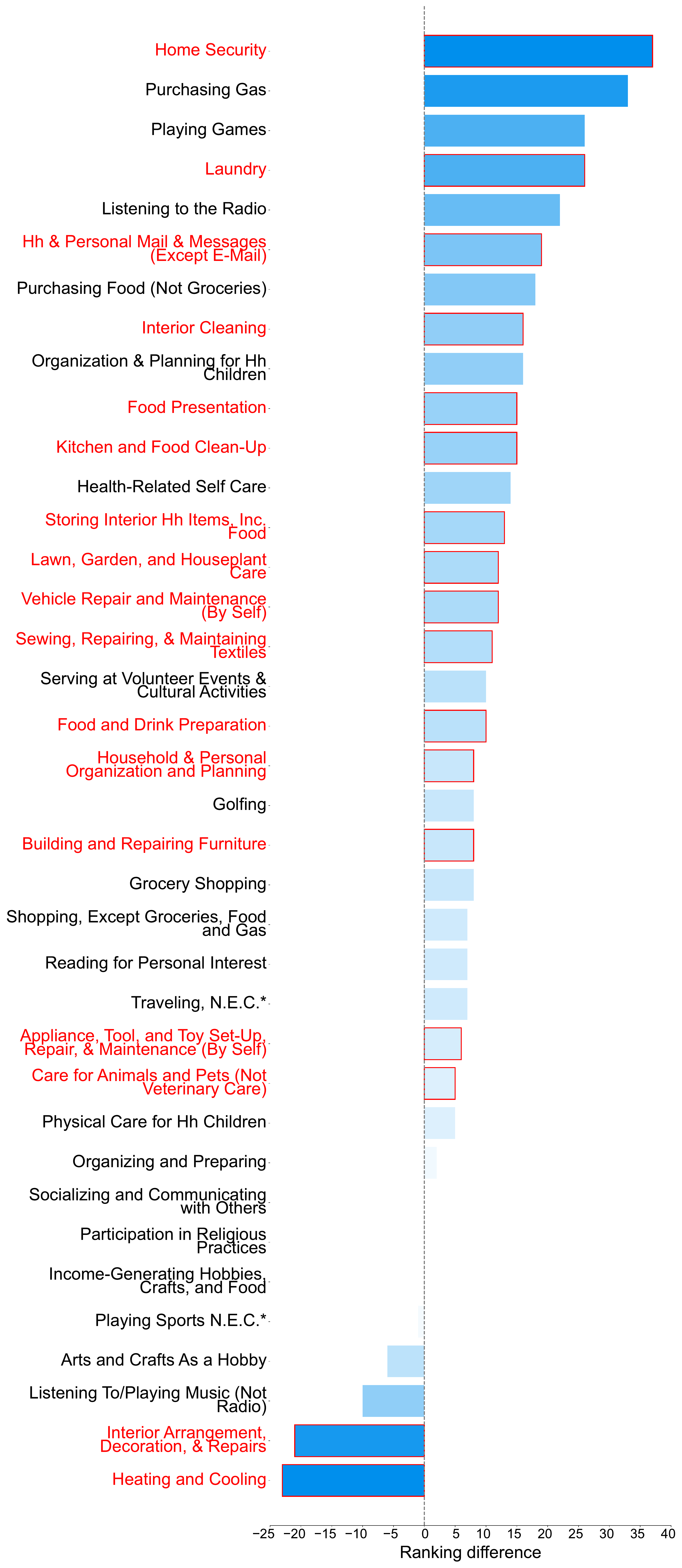}\\
\vspace{1em}
\includegraphics[width=\rankdiffr\textwidth]{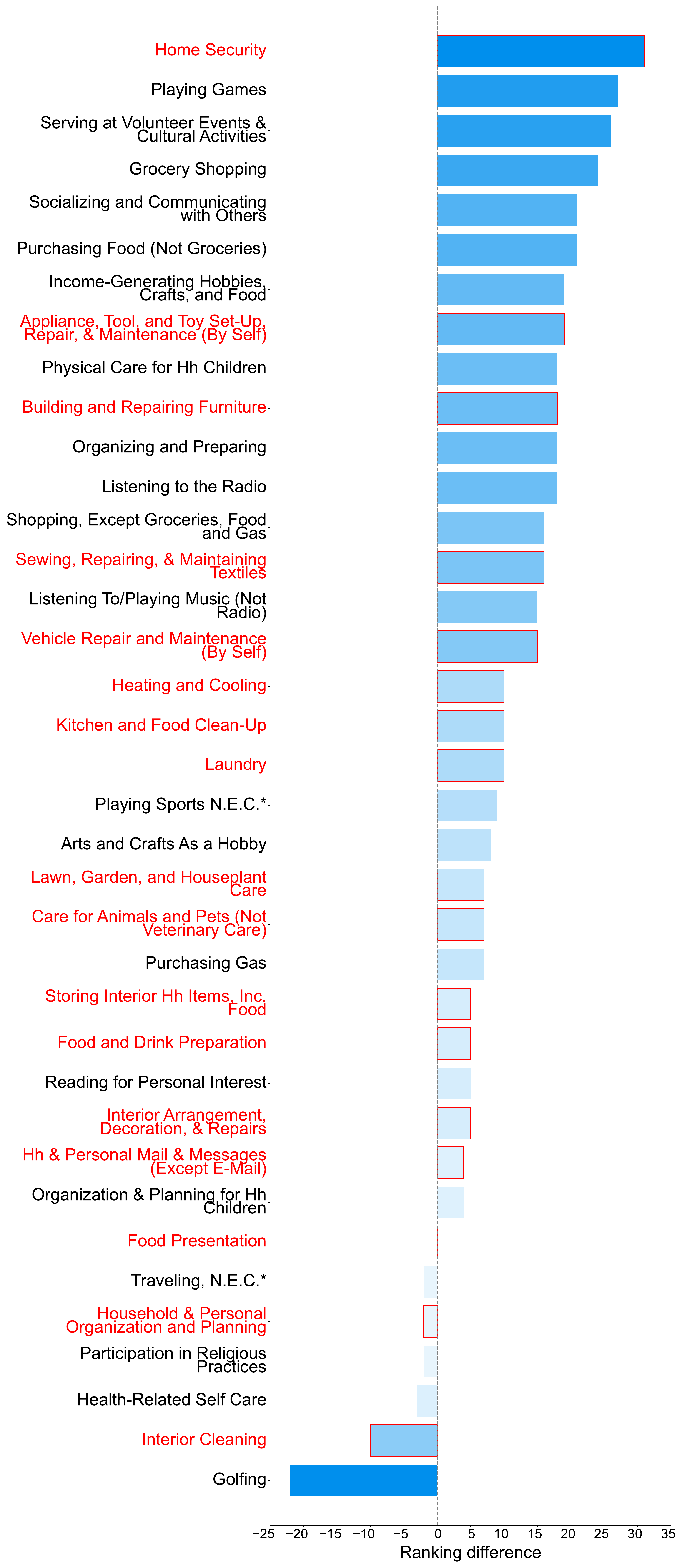}
\includegraphics[width=\rankdiffr\textwidth]{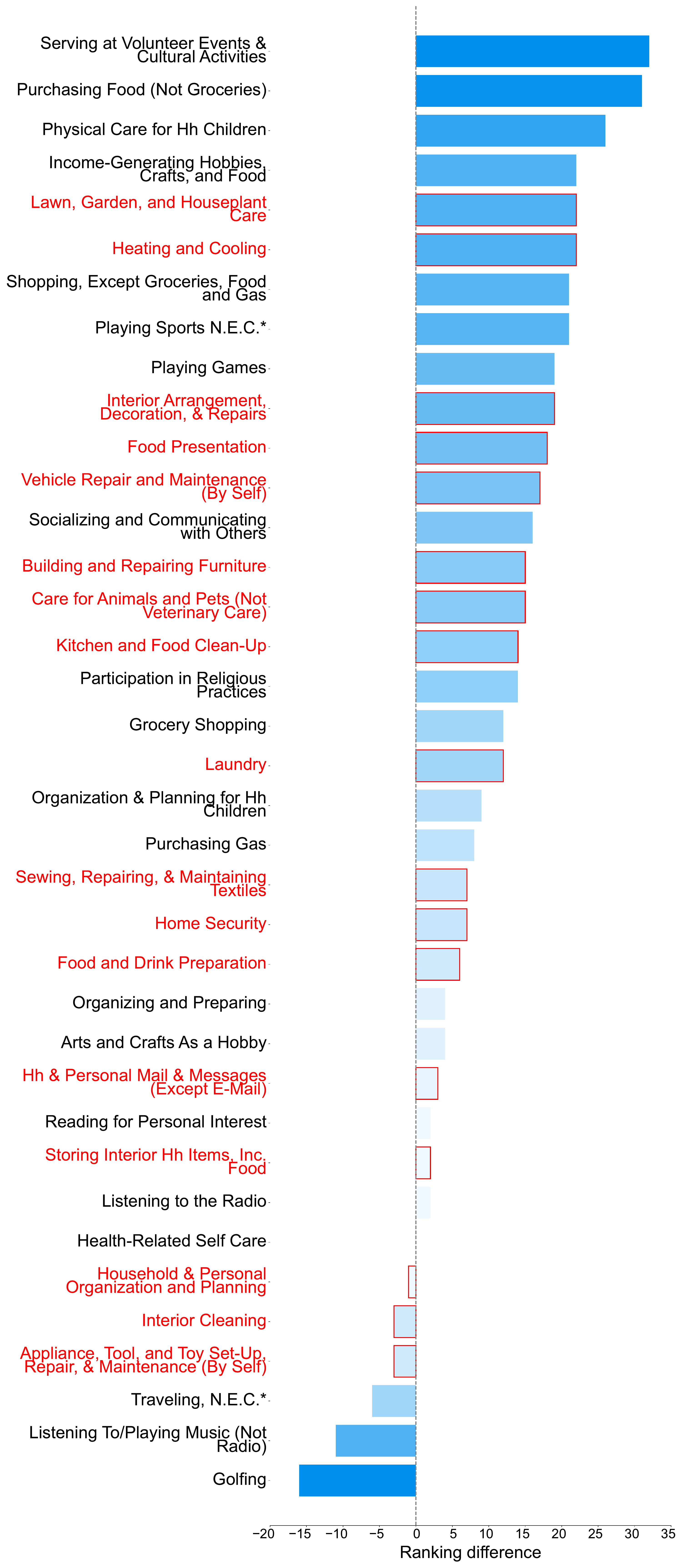}
\includegraphics[width=\rankdiffr\textwidth]{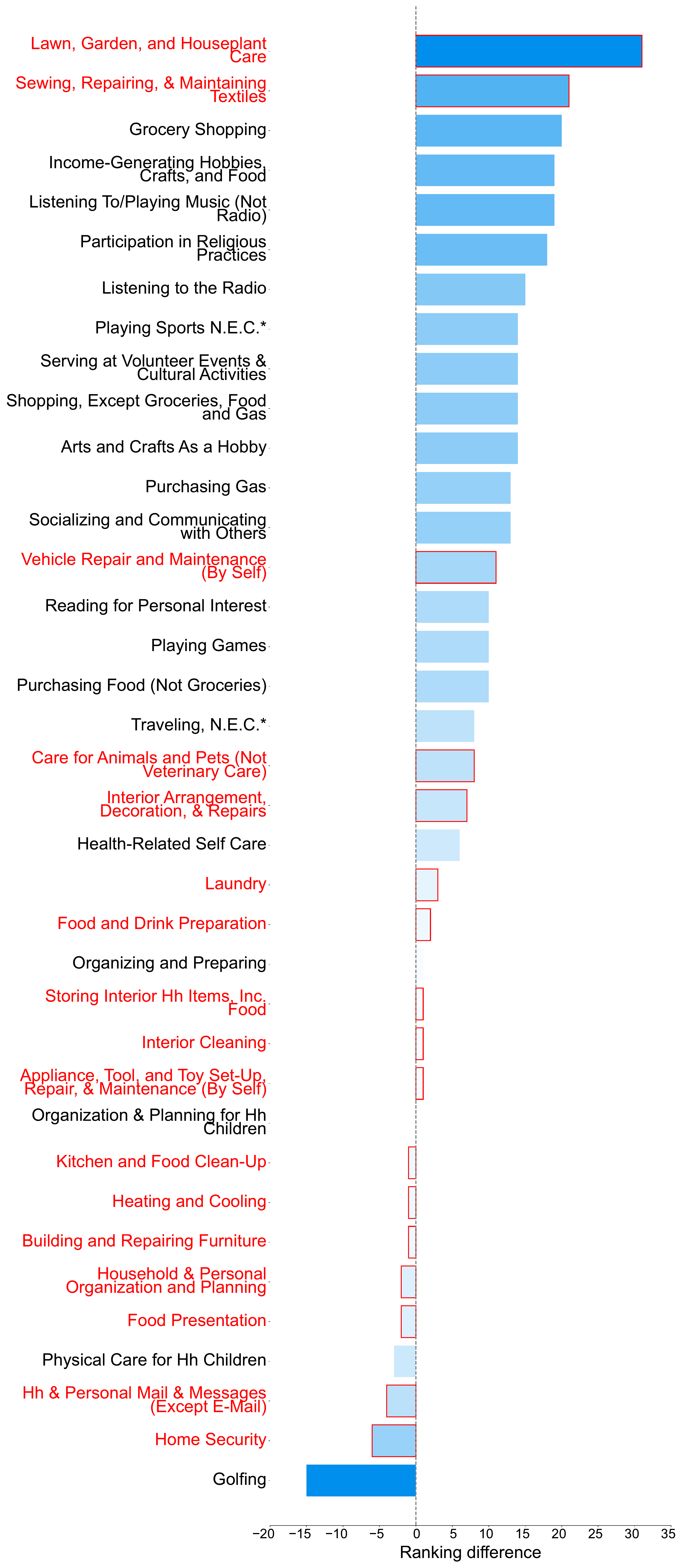}
\includegraphics[width=\rankdiffr\textwidth]{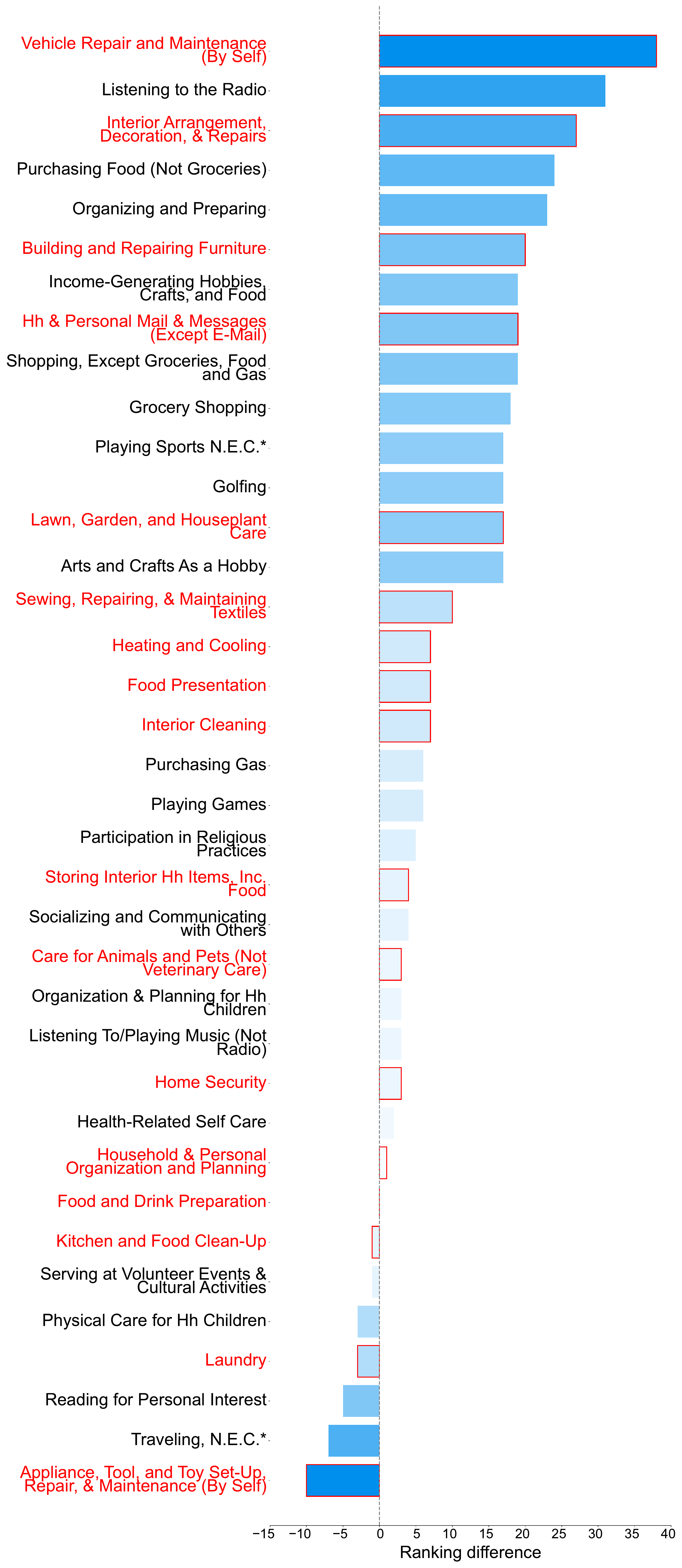}

\caption{\textbf{Difference in ranking} for all activities between the \textbf{general population and low-income} subset; Darker color indicates higher rank differences; Red labels and bar lines indicate activities of the \textit{Household Activities} subset; Differences for Desire for Automation (top row, 1st from left), Time spent (top row, 2nd from left), Happiness (top row, 3rd from left), Meaningfulness (top row, most right), Painfulness (bottom row, 1st from left), Sadness (bottom row, 2nd from left), Stressfulness (bottom row, 3rd from left) and Tiredness (bottom row, most right).}

\end{figure*}

\end{document}